\documentclass{article}
\usepackage{natbib}
\usepackage{epubtk}    
\usepackage{booktabs}  
\usepackage{graphicx}  

\newcommand{\be}{\begin{equation}}
\newcommand{\ee}{\end{equation}}
\newcommand{\bea}{\begin{eqnarray}}
\newcommand{\eea}{\end{eqnarray}}
\newcommand{\dd}{{\,\mathrm{d}}}
\newcommand{\dnu}{$\Delta\nu$}
\newcommand{\teff}{$T_{\rm eff}$}
\newcommand{\logg}{$\log g$}
\newcommand{\numax}{$\nu_{\max}$}
\newcommand{\msun}{\ensuremath{M_{\odot}}}
\newcommand{\rsun}{\ensuremath{R_{\odot}}}
\newcommand{\lsun}{\ensuremath{L_{\odot}}}

\begin{document}

\title{Global Seismology of the Sun}

\author{\epubtkAuthorData{Sarbani Basu}{
Department of Astronomy, Yale University \\
PO Box 208101, New Haven, CT 06520-8101, USA}
{sarbani.basu@yale.edu}
{
}
}

\date{}
\maketitle

\begin{abstract}
The seismic study of the Sun and other stars offers a unique window into the
interior of these stars. Thanks to helioseismology, we know the structure of
the Sun to admirable precision. In fact, our knowledge is good enough to
use the Sun as a laboratory. We have also been able to study the dynamics of the
Sun in great detail. Helioseismic data  also allow us to probe the
changes that take place in the Sun as solar activity waxes and wanes.
The seismic study of stars other than the Sun is a fairly new endeavour,
but we are making great strides in this field.
In this review I discuss some of the techniques used in helioseismic analyses
and the results obtained using those techniques. In this review I focus
on results obtained with global helioseismology, i.e., the study of the Sun using its normal modes
of oscillation. I also briefly touch upon asteroseismology, the seismic study of stars
other than the Sun, and discuss how seismic data of others stars are interpreted.
\end{abstract}

\epubtkKeywords{Sun: Interior, Sun: Helioseismology, Sun: Rotations, Stars: Asteroseismology}

\newpage


\section{Introduction}
\label{sec:intro}

The Sun and other stars oscillate in their normal modes. The analysis and
interpretation of the properties of these modes in terms of the underlying
structure and dynamics of the stars is referred to as global seismology. Global
seismology has given us an unprecedented window into the structure and dynamics
of the Sun and stars.  Arthur Eddington began his book \textit{The Internal
Constitution of the Stars} lamenting the fact that the deep interior of the Sun
and stars is more inaccessible than the depths of space since we so not have an
``appliance'' that can \textit{``$\ldots$ pierce through the outer layers of a
star and test the conditions within''}. Eddington went on to say that perhaps
the only way of probing the interiors of the Sun and stars is to use our
knowledge of basic physics to determine what the structure of a star should be.
While this is still the dominant approach in the field of stellar astrophysics,
in global seismology we have the means of piercing the outer layers of a star
to probe the structure within.

The type of stellar oscillations that are used in helio- and asteroseismic
analyses have very low amplitudes. These oscillations are excited by the
convective motions in the outer convection zones of stars.  Such oscillations,
usually referred to as solar-like oscillations, can for most purposes be
described using the theory of linear, adiabatic, oscillations.  The behaviour
of the modes on the stellar surface is described in terms of spherical
harmonics since these functions are a natural description of the normal modes
of a sphere.  The oscillations are labelled by three numbers, the radial order
$n$, the degree $\ell$ and the azimuthal order $m$. The radial order $n$ can be
any whole number and is the number of nodes in the radial direction.  Positive
values of $n$ are used to denote acoustic modes, i.e., the so-called p~modes (p
for pressure, since the dominant restoring force for these modes is provided by
the pressure gradient). Negative values of $n$ are used to denote modes for
which buoyancy provides the main restoring force. These are usually referred to
as g~modes (g for gravity).  Modes with $n=0$ are the so-called fundamental or
f~modes.  These are essentially surface gravity modes whose frequencies depend
predominantly on the wave number and the surface gravity.  The degree $\ell$
denotes the number of nodal planes that intersect the surface of a star and $m$
is the number of nodal planes perpendicular to equator.

For a spherically symmetric star, all modes with the same degree $\ell$ and
order $n$ have the same frequency. Asphericities such as rotation and magnetic
fields lift this degeneracy and cause `frequency splitting' making the
frequencies $m$-dependent. It is usual to express the frequency $\nu_{n\ell m}$
of a mode in terms of `splitting coefficients':
\begin{equation}
{\omega_{n\ell m}\over2\pi}=\nu_{n\ell m}=\nu_{n\ell }+\sum_{j=1}^{j_{\max}} a_j(n\ell )\mathcal{P}^{n\ell }_j(m),
\label{eq:split}
\end{equation}
where, $a_j$ are the splitting or `$a$' coefficients and $\mathcal{P}$ are
suitable polynomials.  For slow rotation the central frequency $\nu_{n\ell }$
depends only on structure, the odd-order $a$ coefficients depend on rotation,
and the even-order $a$ coefficients depend on structural asphericities,
magnetic fields, and second order effects of rotation. 

The focus of this review is global helioseismology -- its theoretical
underpinnings as well as what it has taught us about the Sun. The last section
is devoted to asteroseismology; the theoretical background of asteroseismology
is the same as that of helioseismology, but the diverse nature of different
stars makes the field quite distinct. This is, of course, not the first review
of helioseismology. To get an idea of the changing nature of the field, the
reader is referred to earlier reviews by \citet{mjt1998}, \citet{jcd2002},
\citet{jcd2004} and \citet{dog2013S}, which along with descriptions of the then
state-of-art of the field, also review the early history of helioseismology.
Since this review is limited to \emph{global} seismology only, readers are
referred to the \textit{Living Reviews in Solar Physics} contribution of
\citet{gizonandbirch2005} for a review of \emph{local} helioseismology.

Solar oscillations were first discovered by \citet{leightonetal1962} and
confirmed by \citet{evans1962}. Later observations, such that those by
\citet{frazier1968} indicated that the oscillations may not be mere surface
phenomena. Subsequently \citet{ulrich1970} and \citet{leibacherandstein1971}
proposed that the observations could be interpreted as global oscillation modes
and predicted that oscillations would form ridges in a wave-number v/s
frequency diagram.  The observations of \citet{deubner1975} indeed showed such
ridges.  \citet{rhodesetal1977} reported similar observations.  Neither the
observations of \citet{deubner1975}, nor those of \citet{rhodesetal1977}, could
resolve individual modes of solar oscillations. Those had to wait for
\citet{clav1979} who, using Doppler-velocity observations integrated over the
solar disk, were able to resolve the individual modes of oscillations
corresponding to the largest horizontal wavelengths, i.e., the truly global
modes. They found a series of almost equidistant peaks in the power spectrum,
just as was expected from theoretical models.

Helioseismology, the study of the Sun using solar oscillations, as we know it
today began when \citet{duvallandharvey1983}
determined frequencies of a reasonably large number of solar oscillation
modes covering a wide range of horizontal wavelengths. Since then,
many more sets of solar oscillation frequencies have been published. A large
fraction of the early work in the field was based on the frequencies determined
by \citet{libbrechtetal1990} from observations made at the Big Bear Solar Observatory (BBSO).

Accurate and precise estimates of solar frequencies require long, uninterrupted observations 
of the Sun, that are possible only with a network of telescopes. The 
Birmingham Solar Oscillation Network \citep[BiSON;][]{elsworthetal1991,chaplinetal2007}
and the International Research on the Interior of the Sun \citep[IRIS;][]{fossat1991} were two
of the first networks. Both these networks however, did Sun-as-a-star (i.e., one pixel) observations,
as a result they could only observe the low-degree modes with $\ell$ of 0--3. To overcome this
limitation, resolved-disc measurements were needed. 
This resulted in the construction of the Global Oscillation Network Group \citep[GONG:][]{hilletal1996}.
This ground-based network has been making full-disc observations of the Sun since 1995.
There were concurrent developments in space based observations and the Solar and Heliospheric Observatory
\citep[\textit{SoHO}:][]{domingoetal1995} was launched in 1995. Among \textit{SoHO}'s observing programme were 
three helioseismology related ones,
the `Solar Oscillations Investigation' (SOI) using the Michelson Doppler Imager\citep[MDI:][]{scherreretal1995}, 
`Variability of solar Irradiance and Gravity Oscillations' \citep[VIRGO:][]{lazreketal1997} 
and `Global Oscillations at Low Frequencies' \citep[GOLF:][]{gabrieletal1997}. 
Of these, MDI was capable of observing intermediate and high degree modes. BiSON and GONG continue
to observe the Sun from the ground; MDI stopped acquiring data in April 2011. MDI has been succeeded by the
Heliospheric and Magnetic Imager \citep[HMI:][]{hmi,scherreretal2012} on board the Solar Dynamics 
Observatory \citep[SDO:][]{pesnelletal2012}. In addition to obtaining better data, there have
also been improvements in techniques to determine mode frequencies from the data
\citep[e.g.,][]{tim2008, sylvain2013} which has facilitated detailed helioseismic analyses.
Descriptions of some of the early developments in the field can be found in the
proceedings of the workshop \textit{``Fifty Years of Seismology of the Sun and Stars''} \citep{jainetal2013}.

Asteroseismology, the seismic study of stars other than the Sun, took longer to develop because of
inherent difficulties in ground-based observations.
Early attempts were focused on trying to observe pulsations of $\alpha$~Cen.
Some early ground-based attempts did not find any convincing evidence for
solar-like pulsations \citep[e.g.,][]{brownandgilliand1990}, though others could place
limits \citep[e.g.,][]{pottasch1992,edmunds1995}. It took many more attempts before
pulsations of $\alpha$~Cen were observed and the frequencies measured
\citep{bouchyandcarrier2001,beddingetal2004,kjeldsenetal2005}.
Other stars were targeted too
\citep[e.g.,][]{kjeldsenetal2003,carrieretal20051,carrieretal20052, 
beddingetal2007,arentoftetal2008}. 
The field did not grow till space-based missions were were available.
The story of space-based asteroseismology started with the
ill-fated Wide-Field Infrared Explorer (WIRE). The satellite failed because coolants
meant to keep the detector cool evaporated, but \citet{buzasi2000} realised that
the star tracker could be used to monitor stellar variability and hence to look for stellar oscillations.
This followed the observations of $\alpha$~UMa and $\alpha$~Cen A \citep{buzasietal2000,schouandbuzasi2001,
fletcheretal2006}.
The Canadian mission Microvariability and Oscillations of Stars \citep[MOST:][]{walkeretal2003} was
the first successfully launched mission dedicated to asteroseismic studies. Although it was not
very successful in studying solar type stars, it was immensely successful in
studying giants, classical pulsators, and even star spots and exo-planets.
The next major step was the ESA/French CoRoT mission \citep{baglinetal2006,corot}.
CoRot observed many giants and showed giants also show non-radial pulsations \citep{deridderetal2009};
the mission observed some subgiants and main sequence stars as well 
\citep[see e.g.,][]{deheuvelsetal2010,ballotetal2011,deheuvels2014}.
While CoRoT and MOST showed that the seismic study of other stars was feasible, the field
began to flourish after the
 launch of the \textit{Kepler} mission
\citep{kochetal2010} and the demonstration that \textit{Kepler} could indeed observe
stellar oscillations \citep{gillilandetal2010} and that the data could be used to derive
stellar properties \citep{chaplinetal2010}. Asteroseismology is going through a phase 
in which we are still learning the best ways to analyse and interpret the data, but
with two more asteroseismology missions being planned -- the
Transiting Exoplanet Survey Satellite (TESS; launch 2017), and PLATO (launch 2024) -- this field is 
going to grow even more rapidly. 
We discuss the basics of asteroseismology in Section~\ref{sec:stars} of this review,
however, only a dedicated review can do proper justice to the field.

This review is organised as follows: Since it is not possible to perform helioseismic
(or for that matter asteroseismic) analyses without models, we start with the
construction of solar and stellar models in Section~\ref{sec:ssm}. We then proceed to
derive the equations of stellar oscillation and describe some the properties of
the oscillations in Section~\ref{sec:theory}. A brief history of solar
models is given in Section~\ref{sec:history}. We show what happens when we try to
compare solar models with the Sun by comparing frequencies in Section~\ref{sec:surf}. The
difficulty in making such comparisons leads us to Section~\ref{sec:inv} where we show how
solar oscillation frequencies may be inverted to infer properties of the solar interior. 
The next three sections are devoted to results. In Section~\ref{sec:struc} we describe
what we have learnt about spherically symmetric part of solar structure. Deviations
from spherical symmetry -- in terms of dynamics, magnetic fields and structural
asymmetries -- are described in Section~\ref{sec:rot}. The solar-cycle
related changes in solar frequencies and the deduced changes in the solar
interior are discussed in Section~\ref{sec:cycle}. Most of the results
discussed in this review were obtained by analysing the
frequencies of solar oscillation. There are other observables though,
such as line-width and amplitude, and these carry information on
how modes are excited (and damped). The issue of mode-excitation is
discussed briefly in Section~\ref{sec:excite}. Finally, in Section~\ref{sec:stars},
we give a brief introduction to the field of asteroseismology.

\newpage

\section{Modelling Stars}
\label{sec:ssm}

In order to put results obtained from seismic analyses in a proper context, we first
 give a short overview of the process of constructing models and of the inputs used
to construct them.
Traditionally in astronomy, we make inferences
by comparing properties of models with data, which in the seismic context means
comparing the computed frequencies with the observed ones.
Thus, seismic investigations of the Sun and other stars start with the construction
models and the calculation of their oscillation frequencies.
In this section we briefly cover the field of modelling. We also discuss how solar 
models are constructed in a manner that is different from the construction of models of other stars. 
There are many excellent textbooks that describe stellar structure and evolution and
hence, we only describe the basic equations and inputs. Readers
are referred to books such as \citet{kippenhahnandweigert2012}, \citet{huang1998}, \citet{hansen},
\citet{weiss}, \citet{maeder}, etc. for details.

\subsection{The equations}
\label{subsec:equations}

The most common assumption involved in making solar and stellar models is that stars
are spherically symmetric, i.e., stellar properties are only a function of radius. This
is a good approximation for non-rotating and slowly-rotating stars.
The other assumption that is usually made is that a star does not change its mass 
as it evolves; this assumption is valid except for the very early stages of
star formation and very evolved stages, such at the tip of the red giant branch. 
The Sun for instance, loses about $10^{-14}$ of its
mass per year. Thus, in its expected main-sequence lifetime of about 10 Gyr,
the Sun will lose only about 0.01\% of its mass. The radius of a star on the other hand, is
expected to change significantly. As
a result, mass is used as the independent variable when casting the equations governing
stellar structure and evolution.

The first equation is basically the continuity equation in the
absence of flows, and is thus a statement of the conservation of mass:
\be
{\dd r\over \dd m}= {1\over 4\pi r^2\rho}\,,
\label{eq:drdm}
\ee
where $m$ is the mass enclosed in radius $r$ and $\rho$ the density.

The next equation is a statement of the conservation of momentum in the
quasi-stationary state. In the stellar
context, it represents hydrostatic equilibrium:
\be
{\dd P\over \dd m}=-{Gm\over 4\pi r^4}\,.
\label{eq:hydro}
\ee

Conservation of energy comes next. Stars produce energy in the core. At equilibrium,
energy $l$ flows through a shell
of radius $r$ per unit time as a result of
nuclear reactions in the interior. If $\epsilon$ be the energy released
 per unit mass per second by nuclear reactions, and $\epsilon_\nu$ the energy
lost by the star because of neutrinos streaming out without
depositing their energy, then,
\be 
{\dd l\over \dd m}=\epsilon -\epsilon_\nu.
\label{eq:nuc}
\ee
However, this is not enough. Different layers of a star can expand or contract 
during their evolution; for instance in the sub-giant and red-giant stages the stellar
core contracts rapidly while the outer layers expand. Thus, Eq.~(\ref{eq:nuc})
has to be modified to include the energy used or released as a result of 
expansion or contraction, and one can show that
\be
{\dd l\over \dd m}=\epsilon-\epsilon_\nu -C_P{\dd T\over \dd t}+
{\delta\over\rho}{\dd P\over \dd t}\,,
\label{eq:nucf}
\ee
where $C_P$ is the specific heat at constant pressure,
$t$ is time,  and $\delta$ is given by the equation of state and defined as
\be 
\delta = -\left({\partial\ln\rho\over\partial\ln T}\right)_{P,X_i},
\ee
where $X_i$ denotes composition.
The last two terms of Eq.~(\ref{eq:nucf}) above are often lumped together and called 
$\epsilon_g$ ($g$ for gravity) because they denote the release of gravitational energy.

The next equation determines the the temperature at any point.
In general terms, and with the help of Eq.~(\ref{eq:hydro}), this equation
can be
 written quite trivially as
\be
{\dd T\over \dd m}=-{Gm T \over 4\pi r^4 P}\nabla,
\label{eq:dtdm}
\ee
where
$\nabla$ is the dimensionless ``temperature gradient'' $\dd\ln T/\dd\ln P$.
The difficulty lies in determining what $\nabla$ is, and this depends on whether
energy is being transported by radiation or convection.
We shall come back to the issue of $\nabla$ presently. 

The last set of equations deals with chemical composition as a function
of position and time. There are three ways that the chemical
composition at any point of a star can change: (1) nuclear reactions, 
(2) the changes in the
boundaries of convection zones, and
(3) diffusion and gravitational settling (usually simply referred to as
diffusion) of helium and heavy elements and other mixing processes as well.

The change of abundance because of nuclear reactions can be written as
\be
{\partial X_i\over \partial t}=
{m_i\over\rho}\left[\sum_j r_{ji}-\sum_k r_{ik}\right],
\label{eq:nucx}
\ee
where $m_i$ is the mass of the nucleus of each isotope $i$, $r_{ji}$ is the rate at which
isotope $i$ is formed from isotope $j$, and $r_{ik}$ is the rate at which isotope $i$
is lost because it turns into a different isotope $k$. The rates $r_{ik}$ are
external inputs to models.

Convection zones are chemically homogeneous -- eddies of moving matter carry their
composition with them and when they break-up, the material gets mixed
with the surrounding. This happens on timescales that are very short compared to the time scale of a star's evolution.
If a convection zone
exists in the region between two spherical shells of masses $m_1$ and $m_2$, the average abundance
of any species $i$ in the convection zone is:
\be
{\bar X_i}={1\over m_2-m_1}\int_{m_1}^{m_2} X_i\;\dd m\,.
\label{eq:barx}
\ee
In the presence of convective overshoot, the limits of the integral in Eq.~(\ref{eq:barx}) have to
be change to denote the edge of the overshooting region.
The rate at which ${\bar X_i}$ changes will depend on nuclear reactions in the convection
zone, as well as the rate at which the mass limits $m_1$ and $m_2$ change. One can therefore
write
\bea
\!\!\!\!{\partial{\bar X_i}\over \partial t} & = &
{\partial\over\partial t}\left({1\over m_2-m_1}\int_{m_1}^{m_2} X_i\; \dd m\right)\nonumber\\
&=& {1\over m_2-m_1}\left[ \int_{m_1}^{m_2} {\partial X_i\over\partial t}\;\dd m+
{\partial m_2\over \partial t}(X_{i,2}-{\bar X_i}) -
{\partial m_1\over \partial t}(X_{i,1}-{\bar X_i})\right],\phantom{as}
\label{eq:convx}
\eea
where $X_{i,1}$ and $X_{i,2}$ is the mass fraction of element $i$ at $m_1$ and
$m_2$ respectively.

The gravitational settling of helium and heavy elements can be described by the
process of diffusion and the change in abundance can be found with the help of the
diffusion equation:
\be
{\partial X_i\over \partial t}=D\nabla^2 X_i\,,
\label{eq:diff}
\ee
where $D$ is the diffusion coefficient, and $\nabla^2$ is the Laplacian operator.
The diffusion coefficient hides the complexity of the process and includes, in addition
to gravitational settling, diffusion due to composition and temperature gradients.
All three processes are generally simply called `diffusion'. Other mixing process,
such as those induced by rotation, are often included the same way by modifying
$D$ \citep[e.g.,][]{richardetal1996}.
$D$ depends on the isotope under consideration, however, it is not uncommon for
stellar-evolution codes to treat helium separately, and use a single value
for all heavier elements.

Equations~(\ref{eq:drdm}), (\ref{eq:hydro}), (\ref{eq:nucf}), (\ref{eq:dtdm})
 together with the
equations relating to change in abundances, form the full set of equations
that govern stellar structure and evolution. In most codes, Eqs.~(\ref{eq:drdm}), (\ref{eq:hydro}),
 (\ref{eq:nucf}) and (\ref{eq:dtdm}) are solved for a given $X_i$ at a given time
$t$. Time is then
advanced, Eqs.~(\ref{eq:nucx}), (\ref{eq:convx}) and (\ref{eq:diff}) are
solved to give new $X_i$, and equations
(\ref{eq:drdm}), (\ref{eq:hydro}),
 (\ref{eq:nucf}) and (\ref{eq:dtdm}) are solved again. Thus, we have two independent
variables, mass $m$ and time $t$, and we look for solutions in the interval
$0\le m \le M$ (stellar structure) and $ t \ge t_0$ (stellar evolution).

Four boundary conditions are required to solve the stellar
structure equations. Two (on radius and luminosity) can be applied quite
trivially at the centre. The remaining conditions (on temperature and pressure) need to be
applied at the surface. The boundary conditions at the surface are much
more complex than the central
boundary conditions and are usually determined with the aid of simple
stellar-atmosphere models. As we shall see later, atmospheric
models plays a large role in determining the frequencies of different modes
of oscillation.

The initial conditions needed to start evolving a star depend on where we start the
evolution. If the evolution begins at the pre-main sequence phase, i.e., while the
star is still collapsing, the initial structure is quite
simple. Temperatures are low enough to make the
star fully convective and hence chemically
homogeneous. If evolution is begun at the Zero Age Main Sequence (ZAMS), which is the point
at which hydrogen fusion begins, a ZAMS model must be used.

We return to the question of the temperature gradient $\nabla$ in Eq.~(\ref{eq:dtdm}). If energy is 
transported by radiation (``the radiative zone''), $\nabla$ is calculated assuming
that energy transport can be modelled as a diffusive process, and that yields
\be
\nabla=\nabla_{\rm rad}={3\over 64\pi \sigma G}{\kappa l P\over m T^4}\,,
\label{eq:gradrad}
\ee
where, $\sigma$ is the Stefan--Boltzmann constant and
 $\kappa$ is the opacity which is an external input to stellar models.

The situation in the convection zones is tricky. Convection is, by its very nature,
a three-dimensional phenomenon, yet our models are one dimensional. Even if we 
construct three-dimensional models, it is currently impossible include convection and evolve the
models at the same time. This is because convection takes place over time scales of minutes to hours, while 
stars evolve in millions to billions of years. As a result, drastic simplifications are
used to model convection. Deep inside a star, the temperature gradient is well approximated
by the adiabatic temperature gradient $\nabla_{\rm ad}\equiv (\partial\ln T/\partial\ln P)_s$
($s$ being the specific entropy), which is determined by the equation of state.
This approximation cannot be used in the outer layers where convection is not
efficient and some of the energy is also carried by radiation. In these layers
one has to use an
approximate formalism since there is no ``theory'' of stellar convection as such.
One of the most common formulations used to calculate convective flux in stellar models
 is the so-called ``mixing length theory'' (MLT).
The mixing length theory was first proposed by \citet{prandtl1925}. 
His model of convection was analogous to heat transfer by particles; the
transporting particles are macroscopic eddies and their mean free path is
the ``mixing length''. This was applied to stars by \citet{biermann1948}, \citet{vitense1953}, and
\citet{bohmvitense1958}. Different mixing length formalisms have slightly different assumptions about
what the mixing length is. The main assumption in the usual mixing length formalism is that 
convective eddies move an average distance equal to the mixing length $l_m$ before
giving up their energy and losing their identity. The mixing length is usually defined as
$l_m=\alpha H_P,$ where
$\alpha$, a constant, is the so-called `mixing length parameter',
and $H_P\equiv -\dd r/\dd \ln P$ is the pressure scale height.
Details of how 
$\nabla$ is calculated under these assumption can be found in any stellar structure textbook, such as
\citet{kippenhahnandweigert2012}.
There is no a priori way to determine
$\alpha$, and it is one of the free parameters in stellar models. Variants of MLT are
also used, and among these are the formulation of \citet{cm1991} and
\citet{arnett2010}.

\subsection{Inputs to stellar models}
\label{subsec:inputs}

The equations of stellar structure look quite simple. Most of the complexity
is hidden in the external inputs needed to solve the equations. There are four important inputs
that are needed: the equation of state, radiative opacities, nuclear reaction rates
and coefficients to derive the rates of diffusion and gravitational settling. These
are often referred to as the \emph{microphysics} of stars. 

\subsubsection{The equation of state}

The equation of state specifies the relationship between density, pressure, temperature
and composition. The stellar structure equations are a set of five equations in 
six unknowns, $r$, $P$, $l$, $T$, $X_i$, and $\rho$. None of the equations 
directly solves for the behaviour of $\rho$ as a function of mass and time. We
determine the density using the equation of state. 

The ideal gas equation is good enough to make simple models. However, the
ideal gas law does not apply to all layers of stars since it does not
include the effects of ionisation, radiation
pressure, pressure ionisation, degeneracy, etc. Among the early published
equations of state valid under stellar condition is that
of Eggleton, Faulkner and Flannery \citep[EFF;][]{eggletonetal1973}. This
equation of state suffered from the fact that it did not include corrections
to pressure due to Coulomb interactions, and this led to the development of the so-called
``Coulomb Corrected'' EFF, or CEFF equation of state \citep{jcdwd1992, guentheretal1992}.
However, the CEFF equation of state is not fully thermodynamically consistent.

Modern equations of state are
usually given in a tabular form with important thermodynamic
quantities such as $\nabla_{\rm ad}$, $C_p$ listed as a function of $T$, $P$ (or $\rho$) and composition.
These include the OPAL equation of state \citep[][]{rogersetal1996,rogersandnayfonov2002}
and the so-called MHD (i.e., Mihalas, Hummer \& D\"appen) equation of state
\citep{dappenetal1988, mihalasetal1988, hummerandmihalas1988, gongetal2001}. Both OPAL and MHD 
equations of state suffer from the limitation that unlike the EFF and CEFF
equations of state, the heavy-element mixture
used to calculate them cannot be changed by the user. This has lead to
the development of thermodynamically consistent extensions of the EFF
equation of state that allow users to change the equation of state easily. The SIREFF equation
of state is an example of this \citep{guzikandswenson1997,guziketal2005}.

\subsubsection{Opacities}

We need to know the opacity $\kappa$ of the stellar material
 in order to calculate $\nabla_{\rm rad}$ (Eq.~\ref{eq:gradrad}).
Opacity
is a measure of how opaque a material is to photons. Like modern equations of state, 
opacities are usually available in tabular form as a function of density, temperature and
composition. Among the widely used opacity tables are the OPAL \citep{opal1996} and
OP \citep{badnelletal2005, mendozaetal2007} tables. The OPAL opacity tables include contributions from 19 heavy elements whose
relative abundances (by numbers) with respected to hydrogen are larger than about $10^{-7}$.
The OP opacity calculations include 15 elements. Neither table is very good at low temperature where
molecules become important. As a results these tables are usually supplemented by specialised low-temperature
opacity tables such that those of \citet{kurucz1991} and \citet{fergusonetal2005}.

\subsubsection{Nuclear reaction rates}

Nuclear reaction rates are required to compute
energy generation, neutrino fluxes and composition
changes. The major sources of reaction rates for solar models are
the compilations of \citet{adelbergeretal1998,adelbergeretal2011} and
\citet{anguloetal1999}.

\subsubsection{Diffusion coefficients}

The commonly used prescriptions for calculating diffusion coefficients are
those of \citet{thouletal1994} and \citet{proffittandmichaud1991}.

\subsubsection{Atmospheres}

While not a microphysics input, stellar atmospheric
models are equally crucial: stellar models do not stop at $r=R$, but generally extend into an atmosphere,
and hence the need for these models.
These models are also used to calculate the outer boundary condition.
The atmospheric models are often quite simple and provide a $T$--$\tau$ relation,
i.e., a relation between temperature $T$ and optical depth $\tau$.
The Eddington $T$--$\tau$ relation is quite popular. Semi-empirical
relations such as the Krishna Swamy $T$-$\tau$ relation \citep{ks1966}, the
Vernazza, Avrett and Loeser (VAL) relation \citep{val1981} though applicable
to the Sun are also used frequently to model other stars. A relatively recent development
in the field is the use of  $T$--$\tau$ relations obtained from simulations
of convection in the outer layers of stars.

\subsection{The concept of ``standard'' solar models}
\label{sec:std}

The mass of a star is the most fundamental quantity needed to model a star.
Other input quantities include the initial heavy-element abundance $Z_0$ and the
initial helium abundance $Y_0$. These quantities affect both the structure and
the evolution of star through their influence on the equation of state and 
opacities. Also required is the mixing-length parameter $\alpha$. Once these
quantities are known, or chosen, models are evolved in time, until they
reach the observed temperature and luminosity. The initial guess of the
mass may not result in a models with the required characteristics, and
a different mass needs to be chosen and the process repeated. Once
a model that satisfies observational constraints is constructed, 
the model becomes a proxy for the star; the age of the star is assumed to be
the age of the model, and the radius of the star is assumed to be the radius of the model.
The most important source of uncertainty in stellar models is 
 our inability to model convection properly. MLT requires a
free parameter $\alpha$ that is essentially unconstrained
and introduces uncertainties in the radius of a star of a given mass and heavy-element abundance.
And even if we were able to constrain $\alpha$, it would not 
account for all the properties of convective heat transport and thus would
therefore, introduce errors in the results.

The Sun is modelled in a somewhat different manner since its  global properties
are known reasonably well. 
We have independent estimates of the solar mass, radius, age and luminosity. The
commonly used values are listed in Table~\ref{tab:globsun}. Solar properties are
usually used a references to express the properties of other stars. Small differences
in solar parameters adopted by different stellar codes can therefore, lead to 
differences. To mitigate this problem, the International Astronomical Union
adopted a set of nominal values for the global properties of the Sun. These
are listed in Table~\ref{tab:iausun}. Note that the symbols used to 
denote the nominal properties are different from the usual solar notation.

\begin{table}[htb]
\caption{Global parameters of the Sun}
\label{tab:globsun}
\centering
\begin{tabular}{lcl}
\toprule
Quantity & Estimate & Reference \\
\midrule
Mass ($M_\odot$)$^*$       & $1.98892(1\pm 0.00013)\times 10^{33}\mathrm{\ g}$ & \citet{cohenandtaylor1987}\\
Radius ($R_\odot$)$^\dagger$ & $6.9599(1\pm0.0001)\times 10^{10}\mathrm{\ cm}$ & \citet{allen1973}\\
Luminosity ($L_\odot$)     & $3.8418(1\pm 0.004)\times 10^{33}\mathrm{\ ergs\ s}^{-1}$ & \citet{frohlichandlean1998}\\
~                         & ~                                                        & \citet{bahcalletal1995}\\
Age                       & $4.57(1\pm 0.0044) \times 10^{9}\mathrm{\ yr}$ & \citet{bahcalletal1995}\\
\bottomrule
\end{tabular}

{$^*$} {\small Derived from the values of $G$ and $GM_\odot$}\hfill\\

{$^\dagger$} {\small See a\citet{schouetal1997}, \citet{antia1998},
\citet{brownjcd1998}, \citet{habe2008}
 for more recent discussions about the exact value of the solar radius.}\\
\end{table}
\begin{table}[htb]
\caption{Nominal solar conversion constants as per IAU resolution B3$^\dag$}
\label{tab:iausun}
\centering
\begin{tabular}{lc}
\toprule
Parameter & value\\
\midrule
$1{\mathcal R}^{\rm N}_\odot$         &  $6.957\times 19^8$ m\\
$1{\mathcal S}^{\rm N}_\odot$         &  1361 W m$^{-2}$\\
$1{\mathcal L}^{\rm N}_\odot$         & $3.828\times 10^{26}$ W\\
$1{\mathcal T}^{\rm N}_{{\rm eff}\odot}$   & 5772 K\\
$1{\mathcal GM}^{\rm N}_\odot$ & $1.327\;1244\times10^{20}$ m$^3$s$^{-2}$\\
\bottomrule
\end{tabular}

$\dag$ See \citet{mamajek2015}\ for details\\

\end{table}

To be called a solar model, a $1\msun$ model must have a luminosity of $1\lsun$ and
a radius of $1\rsun$ at the solar age of 4.57~Gyr. The way this is done is by
recognising that there are two constraints that we need to satisfy at 4.57~Gyr, the solar radius
and luminosity, and that the set of equations has two free parameters, the initial helium abundance $Y_0$
and the mixing length parameter $\alpha$. Thus, we basically have a situation of two constraints and two
unknowns. However, since the equations are non-linear, we
need an iterative method to determine $\alpha$ and $Y_0$. The value of
$\alpha$ obtained in this manner for the Sun is often called the ``solar calibrated''
value of $\alpha$ and used to model
other stars. In addition to $\alpha$ and $Y_0$, and very often
initial $Z$ is adjusted to get the observed $Z/X$ in the solar envelope.
The solar model constructed in this manner does not have any free parameters,
since these are determined  to match solar constraints. Such
a model is known as a ``standard'' solar model (SSM).

The concept of standard solar models is very important in solar physics.
Standard solar models are those where only standard input physics
such as equations of state, opacity, nuclear reaction rates, diffusion
coefficients etc., are used. The parameters $\alpha$ and
$Y_0$ (and sometime $Z_0$) are adjusted to match the current solar radius
and luminosity (and surface $Z/X$). No other input
is adjusted to get a better agreement with the Sun.
 By comparing standard
solar models constructed with different input physics with the Sun we can
put constraints on the input physics. One can use helioseismology to
test whether or not the structure of the model agrees with that of the Sun.

There are many physical processes that are not included in SSMs.  These are processes
that do not have a generally recognised way of modelling and rely
on free parameters. These additional free parameters would make any solar model
that includes these processes ``non standard''. 
Among the missing processes are effects
of rotation on structure and of mixing induced by rotation. There are other
proposed mechanisms for mixing in the radiative layers of the Sun,
such as mixing caused by waves generated at the convection-zone base \cite[e.g.,][]{kumaretal1999}
that are not included either. These missing
processes can affect the structure of a model. For example, \citet{turckchieze_etal2004}
found that mixing below the convection-zone base can change both the
position of the convection-zone base (gets shallower) and the
helium abundance (abundance increases). However, their inclusion in stellar models
require us to choose values for the parameters in the formula\ae\ that describe
the processes. Accretion and mass-loss at some
stage of solar evolution can also affect the solar models
\citep{castro2007}, but again, there are no standard formulations for modelling these
effects.

Standard solar models constructed by different groups are not identical. They depend
on the microphysics used, as well as assumption about the heavy element abundance.
Numerical schemes also play a role. For a discussion of the sources
of uncertainties in solar models, readers are referred to Section~2.4 of \citet{sbhma2008}.

\newpage

\section{The Equations Governing Stellar Oscillations}
\label{sec:theory}

To a good approximation, solar oscillations and 
solar-like oscillations in other stars, can be described as linear and adiabatic. In the
case of the Sun,  modes of oscillation
have amplitudes of the order of $10\mathrm{\ cm\ s}^{-1}$ while the sound speed at the surface is more like $10\mathrm{\ km\ s}^{-1}$, 
putting the amplitudes squarely in the linear regime. The condition of adiabaticity
can be justified over most of the Sun -- the oscillation
frequencies have periods of order 5~min, but the thermal time-scale is much larger and hence,
adiabaticity applies. This condition however
breaks down in the near-surface layers
(where thermal time-scales are short)
 resulting in an error in the frequencies. This error adds to what is later described as
the ``surface term'' (see Section~\ref{sec:surf}) 
but which can be filtered out in many cases. We shall retain the approximations of 
linearity and adiabaticity in this section since most helioseismic results have been
obtained under these assumptions after applying a few corrections.

Details of the derivation of the oscillation equations and a description of their properties has been described
well by authors such as \citet{cox1980}, \citet{unnoetal}, \citet{jcdber}, \citet{dog1993}, etc. Here we give a 
short overview that allows us to derive some of the properties of solar and stellar 
oscillations that allow us to undertake seismic studies of the Sun and other stars.

The derivation of the equations of stellar oscillations begins with the basic equations 
of fluid dynamics, i.e.,
the continuity equation and the momentum equation. We use the
 Poisson equation to describe the gravitational field. 
Thus, the basic equations are:
\be
{\partial\rho\over\partial t}+\nabla\cdot(\rho{\vec v})=0,
\label{eq:cont}
\ee
\be
\rho\left({\partial v\over\partial t}+{\vec v}\cdot\nabla {\vec v}\right)=
-\nabla P +\rho\nabla \Phi,
\label{eq:mom}
\ee
and
\be
\nabla^2\Phi=4\pi G\rho,
\label{eq:pois}
\ee
where ${\vec v}$ is the velocity of the fluid element, $\Phi$ is the gravitational potential, and $G$ the
gravitational constant. The heat equation is written in the form
\be
{\dd q \over \dd t}= {1\over \rho(\Gamma_3-1)}\left( {\dd p\over \dd t}
- {\Gamma_1 P\over \rho}{\dd\rho\over \dd t}\right),
\label{eq:therm}
\ee
where
\be
\Gamma_1=\left({\partial\ln P\over\partial\ln\rho}\right)_{\rm ad},
\quad\hbox{and,}\quad \Gamma_3-1=
\left({\partial\ln T\over\partial\ln\rho}\right)_{\rm ad}.
\label{eq:gam}
\ee
In the adiabatic limit Eq.~(\ref{eq:therm}) reduces to
\be
{\partial P\over \partial t}+{\vec v}\cdot\nabla P=
c^2\left({\partial\rho\over \partial t}+{\vec v}\cdot\nabla\rho
\right), \label{eq:adia}
\ee
where $c=\sqrt{\Gamma_1 P/\rho}$ is
the sound speed

Linear oscillation equations are a result of linear perturbations to the fluid equations above.
 Thus, e.g., we can write the perturbations to density as
\be
\rho({\vec r},t)=\rho_0({\vec r})+\rho_1({\vec r}, t),
\label{eq:rho}
\ee
where the subscript 0 denotes the equilibrium, spherically symmetric, quantity which by definition
does not depend on time, and the subscript 1 denotes the perturbation. Perturbations to other quantities 
can be written in the same way.
Note that Eq.~(\ref{eq:rho}) shows the Eulerian perturbation to density, i.e., perturbations at a fixed 
point in space denoted by co-ordinates $(r,\theta,\phi)$. In some cases, notably when there are time derivatives, 
it is easier to use the Lagrangian perturbation, i.e., a perturbation seen by an observer moving with the fluid. 
The Lagrangian perturbation for density is given by
\be
\delta \rho(r,t)=\rho({\vec r}+\vec{\xi }(\vec r,t))-\rho({\vec r})=
\rho_1({\vec r},t)+\vec{\xi}({\vec r},t)\cdot
\nabla \rho,
\label{eq:lrhoturb}
\ee
where ${\vec \xi}$ is the displacement from the equilibrium position.
The perturbations to the other quantities can be written in the
same way. The equilibrium state of a star is generally assumed to be static, and thus the velocity ${\vec v}$ in the
fluid equations appears only after a perturbation has been applied and is nothing but the rate of change of
displacement of the fluid, i.e., ${\vec v}=\dd{\vec \xi}/\dd t$.

Substituting the perturbed quantities in Eqs~(\ref{eq:cont}), (\ref{eq:mom}) and (\ref{eq:pois}), and keeping only 
linear terms in the perturbation, we get
\be
\rho_1+\nabla\cdot(\rho_0{\vec \xi})=0,
\label{eq:conpturb}
\ee
\be
 \rho{\frac{\partial^2{\vec \xi}}{\partial t^2}}=-\nabla P_1+\rho_0 \nabla\Phi_1+\rho_1\nabla\Phi_0\;,
\label{eq:mom1}
\ee
\be
\nabla^2\Phi_1=4\pi G\rho_1.\label{eq:pois1}
\ee

It is easier to consider
the Lagrangian perturbation for the heat equation since under the assumptions that
at
have been made, the time derivative of the various quantities is simply the time derivative of the Lagrangian
perturbation of those quantities.
Thus, from Eq.~(\ref{eq:therm}) we get
\be
{\partial\delta q\over \partial t}=
{1\over\rho_0(\Gamma_{3.0}-1)}\left({\partial\delta P\over \partial t}
-{\Gamma_{1,0} P_0\over\rho_0}{\partial \delta \rho\over \partial t}\right).
\label{eq:thermp}
\ee
In the adiabatic limit, where energy loss is negligible, $\partial\delta q/\partial t=0$ and
\be
P_1+\vec{\xi}\cdot\nabla P={\Gamma_{1,0} P_0\over\rho_0}(\rho_1
+\vec{\xi}\cdot\nabla\rho).
\label{eq:therma}
\ee
The term $\Gamma_{1,0}P_0/\rho_0$ in Eq.~(\ref{eq:therma}) is nothing but the
squared, unperturbed sound speed $c_0^2$.

In the subsequent discussion, we drop the subscript `0' for the equilibrium quantities, and only keep
the subscript for the perturbations. 

\subsection{The spherically symmetric case}
\label{subsec:spherical}

Since stars are usually spherical, though not necessarily spherically symmetric, it is customary to
write the equations in spherical-polar coordinates with the origin define at the centre of the star, 
with $r$ being the radial distance, $\theta$ the co-latitude, and $\phi$
the longitude. The different quantities can be
decomposed into their radial and tangential components.
Thus, the displacement $\vec{\xi}$ can be decomposed as
\be
\vec{\xi}=\xi_r \hat{a}_r+\xi_t\hat{a}_t,
\label{eq:disp}
\ee
where, $\hat{a}_r$ and $\hat{a}_t$ are the unit vectors in the radial and tangential
 directions respectively,
$\xi_r$ is the radial component of the displacement vector and $\xi_t$ the
transverse component.
An advantage of using spherical polar coordinates is the fact that
tangential gradients of the equilibrium quantities do not
exist. Thus, e.g., the heat equation (Eq.~\ref{eq:therma}) becomes
\be
\rho_1={\rho\over \Gamma_1 P}P_1+\rho\xi_r\left({1\over \Gamma_1 P}
{\dd P\over \dd r}
-{1\over \rho}{\dd \rho\over \dd r}\right).
\label{eq:rhodash}
\ee
The tangential component of the equation of motion
(Eq.~\ref{eq:mom}) is
\be
\rho{\partial^2\vec{\xi_t}\over \partial t^2}=-\nabla_t P_1+
\rho\nabla_t\Phi',
\label{eq:ppp}
\ee
or (taking the tangential divergence of both sides),
\be
\rho{\partial^2\over \partial t^2}(\nabla_t\cdot\vec{\xi_t})
=-\nabla^2_t P_1+\rho\nabla^2_t\Phi_1.
\label{eq:taneq}
\ee

The continuity equation (Eq.~\ref{eq:conpturb}) after decomposition
can be used to eliminate the term $\nabla_t\cdot\vec{\xi_t}$ from
Eq.~(\ref{eq:taneq}) to obtain
\be
-{\partial^2\over \partial t^2}\left[
\rho'+ {1\over r^2}{\partial\over \partial r}(\rho r^2\xi_r)\right]
=-\nabla^2_t P_1+\rho\nabla^2_t\Phi_1.
\label{eq:taneqq}
\ee

The radial component of the equation of motion gives
\be
\rho{\partial^2\xi_r\over \partial t^2}=
-{\partial P_1\over \partial r} -\rho_1 g + \rho {\partial \Phi_1\over
\partial r},
\label{eq:radeqq}
\ee
where we have used the fact that gravity acts in the negative $r$ direction.
Finally, the Poisson's equation becomes 
\be
{1\over r^2}{\partial\over \partial r}\left( r^2 {\partial \Phi_1\over
 \partial r}
\right)+ \nabla^2_t\Phi_1=-4\pi G \rho_1.
\label{eq:pois2}
\ee

Note that the there are no mixed radial and tangential derivatives, and the 
tangential gradients appear only as the tangential component
of the Laplacian, and as a result one can show that \citep[e.g.,][]{jcd2003} the tangential part of the
perturbed quantities can be written in terms of eigenfunctions of the tangential
Laplacian operator. Since we are dealing with spherical objects, the spherical
harmonic function are used. 

Furthermore, note that time $t$ does not appear explicitly in
coefficients of any of the derivatives. This implies that the time dependent
part can be separated out from the spatial part, and the time-dependence can
be expressed in terms of $\exp(-i\omega t)$ where $\omega$ can be real (giving
an oscillatory solution in time) or imaginary (a solution that grows or decays).
Thus, we may write: 
\be
\xi_r(r,\theta,\phi,t)\equiv{\xi_r}(r)Y^m_{\ell}(\theta,\phi)
\exp(-i\omega t),
\label{eq:xir}
\ee
\be
P_1(r,\theta,\phi,t)\equiv{P_1}(r)Y^m_{\ell}(\theta,\phi)
\exp(-i\omega t),
\label{eq:pdash}
\ee
etc.

Once we have the description of the variables in terms of an oscillating function of time, and
spherical harmonic functions in the angular directions, we can substitute those in 
Eqs.~(\ref{eq:taneqq}), (\ref{eq:radeqq}) and (\ref{eq:pois2}). Furthermore, 
Eq.~(\ref{eq:rhodash}) can be used to eliminate
the quantity $\rho_1$ to obtain
\be
{\dd \xi_r\over \dd r}=-\left({2\over r}+ {1\over\Gamma_1 P}
{\dd P\over \dd r}\right)\xi_r
+{1\over\rho c^2}\left({S^2_\ell\over\omega^2}-1\right)P_1
-{\ell(\ell+1)\over\omega^2 r^2}\Phi_1,
\label{eq:eqone}
\ee 
where
$c^2=\Gamma_1P/\rho$ is the squared sound speed, and $S^2_\ell$ is the
\emph{Lamb frequency} defined by
\be
S^2_\ell={\ell(\ell+1)c^2\over r^2}.
\label{eq:lamb}
\ee
Equation~(\ref{eq:radeqq}) and the equation of hydrostatic equilibrium give
\be
{\dd P_1\over \dd r}=\rho(\omega^2-N^2)\xi_r+{1\over\Gamma_1 P}{\dd P\over \dd r}P_1
+\rho{\dd \Phi_1\over \dd r},
\label{eq:eqtwoo}
\ee
where, $N$ is \emph{Brunt--V{\"a}is{\"a}l{\"a}} or \emph{buoyancy frequency} defined as
\be
N^2=g\left({1\over\Gamma_1 P}{\dd P\over \dd r}-{1\over\rho}{\dd \rho\over \dd r}\right).
\label{eq:brunt}
\ee
This is the frequency with which a small element of fluid will oscillate
when it is disturbed from its equilibrium position. When $N^2<0$ the fluid
is unstable to convection and in such regions part of the energy will be transported
by convection.
Finally, Eq.~(\ref{eq:pois2}) becomes
\be
{1\over r^2}{\dd \over \dd r}\left(r^2 {\dd \Phi_1\over \dd r}\right)
=-4\pi G\left({P_1\over c^2}+{\rho\xi_r\over g}N^2\right)+
{\ell(\ell+1)\over r^2}\Phi_1.
\label{eq:eqthreee}
\ee

Equations~(\ref{eq:eqone}), (\ref{eq:eqtwoo}) and (\ref{eq:eqthreee})
form a set of fourth-order differential equations and constitute
an eigenvalue problem with eigenfunction $\omega$. Solution of the equations yields the
radial component, $\xi_r$ of the displacement eigenfunction as
well as $P_1$, $\Phi_1$ as well as $\dd \Phi_1/\dd r$. Each eigenvalue is usually
referred to as a ``mode'' of oscillation.

The transverse component
of displacement vector can be written in terms of $P_1(r)$ and
$\Phi_1(r)$ and one can show that
\be
{\vec {\xi_t}}(r,\theta,\phi,t)=\xi_t(r)\left( {\partial Y^m_{\ell}\over
\partial\theta}{\hat{a_\theta}}+{1\over\sin\theta}{\partial Y^m_{\ell}\over
\partial\phi}{\hat{a_\phi}}\right)\exp(-i\omega t),
\label{eq:xit}
\ee
where ${\hat{a_\theta}}$ and ${\hat{a_\phi}}$ are the unit vectors in the
$\theta$ and $\phi$ directions respectively, and
\be
\xi_t(r)=
{1\over r\omega^2}\left({1\over\rho}P_1(r)-\Phi_1(r)\right).
\label{eq:xitp}
\ee

Note that there is no $n$ or $m$ dependence in the Eqs.~(\ref{eq:eqone})\,--\,(\ref{eq:xitp}).
The different 
eigenvalues for a given value of $\ell$ are given the label $n$.
Conventionally $n$ can be any signed integer and
can be positive, zero or negative depending on the type of the mode. In general $|n|$ 
represents the number of nodes the radial eigenfunction has in the radial direction.
As mentioned in the Introduction, values of $n > 0$ are used to specify acoustic
modes, $n < 0$ label gravity models and $n=0$ labels f~modes. Of course, this
labelling gets more complicated when mixed modes (see Section~\ref{sec:stars}) are present,
but this works well for the Sun. It is usual to denote the eigenfunctions with $n$ and $\ell$,
thus, for example, the total displacement is denoted as
$\vec\xi_{n,\ell}$ and the radial and transverse components as
$\xi_{r,n,\ell}$ and $\xi_{t,n,\ell}$ respectively.
The lack of the $m$ dependence in the equations has to
do with the fact that we are considering a spherically symmetric system, and
to define $m$ one needs to break the symmetry. Thus, in a spherically
symmetric system, all modes are of a given value of $n$ and $\ell$ are degenerate 
in $m$. Rotation, magnetic fields and other asphericities lift this degeneracy.

An important property of all modes is their mode inertia defined as
\be
E_{n,\ell}= \int_V\rho\vec\xi_{n,\ell}\cdot\vec\xi_{n,\ell}\;d^3\vec r =\int_0^{R}
\rho[\xi_{r,n,\ell}^2+\ell(\ell+1)\xi_{t,n,\ell}^2]r^2 \dd r.
\label{eq:enlo}
\ee
\citep{jcd2003}.
Modes that penetrate deeper into the star (which as we shall see later are low degree modes) have
higher mode inertia than those that do not penetrate as deep (higher degree modes).
Additionally, for modes of a given degree,
higher frequency modes have larger inertia than lower frequency modes.
For a given perturbation, frequencies of low-inertia modes affected by
the perturbation change more than
those of affected high-inertia modes. Since the normalisation of eigenfunctions can be arbitrary, it is
conventional to normalise $E_{n,\ell}$ explicitly as
\be
E_{n,\ell}= \frac{\int_0^{R}
\rho[|\xi_{r,n,\ell}|^2+\ell(\ell+1)|\xi_{t,n,\ell}|^2]r^2 \dd r}
{M\left[|\xi_{r,n,\ell}(R)|^2+\ell(\ell+1)|\xi_{t,n,\ell}(R)|^2\right]},
\label{eq:enl}
\ee
where, $M$ is the total mass and $R$ is the total radius.

\subsection{Boundary conditions}
\label{subsec:bound}

The discussion of the equations of stellar evolution cannot be completed without discussing
the boundary conditions that are need to actually solve the equations.
For details of the boundary conditions and how they can be obtained,
the reader is referred to \citet{cox1980} and \citet{unnoetal}. We give a 
brief overview.

A complete solution of the equations requires four boundary conditions. An inspection of the
equations shows that they have a singular point at $r=0$, however, it is a singularity that
allows both regular and singular solutions. Given that we are talking of physical
systems, we need to look for solutions that are regular at $r=0$. As is usual,
the central boundary conditions are obtained by expanding the solution around
$r=0$. This reveals that as $r\rightarrow 0$, $\xi_r \propto r$ for $\ell=0$ modes,
and $\xi_r \propto r^{\ell-1}$ for others. The quantities $P_1$ and $\Phi_1$ vary as
$r^\ell$.

The surface boundary conditions are complicated by the fact that the ``surface'' of a star
is not well defined in terms of density, pressure etc., but is determined by the
way the stellar atmosphere is treated. In fact once can show that the calculated frequencies
can change substantially depending on where the one assumes the outer
boundary of the star is \citep[see e.g.,][]{hekker2013}. The usual assumption that is made is
that $\Phi_1$ and $\dd\Phi_1/\dd r$ are continuous at the surface. Under the simple assumption that
$\rho_1$ is zero at the surface, the Poisson equation can be solved to show that for
$\Phi_1$ to be zero at infinity,
\be
\Phi_1=A r^{-1-\ell},
\label{eq:bc1}
\ee
where $A$ is a constant. It follow that 
\be
\frac{\dd \Phi_1}{\dd r}+\frac{\ell+1}{r}\Phi_1=0
\label{eq:bc2}
\ee
at the surface.

Another condition can be applied by assuming that the pressure on the deformed surface 
of the star be zero, in other words that the Lagrangian perturbation of pressure be zero
at the surface, i.e.,
\be
\delta P=P_1+\xi_r\frac{\dd P}{\dd r}=0
\label{eq:bc3}
\ee
at the surface.
This condition implies that $\nabla\cdot{\vec\xi}\sim0$ at the outer boundary. 
Combining this with Eq.~(\ref{eq:eqone}) we get $\xi_r=P_1/g\rho$ at the surface. If $\Phi_1$
is neglected, i.e., the so-called Cowling approximation is used, this reduces
to
\be
\xi_r=\xi_t\omega^2.
\label{eq:unno}
\ee

This is worth noting that the equations of stellar oscillation are generally
solved after expressing all the physical quantities in terms of dimensionless 
numbers. The dimensionless frequency $\sigma$ is expressed as
\be
\sigma^2=\frac{R^3}{GM}\omega^2
\label{eq:undimfreq}
\ee
where, $R$ is the radius of the star and $M$ the mass. Similarly the dimensionless
radius ${\hat r}=r/R$,
dimensionless pressure is given by ${\widehat P}=(R^4/GM^2)P$ and the dimensionless density by
${\hat \rho}=(R^3/M)\rho$. These relations lead us to the dimensionless expressions for
the perturbed quantities. It can be shown very easily that
 ${\hat \xi}_r$ the dimensionless
displacement eigenfunction, ${\widehat P}_1$ the dimensionless pressure perturbation, and
${\widehat\Phi}_1$ the dimensionless perturbation to the gravitational potential can be written as
\be
{\widehat \xi}_r=\frac{\xi_r}{R},\quad
{\widehat P}_1=\frac{R^4}{GM^2}P_1,\quad
{\widehat\Phi}_1=\frac{GM}{R}\Phi_1.
\label{eq:undimxi}
\ee

\subsection{Properties of stellar oscillations}
\label{subsec:props}

Equations~(\ref{eq:eqone}), (\ref{eq:eqtwoo}) and (\ref{eq:eqthreee}) can appear rather 
opaque when it comes to understanding the properties of stellar oscillations. The
way to go about trying to distil the properties is to apply a few simplifying assumptions.

The first such assumption is to assume that the perturbation to the gravitational
potential, $\Phi_1$, can be ignored, i.e., the Cowling approximation can be applied.
It can be shown that this approximation applies to modes of
oscillation where $|n|$ or $\ell$ is large.

Applying the Cowling approximation reduces the equations to
\be
\frac{\dd\xi_r}{\dd r}=-{\left( \frac{2}{r}-\frac{1}{\Gamma_1}\frac{1}{H_p}\right )}\xi_r
+\frac{1}{\rho c^2}{\left( \frac{S^2_\ell}{\omega^2}-1\right)}P_1,
\label{eq:sim1}
\ee
and
\be
\frac{\dd P_1}{\dd r}=\rho(\omega^2 - N^2)\xi_r -\frac{1}{\Gamma_1}\frac{1}{H_p}P_1,
\label{eq:sim2}
\ee
where
\be
{H_p}=-\frac{\dd r}{\dd \ln P}
\label{eq:hp}
\ee
is the pressure scale height.

Another assumption made is that we are looking far
away from the centre (and we shall see soon that this applies to
modes of high $\ell$). Under this condition 
the term $2/r$ in the right-hand side of Eq.~(\ref{eq:sim1})
can be neglected. The third assumption is that for high $|n|$ oscillations,
the eigenfunctions vary much more rapidly than the equilibrium quantities.
This assumption applies away from the stellar surface. The implication
of this assumption is that terms containing $H^{-1}_p$ in Eqs.~(\ref{eq:sim1})
and (\ref{eq:sim2}) can be neglected when compared with the quantities
on the left hand sides of the two equations. Thus, Eqs.~(\ref{eq:sim1}) and (\ref{eq:sim2})
reduce to 
\be
\frac{\dd \xi_r}{\dd r}= \frac{1}{\rho c^2}P_1,
\label{eq:simp1}
\ee
and
\be
\frac{\dd P_1}{\dd r}=\rho(\omega^2 - N^2)\xi_r.
\label{eq:simp2}
\ee
These two equations can be combined to form one second-order differential
equation
\be
\frac{\dd^2 \xi_r}{\dd r^2}=\frac{\omega^2}{c^2}\left(1-\frac{N^2}{\omega^2}\right)
\left(\frac{S^2_\ell}{\omega^2}-1\right)\xi_r.
\label{eq:crude}
\ee
Equation~(\ref{eq:crude}) is the simplest possible approximation to equations of
non-radial oscillations, but it is
enough to illustrate some of the key properties.
One can see immediately that the equation does not always
have an oscillatory solution. The solution is oscillatory when
(1) $\omega^2 < S^2_\ell$, and $\omega^2 < N^2$, or when (2)
$\omega^2 > S^2_\ell$, and $\omega^2 > N^2$. The solution is exponential
otherwise.

Eq.~(\ref{eq:crude}) can be written in the form
\be
\frac{\dd^2 \xi_r}{\dd r^2}=K(r)\xi(r),
\ee
where
\be
K(r)=\frac{\omega^2}{c^2}\left(1-\frac{N^2}{\omega^2}\right)
\left(\frac{S^2_\ell}{\omega^2}-1\right).
\label{eq:kr}
\ee

In Figure~\ref{fig:lamb} we show $N^2$ and $S^2_\ell$ plotted as a function of
depth for a model of the present-day Sun. Such a figure is often
referred to as a ``propagation diagram.'' The figure shows that for modes
for which the first condition is true, i.e., $\omega^2 < S^2_\ell$, and $\omega^2 < N^2$,
 are trapped mainly
in the core (since $N^2$ is negative in convection zones, and the Sun has an
envelope convection zone). These
are the \emph{g~modes} and their restoring force is gravity through buoyancy.
Modes that satisfy the second condition, i.e., $\omega^2 > S^2_\ell$, and $\omega^2 > N^2$,
 are oscillatory in the outer
regions, though low-degree modes can penetrate right to the centre.
These are the
\emph{p~modes} and their restoring force is predominantly pressure.
One can see that p~modes of different degrees penetrate to different depths
within the Sun. High degree modes penetrate to shallower
depths than low degree modes. For a given degree, modes of higher
frequency penetrate deeper inside the star than modes of lower frequencies. 
Thus, modes of different degrees
 sample different layers of a star. Note that high-$\ell$ modes are
concentrated in the outer layers, justifying to some extent the
neglect of the $2/r$ term in Eq.~(\ref{eq:sim1}).

\epubtkImage{}{%
\begin{figure}[htb]
\centerline{\includegraphics[height=15pc]{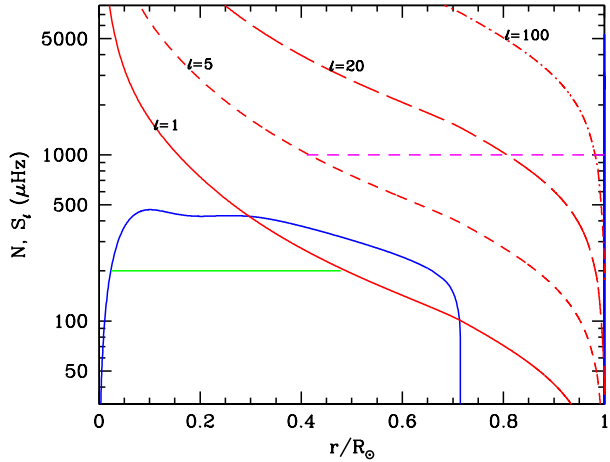}}
\caption{The propagation diagram for a standard solar model. The blue line is the
buoyancy frequency, the red lines are the Lamb frequency for different degrees.
The green solid horizontal line shows the region where a $200\ \mu\mathrm{Hz}$ g~mode can propagate. The pink 
dashed horizontal line
shows where a $1000\ \mu\mathrm{Hz}$ $\ell=5$ p~mode can propagate.
}
\label{fig:lamb}
\end{figure}}

\subsubsection{P modes}
\label{para:pmodes}

As mentioned above p~modes have frequencies with $\omega^2 > S^2_\ell$ and 
$\omega^2> N^2$. The modes are trapped between the surface and a lower or
inner turning point $r_t$ given by $\omega^2=S^2_\ell$, i.e.,
\be
\frac{c_2^2(r_t)}{r_t^2}=\frac{\omega^2}{\ell({\ell+1})}.
\label{eq:rt}
\ee
This equation can be used to determine $r_t$ for a mode of given $\omega$ and $\ell$.

For high frequency p~modes, i.e., modes with $\omega\gg N^2$, $K(r)$ in Eq.~(\ref{eq:kr})
can be approximated as
\be
K(r)\simeq\frac{\omega^2-S^2_\ell(r)}{c^2(r)},
\label{eq:pkr}
\ee
showing that the behaviour of high-frequency p~modes is determined predominantly by the
behaviour of the sound-speed profile, which is not surprising since these are pressure, i.e.,
sound waves. The dispersion relation for sound waves is $\omega^2=c^2|k|^2$ where
${\vec k}$ is the wave-number that can be split into a radial and a horizontal
parts, and $k^2=k_r^2+k_h^2$, where $k_r$ is the radial wavenumber, and $k_h$ the horizontal
one.
At the lower turning point, the wave has no
radial component and hence, the radial part of the wavenumber, $k_r$, vanishes,
which leads to
\be
k^2_r=\frac{\omega^2-S^2_\ell(r)}{c^2(r)},
\label{eq:kradial}
\ee
which immediately implies (and which can be derived rigorously) that
\be
k^2_t=\frac{\ell(\ell+1)}{r^2}.
\label{eq:khoriz}
\ee

A better analysis of the equations \citep[see][]{jcd2003} shows that since we are talking
of normal modes, there are further conditions on $K(r)$. In particular, the requirement
that the modes have a lower turning point $r_t$ and an upper turning point at $r_u$ requires
\be
\int_{r_t}^{r_u} K(r)^{1/2}\dd r = \left(n-\frac{1}{2}\right)\pi.
\label{eq:wkkb}
\ee
We have the approximate expression for $K$ (Eq.~\ref{eq:pkr}) but the analysis that lead to it 
had no notion of an upper turning point. We just assume that the upper
turning point is at $r=R$. Thus, the ``reflection'' at the upper turning point
does not necessarily produce a phase-shift of $\pi/2$, but some unknown shift which we call
$\alpha_p\pi$. In other words
\be
\int_{r_t}^R K(r)^{1/2}\dd r=\int_{r_t}^R (\omega^2-S^2_\ell)^{1/2}\dd r = (n+\alpha_p)\pi.
\label{eq:begduv}
\ee
Since $\omega$ does not depend on $r$, Eq.~(\ref{eq:begduv}) can be rewritten
as
\be
\int_{r_t}^R {\left( 1-\frac{L^2}{\omega^2}\frac{c^2}{r^2}\right )}^{1/2}\frac{\dd r}{c} = 
\frac{(n+\alpha_p)\pi}{\omega},
\label{eq:duv1}
\ee
where $L=\sqrt{\ell(\ell+1)}$, though a better approximation is that $L=\ell+1/2$. 
The LHS of the equation is a function of $w\equiv\omega/L$, and the
the equation is usually written as
\be
F(w)=\frac{(n+\alpha_p)\pi}{\omega},
\label{eq:duvall}
\ee
where
\be
F(w)=\int_{r_t}^R {\left( 1-\frac{L^2}{\omega^2}\frac{c^2}{r^2}\right )}^{1/2}\frac{\dd r}{c}.
\label{eq:fw}
\ee
Eq.~(\ref{eq:duvall}) is usually referred to as the `Duvall Law'. 
\citet{duvall1982} plotted $(n+\alpha_p)\pi/\omega$ as a function of $w=\omega/L$ and
showed that all observed frequencies collapse into a single function of $w$. 
As we can see from Eq.~(\ref{eq:rt}), $w$ is related to the lower turning point of
a mode since
\be
w\equiv\frac{\omega}{\sqrt{\ell({\ell+1}})}=\frac{c^2(r_t)}{r_t}.
\label{eq:wrt}
\ee
A version of the Duvall-law
figure with more modern data is shown in Figure~\ref{fig:duvall}. As we shall
see later, the Duvall Law can be used to determine the solar sound-speed profile
from solar oscillation frequencies.

\epubtkImage{}{%
\begin{figure}[htb]
\centerline{\includegraphics[height=15pc]{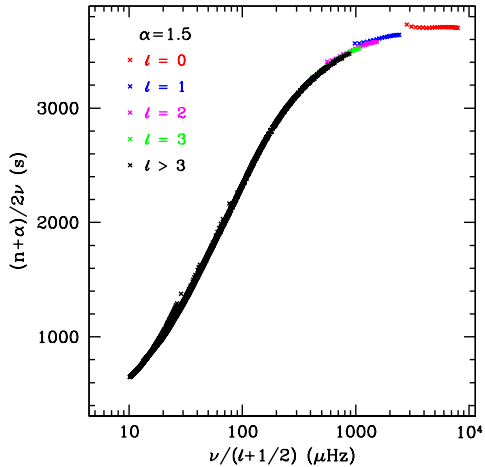}}
\caption{The Duvall Law for a modern dataset. The frequencies used in this
figure are a combination of
BiSON and MDI frequencies, specifically we have used modeset BiSON-13 of \citet{basuetal2009}. Note that 
all modes fall on a narrow curve for $\alpha_p=1.5$. The curve would have been narrower had $\alpha_p$ 
been allowed to be a function of frequency.
}
\label{fig:duvall}
\end{figure}}

A mathematically rigorous asymptotic analysis of the equations \citep[see e.g.,][]{tassoul1980}
shows that the frequency
of high-order, low-degree p~modes can be written as 
\be
\nu_{n,\ell}\simeq(n+\frac{\ell}{2}+\alpha_p)\Delta\nu,
\label{eq:tassoul}
\ee
where
\be
\Delta\nu={\left(2\int_0^R \frac{\dd r}{c}\right)}^{-1}
\label{eq:dnu}
\ee
is twice the sound-travel time of the Sun or star. The time it takes sound to travel from
the surface to any particular layer of the star is usually referred to as the ``acoustic depth.''
$\Delta\nu$ is often referred to as the ``large frequency spacing'' or ``large frequency separation''
and is the frequency difference between two modes of the same $\ell$ but consecutive values of $n$.
Eq.~(\ref{eq:tassoul}) shows us that p~modes are equidistant in frequency.
It can be shown that the large spacing scales as the square root of density \citep{ulrich1986,jcd1988}.
The phase shift $\alpha_p$ however, is generally frequency dependent and thus the spacing
between modes is not strictly a constant.

A higher-order asymptotic analysis of the
equations \citep{tassoul1980,tassoul1990} shows that
\be
\nu_{n,\ell}\simeq\left( n+\frac{\ell}{2}+\frac{1}{4}+\alpha_p\right)\Delta\nu
-(AL^2-\delta)\frac{\Delta\nu^2}{\nu_{n,\ell}},
\label{eq:tassoul1}
\ee
where, $\delta$ is a constant and
\be
A={\frac{1}{4\pi^2\Delta\nu}}
\left[ \frac{c(R)}{R} - \int_0^R\frac{\dd c}{\dd r}\frac{\dd r}{r}.
\right]
\label{eq:A}
\ee
When the term with the surface sound speed is neglected we get
\be
\nu_{n\ell}-\nu_{n-1,\ell+2}\equiv\delta\nu_{n,\ell}\simeq -(4\ell+6)\frac{\Delta\nu}{4\pi^2\nu_{n,\ell}}
\int_0^R\frac{\dd c}{\dd r}\frac{\dd r}{r}.
\label{eq:smallsep}
\ee
$\delta\nu_{n\ell}$ is called the small frequency separation. From
Eq.~(\ref{eq:smallsep}) we see that $\delta\nu_{n\ell}$ is sensitive to the gradient of
the sound speed in the inner parts of a star. The sound-speed
gradient changes with evolution as hydrogen is replaced by heavier helium making
$\delta\nu_{nl}$ is a good diagnostic of the evolutionary stage of a
star.
For main sequence stars, the average value of $\delta\nu$ decreases 
monotonically with the central hydrogen abundance
and can be used in various forms to calibrate age if metallicity is known
\citep[][etc]{jcd1988,mazumdariwr2003,mazumdar2005}.

\subsubsection{G modes}
\label{para:gmodes}

G~modes are low frequency modes with $\omega^2 < N^2$ and $\omega^2 < S^2_\ell$. The 
turning points of these modes are defined by $N=\omega$. Thus, in the case of the Sun
we would expect g~modes to be trapped between the base of the convection zone and the
core.

For g~modes of high order, $\omega^2 \ll S^2_\ell$ and thus
\be
K(r)\simeq \frac{1}{\omega^2}(N^2-\omega^2)\frac{\ell(\ell+1)}{r^2}.
\label{eq:kgmode}
\ee
In other words, the properties of g~modes are dominated by the buoyancy frequency $N$.
The radial wavenumber can be shown to be
\be
k^2_r=\frac{\ell(\ell+1)}{r}{\left(\frac{N^2}{\omega^2}-1\right)}.
\label{krgmode}
\ee

An analysis similar to the one for p~modes show that for g~modes frequencies are determined
by
\be
\int_{r_1}^{r_2} L\left(\frac{N^2}{\omega^2}-1\right)^{1/2}\frac{\dd r}{r}=\left(n-\frac{1}{2}\right)\pi,
\label{eq:gmd}
\ee
where, $r_1$ and $r_2$ mark the limits of the radiative zone. Thus
\be
\int_{r_1}^{r_2} \left(\frac{N^2}{\omega^2}-1\right)^{1/2}\frac{\dd r}{r}=\frac{(n-1/2)\pi}{L}= G(w),
\label{eq:gduv1}
\ee
an expression which is similar to that for p~modes (Eq.~\ref{eq:duvall}) but showing that
the buoyancy frequency plays the primary role this case.

A complete asymptotic analysis of g~modes \citep[see][]{tassoul1980} shows that the
frequencies of high-order g~modes can be approximated as
\be
\omega=\frac{L}{\pi(n+\ell/2+\alpha_g)}\int_{r_1}^{r_2}N\frac{\dd r}{r},
\label{eq:gmodes}
\ee
where $\alpha_g$ is a phase that varies slowly with frequency.
This shows that while p~modes are equally spaced in frequency, g~modes are equally spaced
in period.

\subsubsection{Remaining issues}
\label{para:upper}

The analysis presented thus far does not state anything explicitly about the upper turning point
of the modes. It is assumed that all modes get reflected at the surface. This is not really the
case and is the reason for the inclusion of the unknown phase factors $\alpha_p$ and $\alpha_g$
in Eqs.~(\ref{eq:begduv}) and (\ref{eq:tassoul}). The other limitation is that we do not see
any f~modes in the analysis.

The way out is to do a slightly different analysis of the equations without neglecting the
pressure and density scale heights but assuming that curvature can be neglected.
Such an analysis was presented by \citet{deubnerandgough1984} who followed the analysis of
\citet{lamb1932}. They showed that under the Cowling approximation, one could approximate the
equations of adiabatic stellar oscillations to
\be
\frac{\dd^2\Psi}{\dd r^2}+K^2(r)\Psi=0,
\label{eq:lambp}
\ee
where $\Psi=\rho^{1/2}c^2\nabla\cdot{\vec \xi}$, and the wavenumber $K$ is given by
\be
K^2(r)=\frac{\omega-\omega^2_c}{c^2}+\frac{\ell(\ell+1)}{r^2}{\left( \frac{N^2}{\omega^2-1}\right)},
\label{eq:lambk}
\ee
with
\be
\omega^2_c=\frac{c^2}{4H_\rho^2}\left(1-2\frac{\dd H_\rho}{\dd r}\right),
\label{eq:cutoff}
\ee
where $H_\rho$ is the density scale height given by $H_\rho=-\dd r/\dd\ln\rho$. The
quantity $\omega_c$ is known as the ``acoustic cutoff'' frequency. The
radius at which $\omega=\omega_c$  defines the upper
limit of the cavity for wave propagation and that radius is usually called
the upper turning point of a mode. For isothermal atmospheres the acoustic
cutoff frequency is simply $\omega_c=c g \rho/2P$ \citep[see e.g.,][]{balmforthandgough1990}.
Figure~\ref{fig:acous} shows the acoustic cutoff frequency for a solar model. Note that the
upper turning point of low-frequency modes is much deeper than that of high-frequency modes. 
The effect of the location of the upper turning point of a mode is seen in the 
corresponding eigenfunctions as well -- the amplitudes of the eigenfunctions decrease towards the
surface.  This
results in higher mode-inertia normalised to the surface displacement (see Eq.~\ref{eq:enl})
for low-frequency modes compared to their high-frequency
counterparts at the same value of $\ell$.

\epubtkImage{}{%
\begin{figure}[htb]
\centerline{\includegraphics[height=15pc]{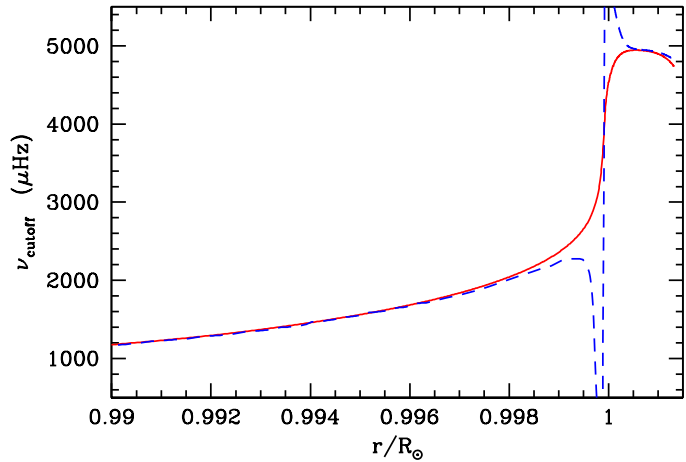}}
\caption{The acoustic cut-off frequency of a solar model calculated as per Eq.~(\ref{eq:cutoff})
is shown as the blue dashed line.
The red curve is the cut-off assuming that the model has an isothermal atmosphere. Note that the
lower frequency modes would be reflected deeper inside the Sun than higher frequency modes.
}
\label{fig:acous}
\end{figure}}

Eq.~(\ref{eq:lambp}) can be solved under the condition $\Psi=0$. These are the f~modes.
It can be shown that the f-mode dispersion 
relation is
\be
\omega^2\simeq gk,
\label{eq:fmode}
\ee
$k$ being the wavenumber. Thus, f-mode frequencies are almost independent of the stratification of
the Sun.  As a result, f-mode frequencies have not usually been used to
determine the structure of the Sun. However, these have been used to
draw inferences about the solar radius \citep[e.g.,][]{schouetal1997,antia1998,sandrine2007}

\newpage

\section{A Brief Account of the History of Solar Models}
\label{sec:history}

The history of solar models dates back to the 1940s.
Among the first published solar models is that of \citet{schwarz1946}. This model was constructed
at a time when it was believed that the CNO cycle was the source of solar energy. By construction, 
this model did not
have a convective envelope, but had a convective core instead. This model
does not, of course, fall under the rubric of the standard solar model -- that concept 
was not defined 
till much later. As the importance of the $p$-$p$ chain came to be recognised, \citet{schwarz1957} 
constructed models with the $p$-$p$ chain as the source 
of energy; this model included a convective envelope. These models showed how the model
properties depended
on the heavy-element abundance and how the initial helium abundance
could be adjusted to construct a model that had the correct luminosity. The central temperature and density of
the models fall in the modern range, however, the adopted heavy element abundance
is very different from what is observed now. In the intervening years, several others
had constructed solar models assuming radiative envelopes and homogeneous
compositions \citep[see e.g.,][]{epstein1953, naur1954, ogden1955}; of course we now
know that none of these 
models represent the Sun very well.

The 1960s saw a new burst of activity in terms of construction of solar models. The development of
new numerical techniques such as the Henyey method for solving the stellar structure equations
\citep{henyey1959} made calculations easier. An added impetus was provided by the development
of methods to detect solar neutrinos \citep[e.g.,][]{davis1955}. This resulted in the construction
of models to predict neutrino fluxes from the Sun, e.g., \citet{pochoda1964}, \citet{sears1964} 
and \citet{bahcall1963}. 
 This was a time when investigations were carried out to examine how changes to input parameters
change solar-model predictions
\citep[e.g.,][]{demarque1964, ezer1965, bahcall1968a, bahcall1968b, Iben1968,salpeter1969,torres1969}. 
This was also the period when nuclear reaction rates and radiative opacities were modified steadily.

The 1970s and early 1980s saw the construction of solar models primarily with the aim of
determining neutrino fluxes. This is when the term ``standard solar model''
was first used \citep[see e.g.,][]{bah1972}. It appears that the origin of
the term was influenced by particle physicists working on solar neutrinos
who, even at that time, had a standard model of particle physics (Pierre Demarque, \textit{private
communication}).  The term ``non-standard'' models also came into play at this
time. 
An example of an early non-standard model, and classified as such by \citet{bah1972},
is that of \citet{ezer1965} constructed with a time-varying gravitational
constant $G$.
Improvements in inputs to solar models led to
many new solar models being constructed. 
\citet{bahcall1982}, \citet{jnbulrich1988} and \citet{tc1988} for instance looked at what happens to standard 
models when different microphysics inputs are changed. For a history of solar
models from the perspective of neutrino physics, readers are referred to \citet{jnb2003}.

The 1980s was when helioseismic data began to be used to examine what can be said of solar models,
and by extension, the Sun. \citet{jcddog1980} compared frequencies of models to observations
to show that none of the models examined was an exact match for the Sun.
\citet{jnbulrich1988} compared the global seismic parameters of many models.
During this time investigators also started examining how the p-mode frequencies of models
change with model inputs. For instance, \citet{jcd1982s} and \citet{guenther1989} examined how the frequencies of
solar models changed with change in opacity. This was also when the first solar models with
diffusion of heavy elements were constructed \citep[see e.g.,][]{coxetal1989}. Ever since
it was demonstrated that the inclusion of diffusion increases the match of solar models with the
Sun \citep{jcdetal1993}, diffusion has become a standard ingredient of standard solar models.

Standard solar models have been constructed and updated continuously as different
microphysics inputs have become available. Descriptions of many standard models have been
published. Helioseismic tests of these models have helped examine the inputs to these
models. Among published models are those of \citet{jnbpin1992}, \citet{jcdetal1996}, \citet{guzikandswenson1997},
\citet{bahcalletal1995},
\citet{dave1996}, \citet{dave1997}, \citet{bahcalletal1998}, \citet{brun1998}, 
\citet{basuetal2000b} \citet{neuforge2001a, neuforge2001b},
\citet{couvidat2003}, \citet{bahcalletal2005, bahcalletal2005b}, \citet{bahcalletal2005c}, etc.

Many non-standard models have been constructed with a 
variety of motives. For instance, \citet{ezer1965}, as well as \citet{roeder1966}, constructed solar models
with a time-varying value of the gravitational constant $G$ following
the Brans--Dicke theory. More modern solar models with time-varying
$G$ were those of \citet{demarque1994} and \citet{guenther1995} who were investigating whether
solar oscillation frequencies could be used to constrain the time-variation of $G$. \citet{jcdetal2005}
on the other hand, tried to examine whether helioseismic data can constrain the value of $G$ given that
$G\msun$ is known extremely precisely.
Another set of non-standard models are ones that include early mass loss in the
Sun. The main motivation for these models is to solve the so-called ``faint Sun paradox''. Models in this
category include those of \citet{guziketal1987} and \citet{sack2003}.

A large number of non-standard models were constructed with the sole purpose of reducing the predicted
neutrino flux from the models and thereby solving the solar neutrino problem (see Section~\ref{subsec:neutrino} for
a more detailed discussion of this issue). 
These include models with extra mixing \citep[][etc.]{bah1968, scahtz1985, rox1985, richard1997}.
And some models were constructed to have low metallicity in the core with accretion of high-$Z$ materials
to account for the higher metallicity at the surface \citep[e.g.,][]{jcdetal1979, winnick2002}.
Models that included effects of rotation were also constructed \citep{pinetal1989}.

More recently, non-standard solar models have been constructed to 
try and solve the problems solar models face if they are constructed with solar
abundances as advocated by \citet{ags05} and \citet{agss09}.
This issue has been reviewed in detail by \citet{sbhma2008} and more recent
updates can be found in \citet{sbhma2013}, and \citet{sbetal2014}. We discuss
this issue in Section~\ref{subsec:abun}.

There is another class of solar models, the so-called `seismic models' that have also been
constructed. These models are constructed with helioseismically derived constraints in
mind. We discuss those in Section~\ref{subsec:seismic}.

\newpage

\section{Frequency Comparisons and the Issue of the `Surface Term'}
\label{sec:surf}

Like other fields in astronomy, as helioseismic data became available, researchers
started testing how good their solar models were. This was done by comparing the computed
frequencies of the models with the observed solar frequencies. Examples of this
include \citet{jcddog1980}, \citet{jcddog1981}. \citet{jcdlebreton1988}, etc. This practice continued till quite
recently \citep[e.g.,][etc.]{coxetal1989, guzikandcox1991, guentheretal1992, guzikandcox1993, sack2003}.
Such comparisons are of course
quite common in the field of classical pulsators where $O-C$ (i.e., observed minus computed
frequency) diagrams are considered standard. However, in the case of solar 
models such a comparison has pitfalls, as we discuss below. 

\epubtkImage{}{%
\begin{figure}[htb]
\centerline{\includegraphics[height=15pc]{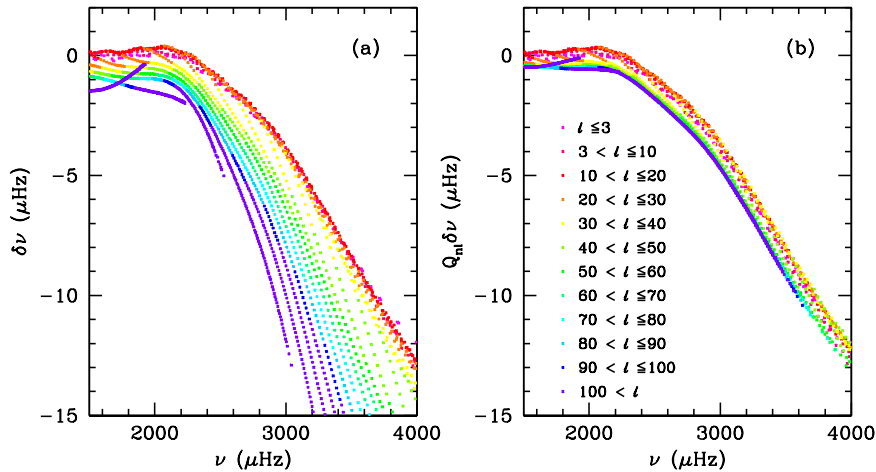}}
\caption{The differences in frequencies of the Sun (mode set BiSON-13 of \citealp{basuetal2009}) and that of standard solar model known as Model~S\citep{jcdetal1996}.
Panel (a) shows the raw frequency differences; Panel (b)
shows scaled differences, where the scaling factor $Q_{n\ell}$ corrects for the fact that
modes with lower inertia change more for a given perturbation compared to modes
with higher inertia.
}
\label{fig:modelS}
\end{figure}}

In Figure~\ref{fig:modelS} we show the differences in frequencies of the Sun and the
solar model Model~S of \citet{jcdetal1996}. We see in Figure~\ref{fig:modelS}(a) that the 
predominant frequency difference is a function of frequency, however, there is a clear 
 dependence on degree. This $\ell$ dependence is mainly due to the fact that higher $\ell$ modes
have lower mode inertia and hence, perturbed easily compared to modes of lower-$\ell$ (and hence
higher inertia). 
This effect can be corrected for by scaling the
frequency differences with their mode inertia. In practice, to ensure that both
raw and scaled difference have the same order of magnitude, the differences
are scaled with the quantity
\be
Q_{n\ell}=\frac{E_{n,\ell}(\nu)}{E_{\ell=0}(\nu)},
\label{eq:qnl}
\ee
where
$E_{n,\ell}(\nu)$ is the mode inertia of a mode of degree $\ell$, radial order
$n$ and frequency $\nu$ defined in Eq.~(\ref{eq:enl}), and $E_{\ell=0}(\nu)$ is the
mode inertia of a radial mode at the same frequency. The quantity $E_{\ell=0}(\nu)$ is obtained 
by interpolation between
the mode inertias of radial modes of different orders (and hence different frequencies).
The scaled frequency-differences between the Sun and Model~S can be
seen in Figure~\ref{fig:modelS}(b). Note that most of the $\ell$-dependence disappears and one is
left with frequency differences that are predominantly a function of frequency. If we were 
using these difference to determine whether Model~S is a good model of the Sun, we
could not draw much of a conclusion.

\epubtkImage{}{%
\begin{figure}[htb]
\centerline{\includegraphics[height=15pc]{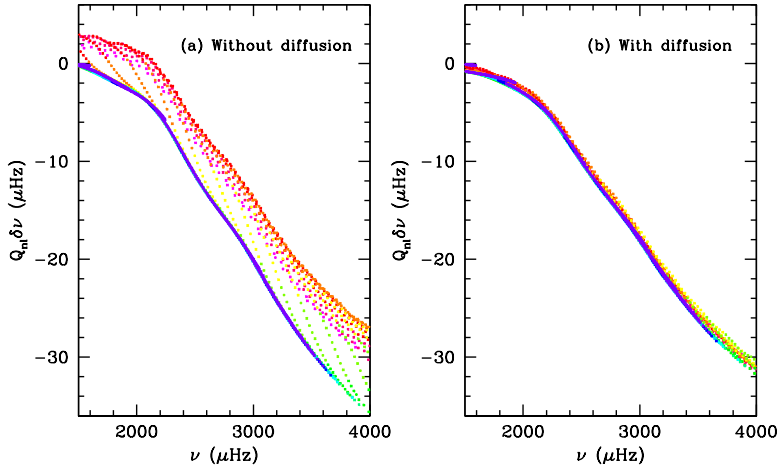}}
\caption{The scaled frequencies differences between the Sun and two standard models, one
without diffusion (Panel a) and one with (Panel b). The BiSON-13 mode set \citep{basuetal2009}
is used for the Sun. The two models are models NODIF and STD of \citet{basuetal2000b}.
Colours are the same as in Figure~\ref{fig:modelS}.
}
\label{fig:difnodif}
\end{figure}}

Similar issues are faced when we compare frequency-differences of different
models with respect to the Sun. As an example in Figure~\ref{fig:difnodif} we show the
scaled frequency difference between the Sun and two models, one that does not include
the diffusion and gravitational settling of helium and heavy elements and one
that other; the remaining physics inputs are identical and the models
were constructed with the same code. One can see that in
both cases the predominant frequency difference is a function of frequency. There is a
greater $\ell$ dependence in the model without diffusion, and this we know now is a result
of the fact that the model without diffusion has a very shallow convection zone compared with 
the Sun. However, quantitatively, it is not possible to judge which model is better from
these frequency differences alone. The $\chi^2$ per degree of freedom calculated for
the frequency differences are extremely large (of the order of $10^5$) given the
very small uncertainties in solar frequency measurements.  Consequently, to try and
draw some conclusion about which model is better, we look
at the root-mean-square frequency difference instead.
The root-mean-square frequency difference for the
the model with diffusion is $17.10\ \mu\mathrm{Hz}$, while that for the model without diffusion
is $17.03\ \mu\mathrm{Hz}$. Thus, if the purpose of comparing frequencies was to determine
which model is better compared with the Sun, we have failed.
As we shall see later in Section~\ref{subsec:details}, the model with diffusion is
in reality the better of the two models in terms of match with the
Sun.

\epubtkImage{}{%
\begin{figure}[htb]
\centerline{\includegraphics[height=25pc]{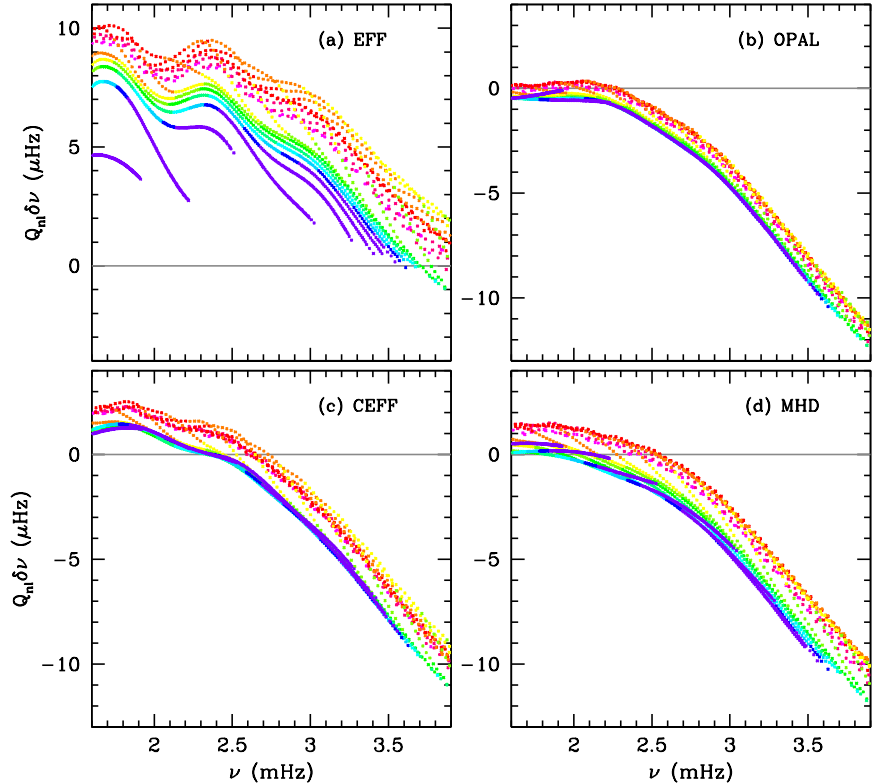}}
\caption{The scaled frequency differences between the Sun and four models 
constructed with different equations of state but otherwise identical inputs. The
OPAL model is Model~S of \citet{jcdetal1996}. 
Colours are the same as in Figure~\ref{fig:modelS}.
(Models courtesy of J.~Christensen-Dalsgaard.)
}
\label{fig:eos}
\end{figure}}

Similar issues are faced if we are to compare other physics, such as the equation of
state. In Figure~\ref{fig:eos} we show the frequency difference between the Sun four models, 
each constructed with identical inputs, except for the equation of state. The EFF model
has an RMS frequency difference of 5.5, the CEFF model of 4.8, the MHD model has an
RMS difference of 5.3 and the OPAL model (this is Model~S) has an RMS difference 
of 6.0. This we could be tempted to say that the CEFF model is the best and OPAL the
worst, and the CEFF
equation of state is the one we should use. However, as we shall see later, that
is not the case: while CEFF is not too bad, OPAL is actually the best. 
In fact the spread in the differences of modes of different degrees
is the clue to which model is better -- generally speaking, the larger the spread, the worse the 
model. The models shown in Figure~\ref{fig:difnodif} were constructed using a different code
and different microphysics and atmospheric inputs than the models shown in Figure~\ref{fig:eos},
this gives rise to the large difference in the value of the RMS frequency differences for the
different sets of models, again highlighting problems that make interpreting frequency
differences difficult.

The reason behind the inability to use frequency comparisons to distinguish which
of the two models is two-fold. The first is that the frequencies of all the models were
calculated assuming that the modes are fully adiabatic, when in reality adiabaticity
breaks down closer to the surface. This of course, can be rectified by
doing non-adiabatic calculations, as have been done by \citet{jcdetal1975}, \citet{guzikandcox1991},
\citet{guzikandcox1992}, \citet{guenther1994}, \citet{rosenthaletal1995}, etc. However,
this is not completely straightforward since there is no consensus on how non-adiabatic
calculations should be performed. The main uncertainty involves accounting for the influence of
convection, and often damping by convection is ignored. 
The second reason is that there are large uncertainties
in modelling the near-surface layers of stars. These uncertainties again include the
treatment of convection. Models generally use the mixing length approximation or 
its variants and as a result, the region of inefficient convection
close to the surface is modelled in a rather crude manner. The models do not include the dynamical
effect of convection and pressure support due to turbulence either; this is important
again in the near surface layers. The treatment of stellar atmospheres is 
crude too, one generally uses simple atmospheric models such as the Eddington $T$-$\tau$ relation 
or others of a similar nature, and these are often, fully radiative, grey atmospheres and do not
include convective overshoot from the interior. Some of the microphysics inputs can
be uncertain too, in particular low-temperature opacities that have to include molecular
lines as well as lines of elements that cause line-blanketing. Since all these
factors are relevant in the near-surface regions, their combined effect is usually
called the ``surface effect'' and the differences in frequencies they
introduced is referred to as the ``surface term.'' Even if we could calculate non-adiabatic
frequencies properly and alleviate some of the problems, the issues with modelling
will still give rise to a surface term.

The surface term also hampers the inter-comparison of models. Figure~\ref{fig:modeldif} shows the
frequency differences between Model~S and two other standard solar models, model BP04
of \citet{bahcalletal2005} and model BSB(GS98) (which we simply refer to as BSB)
of \citet{bahcalletal2006}. Also shown are the
relative sound-speed and density differences between Model~S and the two models.
As can be seen while the structural differences between Model~S and two models 
are similar, the frequency differences are very different. Most of the frequency differences can be attributed
to how the outer layers of the models were constructed.

\epubtkImage{}{%
\begin{figure}[htb]
\centerline{\includegraphics[height=15pc]{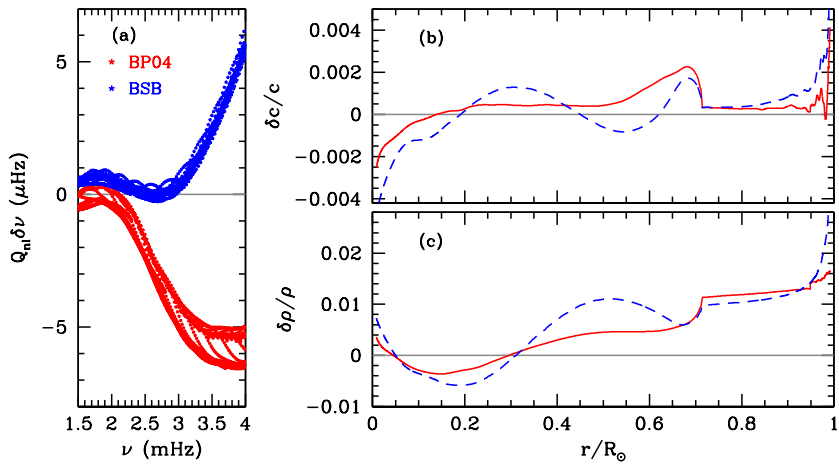}}
\caption{Panel (a): The scaled frequency differences between Model~S and two other solar
models. Panel (b) The relative sound-speed differences, and Panel (c) the relative density
differences between the same models. All differences are in the sense (Model~S\,--\,other~model).
}
\label{fig:modeldif}
\end{figure}}

For low and intermediate degree modes, the surface term depends only on the frequency
of the mode once mode-inertia is taken into account. A simple way to visualise this would
be to look at modes close to their upper turning points. In that regions, these
modes travel almost radially, regardless of their degree. The only difference
is brought about by the fact that modes of different frequencies are reflected at
different layers in the near surface layers as was shown in Figure~\ref{fig:acous}. 
The effect of near-surface uncertainties is the least in low-frequency modes
that have upper turning points deeper insider a star, and increases with 
mode-frequency since increasing mode frequencies push the upper turning point to 
progressively shallower layers were the uncertainties dominate. This is why
the scaled frequency differences between Sun and the models increase with frequency.
The fact that in Figure~\ref{fig:eos} the EFF model shows large difference
with respect to the Sun at low frequencies points to serious issues with the
model in the deeper layers. However, this is not full picture. 
\citet{jcdmjt1997} investigated the reasons for the small frequency shifts
in low-frequency modes that result from near-surface changes and challenged the
view that the deeper upper-turning points of the low-frequency modes caused the
small difference at low frequencies. They showed that the small shifts could be a
result of near-cancellation of different contributions, which are
individually much larger than the results shifts. The small shifts can be more
easily explained if Lagrangian differences (i.e., differences at fixed mass), rather than 
 Eulerian differences (differences at fixed radius) are considered; they found that 
the Lagrangian differences are indeed confined closer to the surface and explain the
frequency shifts more naturally.

The surface term is not just any function of frequency, it is a \emph {smooth} function
of frequency that can be modelled as a low-degree polynomial.
 \citet{dog1990} showed that any localised feature in the sound speed
inside a star introduces an oscillatory term in frequencies as a function of radial
order $n$ which is proportional to
\be
\sin(2\tau_m \omega_{n,\ell}+\phi),
\label{eq:glitch}
\ee
where, $\tau_m$ is the acoustic depth (i.e., the time it takes for
sound to travel from the surface to a given layer) of the localised feature (usually
referred to as an ``acoustic glitch''), and 
\be
\tau_m=\int_{r_m}^R \frac{\dd r}{c},
\label{eq:tau}
\ee
where $r_m$ is the radial position of the feature. As can be seen from Eq.~(\ref{eq:glitch}), the
smaller the acoustic depth of the glitch (i.e., the shallower the layer in
which the glitch occurs), the larger the `wavelength' of the frequency
modulation.
 Our inability to model the near-surface layers of the Sun properly results
in near-surface acoustic glitches in our models when compared with the Sun. Because these glitches arise
at shallow depths, i.e., at small values of $\tau_m$, the effect on the frequency does
not look like a sinusoidal modulation, but merely a smooth function that can be described
as a low-order polynomial.

\epubtkImage{}{%
\begin{figure}[htb]
\centerline{\includegraphics[height=15pc]{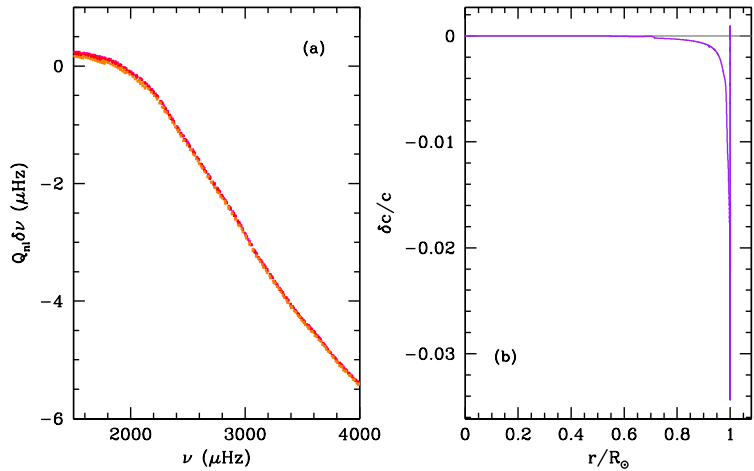}}
\caption{(a) The scaled frequencies differences between a solar model constructed
with the Eddington $T$--$\tau$ relation and one with the Krishna Swamy $T$--$\tau$ relation.
Only differences of modes with $\ell <= 50$ are shown.
(b) The relative sound-speed difference between the two models.
The differences are in the sense (Eddington\,--\,Krishna Swamy).
}
\label{fig:eddiks}
\end{figure}}

The smoothness of the surface term can be demonstrated by making models with different
inputs that affect the near-surface layers. In Figure~\ref{fig:eddiks} we show the 
scaled frequency differences between two standard solar models, one constructed with
the Eddington $T$--$\tau$ relation and the other the Krishna Swamy $T$--$\tau$ relation. Also
shown is the the relative difference between the sound-speed profiles between the two
models. Note that the sound-speed differences significant only close to the surface, and
that the frequency differences are smooth. Similar results are obtained when we
have models with different formulations of convection but otherwise identical
inputs (Figure~\ref{fig:mltcm}).

\epubtkImage{}{%
\begin{figure}[htb]
\centerline{\includegraphics[height=15pc]{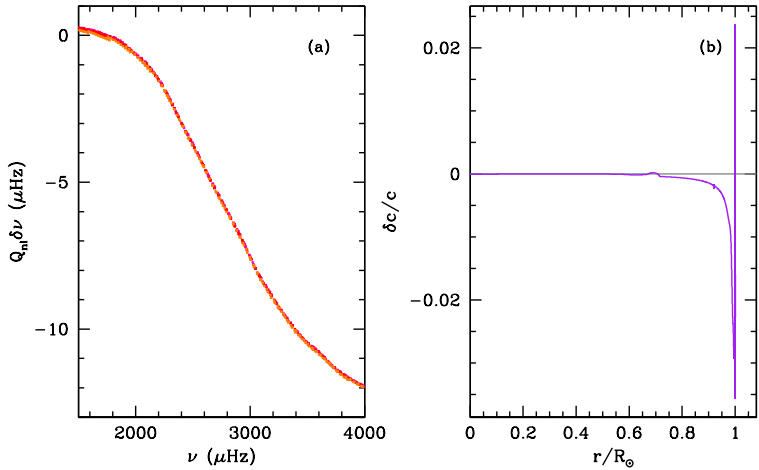}}
\caption{(a) The scaled frequencies differences between a solar model constructed
with the conventional mixing length treatment and one with the Canuto--Mazzitelli (CM) formulation.
Only differences of modes with $\ell <= 50$ are shown.
(b) The relative sound-speed difference between the two models.
The differences are in the sense (MLT\,--\,CM).
}
\label{fig:mltcm}
\end{figure}}

While the surface term is a nuisance, it is by no means devoid of information. Asymptotically, the
influence of the upper layers of a star on p~mode frequencies can be described as a phase function
which  essentially depends on the frequency of the modes. \citet{fernando1992,fernando1994a, fernando1994b}
did  extensive studies of the surface term and derived kernels for the term relating
it to near-surface differences in structure. \citet{fernando1994b} even used it to determine the
convection zone helium-abundance of the Sun. \citet{hmasb1994} and \citet{sbhma1995} also used the
surface term, though in a different manner, to determine the solar helium abundance; they however,
found that the effect of the equation of state on the surface term resulted in systematic
uncertainties in the estimate of the solar convection-zone helium abundance.

Since one of the major sources of the surface term is the treatment of convection, there have
been attempts to change and improve the treatment of convection in solar models. \citet{rosenthaletal1995}
showed that the non-local mixing length formulation of \citep{dog1977} as applied
by \citet{balmforth1992} reduces the surface term
with respect to the Sun. These non-local formulations however, often use more than one free parameter,
which is unsatisfactory. 
The most promising avenue however, seems to be to inclusion of dynamical effects from numerical simulations
in models. \citet{demarque1997} showed that by parametrizing the structure of the super-adiabatic layer
seen in simulations and applying that to solar models improves frequencies. 
Instead of parametrizing the effects of convective dynamics, \citet{rosenthaletal1998,rosentaletal1999}
attached their simulation results to 
 models and showed that the surface term improves. This approach was also used by
\citet{piauetal2014} who, in addition to normal hydrodynamic simulations, also
looked at the result of using magnetohydrodynamic simulations. This way of
``pasting'' the simulation results on a solar model however, does not allow one to construct evolutionary
models of the Sun. \citet{lietal2002} tried a different approach. They explicitly
include the effects of turbulence in stellar structure equations and determined
the quantities from simulations. However, this too has the problem that models can only be constructed
for the present day Sun. Simulations of convection show that properties
of stellar convection change with $\log g$ and $T_{\rm eff}$ \citep[][etc.]{ludwigetal1999, robinsonetal2004, tanner2013a, 
trampedachetal2013, magicetal2013}, thus a consistent treatment of realistic convection properties in
solar models constructed from the zero-age main sequence is difficult.

The difficulty in drawing conclusions about models and their constituent microphysics by comparing
frequencies have led to the development of other techniques, such as inversions, to determine
solar structure and physics of the solar interior. Most helioseismic results are a result of inversions
or similar techniques. In Section~\ref{sec:inv} we describe the common inversion techniques
and some of the major results are described in Section~\ref{sec:struc}

\newpage

\section{Inversions to Determine Interior Structure}
\label{sec:inv}

The dominance of the surface effects in frequency differences between the Sun and
solar models led to efforts in
the direction of `inverting' solar oscillation frequencies. The idea is simple: the frequencies
of solar oscillations depend on the internal properties of the Sun in a known manner, thus if we
had enough frequencies, we could determine solar structure from the frequencies. 

Initial inversions for solar sound speed were done using the asymptotic expression
given by Duvall Law (Eq.~\ref{eq:duvall}). However,
limitations of the asymptotic expression and improvements in computational resources
led to the development of inversion techniques using the full set of oscillation 
equations. We discuss both below. Earlier discussions of these techniques can be found
in \citet{dogmjt1991}, \citet{mjt1998I} and \citet{jcd2002}.

\subsection{Asymptotic inversions}
\label{subsec:asymp}

The first inversions were based on the Duvall Law. \citet{dog1984R} had shown that
with a little mathematical manipulation, Eq.~(\ref{eq:fw}) can be re-written as
\be
w^3\frac{\dd F}{\dd w}=\int_w^{a_s}\left(1-\frac{a^2}{w^2}\right)^{-1/2}\;\frac{\dd\ln r}{\dd\ln a}\dd a,
\label{eq:dogfw}
\ee
where $a\equiv c(r)/r$, and $a_s= a(R)$. This is an equation of the Abel type for which an 
analytic inverse exists:
\be
r=R\exp\left[ \frac{-2}{\pi}\int_{a_s}^a \left(\frac{1}{w^2}-\frac{1}{a^2}\right)^{-1/2} \frac{\dd F}{\dd w}\dd w\right].
\label{eq:doginv}
\ee
The function $F(w)$ required in this equation can be obtained from observations using the Duvall Law
by fitting a function (like a spline) though the points. The surface term can be accounted for by making
$\alpha_p$ in Eq.~(\ref{eq:duvall}) a function of frequency, though early inversions did not do
so \citep[see][]{jcdetal1985N}. One of the issues in this inversion is that it requires the value of
$F(w)$ at the surface. Most observed modes have their lower-turning points well below the
surface and hence $F(w)$ needs to be extrapolated. Inversions of frequencies of solar models also show
that results near the core are unreliable.
\citet{brodskyandvorontsov1987,brodskyandvorontsov1988} also used the Duvall Law as the starting
point, but their way of determining $F(w)$ was different. They assumed that mode frequencies
are a continuous function of $n$ and $L$ that can be determined from observations. They also
used a frequency-dependent $\alpha_p$. Asymptotic inversions were also carried out by
\citet{agk1988E} who showed that the solar sound speed could be successfully determined
between radii of $0.4$ and $0.7\rsun$.
\citet{shibahashi1988} and \citet{sekiishibahashi1989} had a different starting point: instead of using
the Duvall Law, they used the quantisation condition obtained by Eq.~(\ref{eq:lambp}). They did however,
assume that $\omega^2\gg N^2$ and neglected the buoyancy frequency.

A breakthrough in the process of asymptotic inversions came when \citet{jcdetal1988E,jcdetal1989M}
showed that more accurate (and precise) inversion results could be obtained if one linearised 
the equation for Duvall Law around a known solar model (a `reference model') and inverted the frequency
differences between the model and the Sun to determine the sound-speed difference
between the model and the Sun. They perturbed Eq.~(\ref{eq:duv1}) to show that, when one retained only
terms linear in the perturbations, one gets
\be
S(w)\frac{\delta\omega}{\omega}=\int^R_{r_t}\left(1-\frac{c^2}{w^2r^2}\right)^{-1/2}\frac{\delta c}{c}
\frac{\dd r}{c} +\pi\frac{\delta\alpha}{\omega},
\label{eq:duvdif}
\ee
where
$S(w)$ is a function of the reference model, and is given by
\be
S(w)=\int^R_{r_t}\left(1-\frac{c^2}{w^2r^2}\right)^{-1/2}\frac{\dd r}{c}\;.
\label{eq:sw}
\ee
Note that $2S(w)$ is the sound travel time along a ray between successive deflections at the
surface. For properties of Eq.~(\ref{eq:duvdif}), see \citet{jcddog1988}.
Note that the first term in the RHS of Eq.~(\ref{eq:duvdif}) is a function of
$w$, while the second term is a function of $\omega$. The scaled frequency
differences $S(w)\delta\omega/\omega$ can therefore be expressed as
\be
S(w)\frac{\delta\omega}{\omega}=H_1(w)+H_2(\omega).
\label{eq:h1h2}
\ee
Thus, the scaled frequency difference depend on the interior sound-speed difference
through $H_1(w)$ and on differences at the surface through a function of $H_2(\omega)$
The function $H_1(w)$ is related to the sound speed by
\be
H_1(w)=\int^{\ln R}_{\ln r_t(w)} \left(1-\frac{a^2}{r^2}\right)^{-1/2}\frac{\delta c}{c}\frac{1}{a}\dd\ln r. 
\label{eq:h1}
\ee
This equation also has a closed-form inverse and one can show that \citep{jcdetal1989M}
\be
\frac{\delta c}{c}=
-\frac{2a}{\pi}\frac{\dd}{\dd\ln r}\int^a_{a_s} (a^2-w^2)^{-1/2}H_1(w)\dd w,
\label{eq:csq1}
\ee
or, alternatively \citep{sbhma1994}
\be
\frac{\delta c}{c}=
-\frac{2r}{\pi}\frac{\dd a}{\dd r}\int^a_{a_s}\frac{w}{(a^2-w^2)^{1/2}}\frac{\dd H_1}{\dd w}\dd w.
\label{eq:csq2}
\ee
$H_1(w)$ and $H_2(\omega)$ can be obtained fitting one function of $w$ and one
of $\omega$ to the frequency differences between the Sun and the reference
model. Normally this is done by fitting splines \citep[see, e.g.,][]{jcdetal1989M,sbhma1994}.

\subsection{Full inversions}
\label{subsec:full}

While asymptotic inversions, in particular differential asymptotic inversions,
were successful in revealing some of the details of the solar interior, their
inherent limitation was that they assumed that solar oscillations could
be explained solely through differences in sound speed alone. The role
of density perturbations was generally ignored. Although some attempts were made to include
higher order terms that did not neglect the density perturbations 
 \citep[][]{svvshibahashi1991,iwrsvv1993}, the methods
did not become popular. Other limitations arise from the fact that the
asymptotic relations generally assume that the eigenfunctions vary more rapidly
than the pressure scale height, but this breaks down close to the surface.
As a result, these days it is more common to do inversions based on the
full set of equations. The increase in computing power also encouraged this 
change.

The equations of stellar oscillation derived earlier cannot be used to invert stellar frequencies 
to determine the interior structure. For that, we need to start with the perturbed form of
the equation of motion, i.e., Eq~(\ref{eq:mom1}), i.e, 
\be
 \rho{\frac{\partial^2{\vec \xi}}{\partial t^2}}=-\nabla P_1+\rho_0 {\vec g_1}+\rho_1{\vec g}.
\ee
As mentioned earlier, the displacement vector can be written as ${\vec \xi}\exp{(-i\omega t)}$. Substituting 
this in the equation, we get
\be
-\omega^2\rho{\vec \xi}=-\nabla P_1+\rho {\vec g_1}+\rho_1{\vec g},
\label{eq:momxi}
\ee
Substituting for $\rho_1$ from the perturbed continuity equation (Eq.~\ref{eq:conpturb}) and 
$P_1$ from Eq~(\ref{eq:therma}), we get
\be
-\omega^2\rho\vec{\xi}
=\nabla(c^2\rho\nabla\cdot\vec{\xi}+\nabla P\cdot\vec{\xi})
-\vec g\nabla\cdot(\rho\vec{\xi}) - G\rho\nabla\left(
{\int_V{\nabla\cdot(\rho\vec{\xi}\dd^3r\over |\vec r -\vec{r'}|}}
\right),
\label{eq:herm}
\ee
where we have expressed the gravitational potential as an integral:
\be
{\vec g_1}=\nabla\Phi_1=\nabla{\int_V\rho_1 d^3 r \over {|\vec r -\vec{r'}|}}=
-\nabla{\int_V{\nabla\cdot(\rho\vec{\xi})\dd^3r\over |\vec r -\vec{r'}|}}.
\label{eq:pgrav}
\ee
Eq.~(\ref{eq:herm}) describes how the mode frequencies $\omega$ depend on the structure of the star and is the 
starting point for inversions. Note that $\xi\equiv\xi_{n,\ell}$ and $\omega\equiv\omega_{n,\ell}$, and 
thus there is one such equation
for each mode described by eigenfunction $\xi_{n,\ell}$.

While Eq.~(\ref{eq:herm}) is the starting point of inversions, it is clear that
the equation cannot be used as it stands. We can measure $\omega$ and we want to
determine $c^2$ and $\rho$ (pressure $P$ is related to
$\rho$ through the equation of hydrostatic equilibrium, and ${\vec g}$ can be derived from $\rho$), we
do not have any means of determining the displacement eigenfunction $\vec{\xi}$ except at the surface.
The way out of this stalemate is to recognise that Eq.~(\ref{eq:herm}) is an eigenvalue equation of the 
form 
\be
\mathcal{L}(\vec{\xi}_{n,\ell})=-\omega^2_{n,\ell}\vec{\xi}_{n,\ell}\,,
\label{eq:herl}
\ee
$\mathcal{L}$ being the differential operator in Eq.~(\ref{eq:herm}).
\citet{chandrasekhar1964} showed that under specific boundary conditions, namely
$\rho=P=0$ at the outer boundary, the eigenvalue problem defined by
Eq.~(\ref{eq:herm}) is a Hermitian one. This allows us to move ahead.

If $\mathcal{O}$ be a differential Hermitian operator, and $x$ and $y$ be two eigenfunctions
of the operator, then
\be
\int x^*\mathcal{O}(y) \dd V = \int y\mathcal{O}(x^*) \dd V,
\label{eq:ex}
\ee
where $^*$ indicates a complex conjugate. This is often written
as 
\be
\langle x,\mathcal{O}(y) \rangle = \langle \mathcal{O}(x),y \rangle,
\label{eq:ex1}
\ee
where the operation $\langle \rangle$ is usually called an inner product. 
The relevant inner product in the case of Eq.~(\ref{eq:herm}) is
\be
\langle {\vec\xi},{\vec\eta} \rangle = \int_V \rho {\vec \xi}^{*}\cdot{\vec \eta} \dd^3{\vec r}
=4\pi\int^R_0 [\xi^*_r(r)\eta_r(r)+L^2\xi^*_t(r)\eta_t(r)] r^2\rho \dd r.
\label{eq:inner}
\ee
Hermitian operators have the property that their eigenvalues are
real and that the variational principle applies. This means that if $a$ be the eigenvalue
corresponding to the eigenfunction $x$, then
\be
a={ \langle x, \mathcal{O}(x) \rangle \over \langle x, x \rangle }.
\label{eq:eig}
\ee
Thus, if the frequency
 $\omega_{n,\ell}$ be the eigenvalue corresponding to the displacement eigenfunction
${\vec\xi_{n,\ell}}$, then
\be
-\omega_{n,\ell}^2= {\int_V \rho {\vec \xi}_{n,\ell}^{*}\cdot\mathcal{L}({\vec \xi}_{n,\ell}) \dd^3{\vec r}
\over \int_V \rho {\vec \xi}_{n,\ell}^{*}\cdot{\vec \xi}_{n,\ell} \dd^3{\vec r}}.
\label{eq:eigen}
\ee
Note that the denominator is the mode inertia [Eq.~(\ref{eq:enlo})].

The Hermitian nature of Eq.~(\ref{eq:herm}) allows us to use the variational
principle to put the equation in the form that can be inverted. Eq.~(\ref{eq:herl}) is
first linearised around a known model, \emph{reference} model.
Thus, if $\mathcal{L}$ be the operator for a model, we assume that the operator for the Sun or
other star can be expressed as $\mathcal{L}+\delta\mathcal{L}$. The corresponding
displacement eigenfunctions are $\vec\xi$ and $\vec\xi+\delta\vec\xi$ respectively and the
corresponding frequencies are respectively $\omega$ and $\omega+\delta\omega$. Thus
\be
(\mathcal{L}+\delta\mathcal{L})({\vec\xi}+\delta{\vec\xi})
= -(\omega+\delta\omega)^2({\vec\xi}+\delta{\vec\xi}),
\label{eq:hermlin}
\ee
which on expansion and retention of only linear terms gives
\bea
\int{\vec \xi}^*\mathcal{L}{\vec \xi} \dd V&+&\int{\vec \xi}^*\delta\mathcal{L}{\vec \xi} \dd V+
\int{\vec \xi}^*\mathcal{L}{\delta\vec \xi} \dd V +\int{\vec \xi}^*\delta\mathcal{L}{\delta\vec \xi} \dd V
= -\omega^2\int{\vec \xi}^*{\vec \xi} \dd V\nonumber\\
&-&\omega^2\int{\vec \xi}^*\delta{\vec \xi} \dd V
-(\delta\omega^2)\int{\vec \xi}^*{\vec \xi} \dd V
-2\omega\delta\omega\int{\vec \xi}^*{\vec \xi} \dd V
\label{eq:linexp}
\eea
Because $\mathcal{L}$ is a Hermitian operator, all but two terms in
Eq.~(\ref{eq:linexp}) cancel out to give
\be
\int{\vec \xi}^*\delta\mathcal{L}{\vec \xi} \dd V = -2\omega\delta\omega\int{\vec \xi}^*{\vec \xi} \dd V,
\label{eq:twoterms}
\ee
or
\be
{\delta\omega\over\omega}=
-{\int_V\rho\vec\xi\cdot\delta\mathcal{L}\vec\xi \dd^3\vec r\over
2\omega^2\int_V\rho\vec\xi\cdot\vec\xi \dd^3\vec r}=-{\int_V\rho\vec\xi\cdot\delta\mathcal{L}\vec\xi \dd^3\vec r\over
2\omega^2 E_{n,\ell}}.
\label{eq:final}
\ee
One such equation can be written for each mode, and these are the equations that we invert.
Since the 
mode inertia appears in the denominator of Eq.~(\ref{eq:final}), the equation indicates
that for a given perturbation, frequencies of modes with lower inertia will
change more than those with higher inertia.

\subsubsection{Inversions kernels}
\label{subsubsec:ker}

In order to make further progress, we need to determine what $\delta\mathcal{L}$ is. For that we need
to return to Eq.~(\ref{eq:herm}) and perturb it, keeping only terms linear in the perturbations.
Thus,
\be
\delta \mathcal{L}\vec\xi=\nabla(\delta c^2\nabla\cdot\vec\xi+\delta\vec g\cdot\vec\xi)
+\nabla\left(\delta\rho\over\rho\right)\;c^2\nabla\cdot\vec\xi
+{1\over\rho}\nabla\rho\delta c^2\nabla\cdot\vec\xi
+\delta \vec g\nabla\cdot
\vec\xi-G\nabla\int_V{\nabla\cdot(\delta\rho\vec\xi)\over|\vec r-\vec r'|}
\dd^3\vec{r'},
\label{eq:deltal}
\ee
where
$\delta c^2$, $\delta{\vec g}$ and $\delta\rho$ are the differences in sound speed, acceleration
due to gravity and density between the reference model and the Sun. The quantity $\delta{\vec g}$
can be expressed in terms of $\delta\rho$.
Following \citet{antiaandbasu1994}, we substitute Eq.~(\ref{eq:deltal}) in Eq.~(\ref{eq:final}) and 
rearrange the terms to get:
\be
{\delta\omega\over\omega} = -{1\over2\omega^2 E}(I_1+I_2 +I_3+I_4),
\label{eq:hma}
\ee
where $E$ is the mode inertia and
\be
I_1=-\int_0^{R}\rho(\nabla\cdot\vec\xi)^2\delta c^2 r^2\;\dd r,
\label{eq:i1}
\ee
\be
I_2=\int_0^{R}\xi_r[\rho\nabla\cdot\vec\xi+\nabla\cdot(\rho\vec\xi)]
\delta gr^2\;\dd r,\quad\hbox{where}\quad \delta g(r)={4\pi G\over r^2}\int_0^r\delta\rho(s) s^2\;\dd s
\label{eq:i2}
\ee
\be
I_3=\int_0^{R}\rho c^2\xi_r\nabla\cdot\vec\xi{\dd \over \dd r}
\left(\delta\rho\over\rho\right) r^2\;\dd r, 
\label{eq:i3}
\ee
and
\be
I_4={4\pi G\over2\ell+1}\int_0^{R}\nabla\cdot(\rho\vec\xi)r^2\;\dd r
\left[{1\over r^{\ell+1}}\int_0^rs^{\ell+2}\left(\rho\nabla\cdot\vec\xi-
\rho{d\xi_r\over \dd s}-{\ell+2\over s}\rho\xi_r\right){\delta\rho\over\rho}\;\dd s\right].
\label{eq:i4}
\ee

One can rewrite Eq.~(\ref{eq:i1}) in terms of $\delta c^2/c^2 $, the relative
difference between the squared sound speed of the Sun and the reference model.
Similarly Eqs.~(\ref{eq:i2})--(\ref{eq:i4}) can be rewritten in terms of the relative
density difference $\delta\rho/\rho$. In some cases this requires us to change
the order of the integrals, as has been explained in \citet{dogmjt1991} and \citet{dog1993}.
Thus, Eq.~(\ref{eq:hma}) for each mode $i \equiv (n,\ell)$ can be written as
\be
{\delta\omega_i\over\omega_i}=\int K^i_{c^2,\rho}(r)
{\delta c^2\over c^2}(r)\;\dd r
+ \int K^i_{\rho,c^2}(r){\delta\rho\over\rho}(r)\;\dd r.
\label{eq:inv1}
\ee
The terms $K^i_{c^2,\rho}(r)$ and $K^i_{\rho,c^2}(r)$ are known functions of the reference model
and represent the change in frequency in response to changes in sound speed and density respectively.
These two functions are known as the ``kernels'' of the inversion.

We show a few sound speed and density kernels in Figure~\ref{fig:ker1} and Figure~\ref{fig:ker2}. 
Note that modes of higher $\ell$ are
restricted closer to the surface, while modes of lower $\ell$ penetrate deeper. This is 
consistent with our earlier discussion of the lower turning points of different modes. 
Also note that sound-speed kernels are positive
at all radii, while density kernels oscillate between positive and negative
values. This oscillation reduces the contribution of the second term of the RHS of Eq.~(\ref{eq:inv1})
and explains why density inversions are more difficult than sound speed inversions.
This also explains why we can invert frequencies reasonably when using the asymptotic form of the oscillation
equations where it is assumed that the frequency differences between a model and the Sun can be
explained by differences in sound speed alone. The small discrepancies in the results obtained
using asymptotic relation is a result of ignoring the small contribution from density.

\epubtkImage{}{%
\begin{figure}[htb]
\centerline{\includegraphics[height=17pc]{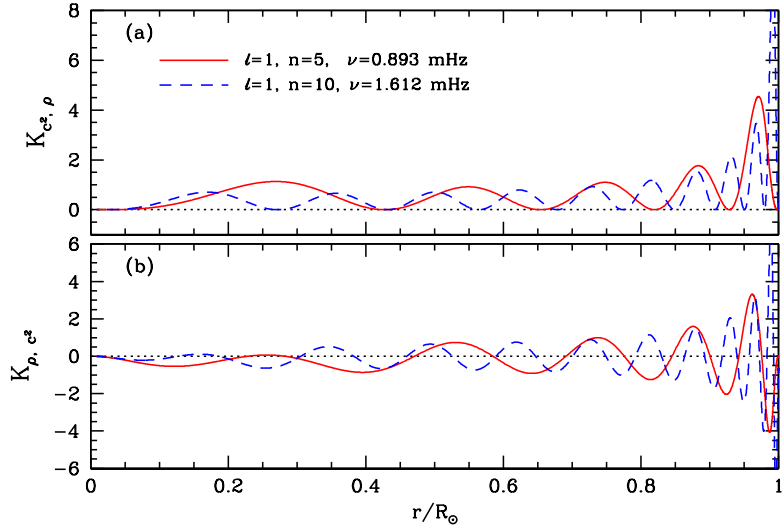}}
\caption{Sound-speed kernels (Panel a) and density kernels (Panel b) for modes of the same degree 
but different radial orders (and hence frequencies).
}
\label{fig:ker1}
\end{figure}}

\epubtkImage{}{%
\begin{figure}[htb]
\centerline{\includegraphics[height=17pc]{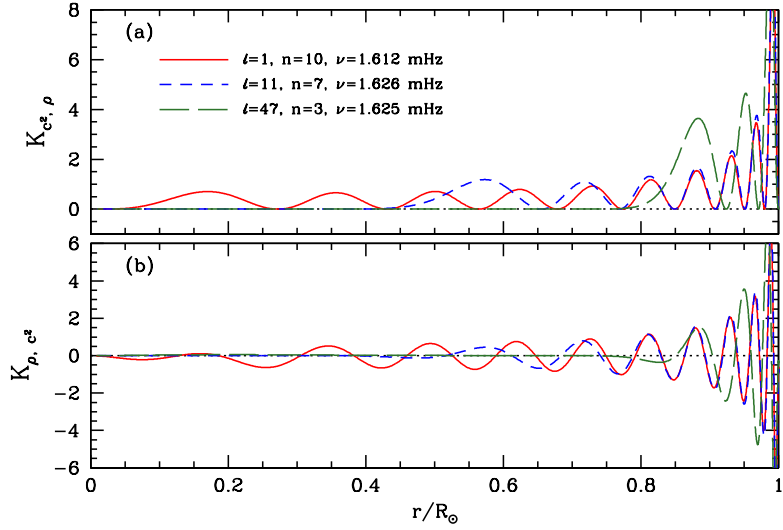}}
\caption{Sound-speed kernels (Panel a) and density (Panel b) for modes of different degrees
but similar frequencies. Note that modes of lower degrees penetrate deeper than modes of
higher degrees.
}
\label{fig:ker2}
\end{figure}}

The kernels for sound speed and density can be converted to those of other quantities \citep[see][]{dog1993}.
It is relatively straightforward to change the $(c^2,\rho)$ kernels to those for $(\Gamma_1,\rho)$.
Since $c^2=\Gamma_1 P/\rho$,
\be
{\delta c^2\over c^2}={\delta \Gamma_1\over \Gamma_1}+{\delta P\over P}-{\delta\rho\over\rho}.
\label{eq:csqgam}
\ee
Substituting Eq.~(\ref{eq:csqgam}) in Eq.~(\ref{eq:inv1}) we get
\be
{\delta\omega_i\over\omega_i}=\int K^i_{c^2,\rho}(r)
{\delta \Gamma_1\over \Gamma_1}\;\dd r+\int K^i_{c^2,\rho}{\delta P\over P}\dd r +
\int ( K^i_{\rho,c^2}- K^i_{c^2,\rho}){\delta\rho\over\rho} \dd r.
\label{eq:inv2}
\ee
Thus, $K_{\Gamma_1,\rho}=K_{c^2,\rho}$. To obtain the $K_{\rho, \Gamma_1}$,
the equation of hydrostatic equilibrium, $dP/\dd r=-g\rho$, has to be used to 
 write $\int K_{c^2,\rho} (\delta P/P) \dd r$
 in terms of $\delta\rho/\rho$ though an integral.
Using that and changing the order of integrations, we can rewrite
\be
\int K_{c^2,\rho} (\delta P/P) \dd r -\int K_{c^2,\rho}(r)({\delta\rho/\rho})\dd r+\int K_{\rho,c^2}({\delta\rho/\rho})\dd r
=\int B(r)({\delta\rho/\rho})\dd r,
\ee
where $B(r)$ is the kernel $K_{\rho,\Gamma_1}$.

Closed-form kernels are not possible for variable pairs other than
$(c^2,\rho)$ and $(\Gamma_1,\rho)$. For instance, going from $(c^2,\rho)$ to $(u, \Gamma_1)$ ($u\equiv P/\rho$),
 we find that $K_{\Gamma_1,u}\equiv K_{c^2,\rho}$ (from Eq~\ref{eq:csqgam}), but the 
second kernel of the
pair, $K_{u,\Gamma_1}$, does not have a closed form solution. It can be
written as
\be
K_{u,\Gamma_1}=K_{c^2,\rho}-P{\dd \over \dd r}\left({\psi\over P}\right),
\label{eq:kug}
\ee
where, $\psi$ is a solution of the equation
\be
{\dd \over \dd r}\left({P\over r^2\rho}{d\psi\over \dd r}\right)-{\dd \over \dd r}\left({Gm\over r^2}\psi\right)
+{4\pi G\rho\over r^2}\psi= -{\dd \over \dd r}\left({P\over r^2}F\right)
\label{eq:convert}
\ee
for $F=K_{\rho, c^2}$.

It has also been common to use the knowledge of the equation of state of stellar matter to
inject some more information into the system and convert $(\Gamma_1,\rho)$ kernels
to kernels of $(u, Y)$, $Y$ being the helium abundance.
Kernels for $Y$ are non-zero only in the ionisation zone, which
makes inversions easier.
The adiabatic index $\Gamma_1$ can be expressed as
\bea
{\delta\Gamma_1\over\Gamma_1}&=&\left({\partial\ln\Gamma_1\over\partial\ln P}\right)_{\rho,Y}{\delta P\over P}
+\left({\partial\ln\Gamma_1\over\partial\ln\rho}\right)_{P,Y}{\delta\rho\over\rho}
+\left({\partial\ln\Gamma_1\over\partial Y}\right)_{\rho,P}\delta Y\nonumber\\
&\equiv&\Gamma_{1,P}{\delta P\over P}+\Gamma_{1,\rho}{\delta\rho\over\rho}+\Gamma_{1,Y}\delta Y,
\label{eq:gamexp}
\eea
where the partial derivatives can be calculated using the equation of state. Once we know this, it is
easy to see that
\be
K_{Y,u}\equiv\Gamma_{1,Y}K_{\Gamma_1,\rho},\quad\hbox{and}\quad
K_{u,Y}\equiv\Gamma_{1,P}K_{\Gamma_1,\rho}-P{\dd \over \dd r}\left({\psi\over P}\right),
\label{eq:uY}
\ee
with $\psi$ being the solution of Eq.~(\ref{eq:convert}) with
$F\equiv(\Gamma_{1,P}+\Gamma_{1,\rho})K_{\Gamma_1,\rho}+K_{\rho,\Gamma_1}$.
In a similar way, it is possible to derive kernels for the stability criterion $A^*$ given
by
\be
A^*(r)=\frac{1}{\Gamma_1}\frac{\dd \ln P}{\dd\ln r}-\frac{\dd\ln\rho}{\dd\ln r},
\label{eq:stab}
\ee
\citep[e.g.,][]{agk1990L, elliott1996}.

\subsubsection{Taking care of the surface term}
\label{subsubsec:sur}

Unfortunately, Eq.~(\ref{eq:inv1}) does not take into account the issue of the surface term
that was discussed earlier in Section~\ref{sec:surf}.
Eq.~(\ref{eq:final}) implies that we can invert the
frequencies provided we know how to model the Sun properly, and provided that the frequencies
can be described by the equations for adiabatic oscillations, which we know from
Section~\ref{sec:surf}, is not the case at all. For modes that are not of very high degree (p modes of
$\ell \simeq 200$ and lower),
the structure of the wavefront near the surface is almost independent of the
degree of the modes. Thus, any additional frequency difference is a function of frequency alone once
the effect of mode inertia has been taken into account. 
This leads us to modify Eq.~(\ref{eq:inv1}) to represent
the difference between the Sun and the reference model as
\be
{\delta \omega_i\over \omega_i}=\int \mathcal{K}_{c^2,\rho}^i(r)
{\delta c^2\over c^2}(r)\;\dd r+
\int \mathcal{K}_{\rho, c^2}^i(r) {\delta \rho\over \rho}(r)\;\dd r+
{F(\omega_i)_{\rm surf}\over E_i}\,,
\label{eq:inv}
\ee
where $F(\omega_i)_{\rm surf}$ is a slowly varying function of frequency that
represents the surface term 
\citep[see e.g.,][and references therein]{dziembowskietal1990,antiaandbasu1994}.

Using a surface correction of the form shown in Eq.~(\ref{eq:inv}) is equivalent
to passing the frequency differences and kernels through a high-pass filter and thus
filtering out a slowly-varying component from the frequency differences and the kernels
\citep{basuetal1996}. \citet{basuetal1996} also showed that such a simple filtering 
is not sufficient when dealing with modes of a degree higher than
about $\ell=200$ since the wavefront at the surface
is not radial for these modes. Indeed, when using higher degree models, often an
$\ell$-dependent second function is added to Eq.~(\ref{eq:inv}). \citet{brodskyandvorontsov1993} 
developed this in the context of asymptotic inversions. This was explored
further by \citet{goughandvorontsov1995}. The asymptotic expansion was adapted to the
full-non-asymptotic inversions by \citet{dimauroetal2002} 
who used a surface term of the form 
\be
F(\omega)=F_1(\omega)+F_2(\omega){w}^2+F_3(\omega){w}^4,
\label{eq:dimauro}
\ee
where as before ${w}\equiv\omega/L$.
\citet{antia1995}, looking at inversion
results between two models, showed that a slightly different form of the
surface term 
\be
F(\nu)\equiv F_0(\nu)+\ell(\ell+1)F_1(\nu)+[\ell(\ell+1)]^2F_2(\nu)\ldots,
\label{eq:antia}
\ee
also improves inversions with high-degree modes.

\subsection{Inversion techniques}
\label{subsec:techn}

For $N$ observed modes, Eq.~(\ref{eq:inv}) represents $N$ equations that need to
be ``inverted'' or solved to determine the sound-speed and density differences.
The data ${\delta \omega_i/\omega_i}$ or equivalently $\delta \nu_i/\nu_i$
are known. The kernels ${K}_{c^2,\rho}^i$ and ${K}_{\rho, c^2}^i$ are known
functions of the reference model. The task of the inversion process is to
estimate ${\delta c^2/c^2}$ and $\delta\rho/\rho$ after somehow accounting for
the surface term $F_{\rm surf}$. 

There is one inherent problem: no matter how many
modes we observe, we only have a finite amount of data and hence a finite
amount of information. The two unknowns, $\delta c^2/c^2$ and $\delta\rho/\rho$, being functions
have an infinite amount for information. Thus, there is really no way that we can
recover the two functions from the data. The best we may hope for is to find localised
averages of the two functions at a finite number of points. How localised the averages
are depends on the data we have and the technique used for the inversion. 
During the inversion process we often have to make additional
assumptions, such as the sound-speed and density profiles
 of the Sun inferred from inversions should be positive, in order to get physically valid solutions.
We might need to put other constraints too. For instance, when we invert oscillation data 
to determine the density profile we need
to ensure that mass is conserved, i.e.,
\be
\int {\delta\rho\over\rho}\rho(r) r^2 \dd r=0.
\label{eq:conserve}
\ee
This is usually done by defining an additional mode with:
\be
\omega=0,\quad {K}_{c^2,\rho}=0,\quad\hbox{and}\quad {K}_{\rho, c^2}=\rho(r)r^2.
\label{eq:extramode}
\ee

One of the first prescriptions for numerically inverting solar oscillation frequencies
using Eq.~(\ref{eq:inv}) was given by \citet{dog1985}. Since then, many groups have inverted
available solar frequencies to determine the structure of the Sun 
\citep[][etc.]{dogkosovichev1988,dogkosovichev1993,dziembowskietal1990,dziembowskietal1991,dziembowskietal1994,
dappenetal1991,antiaandbasu1994,basuetal1996apj,basuetal1997,basuetal2000,basuetal2009}

There are two popular methods of inverting Eq.~(\ref{eq:inv}): (1) The Regularised Least Squares (RLS)
method, and (2) the method of Optimally Localised Averages (OLA). These are complementary techniques
\citep[see][for a discussion]{sekii1997} that have different aims. RLS aims to
find the $\delta c^2/c^2$ and $\delta\rho/\rho$ profiles that give the best fit to the data (i.e.,
give the smallest residuals) while keeping the errors small; the aim of OLA is not to fit the data at all, but to find
linear combinations of the frequency differences in such a way that the corresponding
combination of kernels provides a localised average of the unknown function, again
while keeping the errors small. We discuss the two inversion techniques in some detail below. Other descriptions
may be found in \citet{dogmjt1991} and \citet{basu2014}. Properties of inversion
techniques were investigated by \citet{jcdetal1990MNRAS}; although
their investigation was specifically for inversions of solar dynamics, the general 
principles apply to structure inversions too.

\subsubsection{The Regularised Least Squares technique}
\label{subsubsec:rls}

RLS inversions start with expressing the three unknown functions in Eq.~(\ref{eq:inv}) in
terms of well-defined basis functions. Thus,
\bea
F(\omega)&=&\sum_{i=1}^m a_i\psi_i(\omega),\nonumber\\
{\delta c^2\over c^2}&=&\sum_{i=1}^n b_i\phi_i(r),\nonumber\\
{\delta \rho\over \rho}&=&\sum_{i=1}^n c_i\phi_i(r),
\label{eq:basis}
\eea
where $\psi_i(\omega)$ are suitable basis functions in frequency $\omega$, and
$\phi_i(r)$ are suitable basis functions in radius $r$. Thus, for $N$ observed
modes, Eq.~(\ref{eq:inv}) represent $N$
equations of the form:
\bea
b_1\int\phi_1(r)K^i_{c^2}\dd r+b_2\int\phi_2(r)K^i_{c^2}\dd r+\ldots+b_n\int\phi_n(r)K^i_{c^2}\dd r+\nonumber\\
c_1\int\phi_1(r)K^i_{\rho}\dd r+c_2\int\phi_2(r)K^i_{\rho}\dd r+\ldots+c_n\int\phi_n(r)K^i_{\rho}\dd r+\nonumber\\
a_1\psi_1(\omega_i)/E_i+a_2\psi_2(\omega_i)/E_i+\ldots+a_m\psi_m(\omega_i)/E_i=\delta\omega_i/\omega_i,
\label{eq:expand}
\eea
where, for ease of writing we have denoted $K^i_{c^2,\rho}$ as $K^i_{c^2}$ and $K^i_{\rho, c^2}$ as
$K^i_{\rho}$. What we need to do is find coefficients $a_i$, $b_i$ and $c_i$ such that we get the
best fit to the data $\delta\omega_i/\omega_i$. This is done by minimising
\be
\chi^2=\sum_{i=1}^N{\left({{\delta\omega_i\over\omega_i}-
\Delta\omega_i\over
\sigma_i}\right)}^2,
\label{eq:chi}
\ee
where $\Delta\omega_i$ represents the LHS of Eq.~(\ref{eq:expand}). While such a minimisation
does give us a solution for the three unknown functions, the solutions often are usually extremely
oscillatory in nature (see Figure~\ref{fig:rls_soln}(a)).
The main reason for this is that the data have errors and the oscillatory
solution is a result of these errors propagating into solutions for the unknown function. 
The phenomenon is also related
to the so-called `Gibbs' Phenomenon' in mathematics and is a result of the
fact that we only have a finite amount of data but are trying to recover infinite
information in the form of three functions.
To ensure a more physical profile, `regularisation' or `smoothing' is
applied \citep{craigandbrown1986}. This is implemented
by demanding that we
try to minimise the second derivative of the unknown functions while trying to
minimise the $\chi^2$.
Thus, instead of minimising Eq.~(\ref{eq:chi}), we minimise
\be
\chi_{\rm reg}^2=\chi^2+||L||^2=\sum_{i=1}^N{\left({{\delta\omega_i\over\omega_i}-
\Delta\omega_i\over
\sigma_i}\right)}^2
+\alpha^2\int_0^{R}
\left[\left({d^2\over \dd r^2}{\delta\rho\over\rho}
\right)^2+\left({d^2\over \dd r^2}{\delta c^2\over c^2}\right)^2\right],
\label{eq:regur}
\ee
where $\alpha$ is the regularisation or smoothing parameter.
We could, if we wanted to, have
different smoothing parameters for the the two functions.
 The influence of the regularisation
parameter on the solution can be seen in Figure~\ref{fig:rls_soln}.
Smoothing is not usually applied to the
surface term $F(\omega)$, instead, the number and type of basis functions are chosen is such that
 $F(\omega)$ is a slowly varying function of frequency.

\epubtkImage{}{%
\begin{figure}[htb]
\centerline{\includegraphics[height=13.5pc]{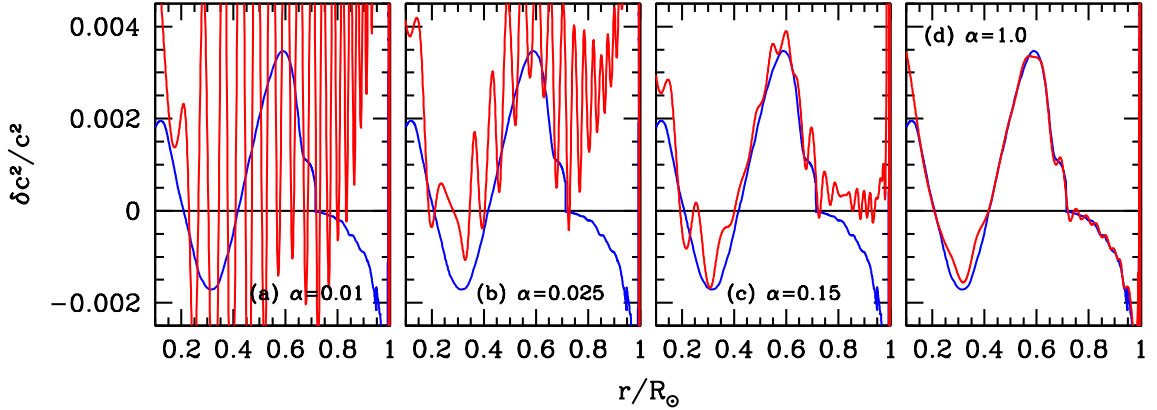}}
\caption{
The relative sound-speed differences between two solar models obtained by inverting
their frequency differences using the RLS technique. In each panel the blue line 
is the true difference between the models and the red line is the inversion result. 
The different
panels show the result obtained using different values of the smoothing parameter $\alpha$.
The reference model used is model BP04 of \citet{bahcalletal2005} and the
test model, or proxy Sun, is model BSB(GS98) of \citet{bahcalletal2006}. We only used those modes that have
been observed in the Sun, specifically mode-set BiSON-13 of
\citet{basuetal2009}. Random errors, corresponding to the uncertainties in the
observed mode-set, were added to the data of the test model.
}
\label{fig:rls_soln}
\end{figure}}

\subsubsection*{Implementing RLS}
\label{para:rla_impl}

There are many ways in which the coefficients $a_i$, $b_i$ and $c_i$ may be determine.
The implementation of \citet{antiaandbasu1994} and \citet{basumjt1996} is discussed below.

Eq.~(\ref{eq:expand}) clearly represents $N$ equations in $k=2n+m$ unknowns. These equations
can thus be represented as:
\be
A{\mathbf x}={\mathbf d},
\label{eq:matrix}
\ee
where $A$ is an $N\times(2n+m)$ matrix, ${\mathbf x}$ is a vector of length $(2n+m)$
consisting of the unknown coefficients $a_i$, $b_i$ and $c_i$, and ${\mathbf d}$ is the
 vector with the data, i.e., $\delta\omega/\omega$. To take data errors into account,
each row $i$ of each matrix is divided by $\sigma_i$, where $\sigma_i$ is the uncertainty
of the $i$th data point. The elements of vector ${\mathbf x}$ can be determined using
Singular Value Decomposition (SVD).

An SVD of matrix $A$ results in the decomposition of $A$ into three matrices,
$A=U\Sigma V^T$. $U$ and $V$ have the property that $U^TU=I$, $V^TV=VV^T=I$, and
 $\Sigma = {\rm diag}(s_1,s_2,\cdots,s_{2n+m})$,
$s_1\ge s_2\ge \cdots s_{2n+m}$ being the singular values of $A$.
Thus, the given set of equations can be reduced to
\be
U\Sigma V^T{\mathbf x}={\mathbf d},
\label{eq:svd}
\ee
and hence,
\be
{\mathbf x}=V\Sigma^{-1}U^T{\mathbf d}.
\label{eq:soln}
\ee
It can be shown that this gives a least-squares solution to the set of equations \citep[e.g.,][]{golub1996}.
Since the equations have already been normalised by the errors, the standard error on any component
$x_i$ of vector ${\mathbf x}$ is given by
\be
e^2_i=\sum_{i=k}^{2n+m}{v^2_{ik}\over s^2_k},
\label{eq:error}
\ee
where $v_{ik}$ are elements of vector $V$.

Smoothing is implemented by replacing the integrals in Eq.~(\ref{eq:regur}) by a sum 
over $M$ uniformly spaced points and
adding the corresponding equations to the system of equations to be solved. Thus, the 
following terms are added:
\be
{\alpha\over \sqrt{M}} q(r_j) {\left({\dd^2\over \dd r^2}{\delta c^2\over c^2}\right)}_{r=r_j}=
{\alpha\over \sqrt{M}} q(r_j)\sum_i^n b_i{\dd^2\over \dd r^2}\phi_i(r_j)=0,
\label{eq:csmooth}
\ee
for $i=1\ldots n$ and $j=1\ldots M$; and
\be
{\alpha\over\sqrt{M}} q(r_j) {\left({\dd^2\over \dd r^2}{\delta\rho\over \rho}\right)}_{r=r_j}=
{\alpha\over \sqrt{M}} q(r_j)\sum_i^n c_i{\dd^2\over \dd r^2}\phi_i(r_j)=0
\label{eq:dsmooth}
\ee
for $i=n+1\ldots 2n$ and $j=M+1\ldots 2M$.
The function $q(r)$ in the above equation allows us to smooth $\delta c^2/c^2$ and $\delta\rho/\rho$ 
preferentially in any given part of the Sun or the star. 
Thus, there are now
$N'=N+2M$ equations in $k=2n+m$ unknowns which can be solved using SVD.

The `art' of doing an inversion lies in choosing the correct parameters. 
There are at least three parameters that need to be specified to do an RLS inversion:
(a) the number of basis functions in frequency used to define the surface term, (b) the
number of basis functions in radius to describe the two functions, and (c) the regularisation
or smoothing parameter $\alpha$. We have the option of increasing the number
of parameters by having different numbers of basis functions the two unknown functions 
(e.g., $c^2$ and $\rho$) and/or different values of $\alpha$ for the two functions.
How these parameters can be determined have been described by \citet{basumjt1996} and \citet{basu2014},
a brief outline is given here.

The number of basis functions needed depends on the type of basis functions used.
It is most advantageous to have well localised basis functions; this ensure that the solution
at one radius is not correlated much with the solution at other radii.
\citet{antiaandbasu1994} and \citet{basumjt1996} used cubic B-splines \citep{deboor2001}.
Unlike normal cubic splines, these functions are localised. They also have
continuous first and second derivatives making it easy to apply smoothing.
The positions where B-splines are
defined are known as knots and each spline is defined over five knots.
 $m+2$ knots need to be defined for $m$ basis functions.
For inversions of solar frequencies, knots in
frequency are usually kept equidistant in frequency. Knots in $r$ are defined so that
they are equidistant in the acoustic depth $\tau$. This takes into account the
fact that modes spend more time in regions of lower sound speed making those
regions easier to resolve.

It is usual to determine the number of frequency knots first. This is done by fixing the
number of $r$ knots to a reasonably large value and then examining the residuals left when the number of
frequency-knots is changed; the aim is to eliminate structure in the residuals
when plotted against frequency. 
To determine the number of knots in $r$, one can use a combination of the reduced
$\chi^2$ obtained for reasonable smoothing,
as well as the condition number (the ratio of the largest to smallest singular
value) of the matrix $A$ in Eq.~(\ref{eq:matrix}). The two quantities have different
behaviours when plotted as a function of the number of knots. There is a 
sharp decrease in $\chi^2_\nu$ as
the number of $r$ knots is increased. The number of knots selected should lie after the decrease,
where the curve flattens out. The behaviour of the condition number is the opposite, and there
is a steep rise beyond a certain number of knots. Experience shows that it is best to select the
number of knots from the part of the curve just after the steep jump in the condition
number. This position does depend on the smoothing, and hence the selection of the number
of knots and the smoothing parameter have to be done in an iterative manner.

The smoothing parameter $\alpha$ is usually determined by inspecting the so-called
L-curve \citep{hansen1992} that shows the the balance between how smooth the solution is and how large the
residuals are by plotting the smoothing constraint $||L||^2$ (Eq.~\ref{eq:regur}) as
a function of the $\chi^2$ per degree of freedom. Increasing the smoothing parameter increases the
mismatch between the data and the solution, i.e., increases $\chi^2$, while decreasing the
oscillations in the solution (i.e., lowering $||L||^2$). Decreasing the smoothing parameter
reduces $\chi^2$, but increases $||L||^2$. The optimum value of smoothing should lie somewhere
in between. 
The smoothing parameter not only influences how smooth a solution is, it also
affects the uncertainties in the solution caused by uncertainties in the
data. The higher the smoothing, the lower the uncertainties. 

The inversion parameters not only depend on the type of basis function, but also on the number
of modes available, their uncertainties, and the distribution of the lower turning points of those modes.
Thus one cannot use the same inversion parameters if the mode-set changes.

\subsubsection{Optimally Localised Averages}
\label{subsubsec:ola}

The method of Optically Localised Averages was originally developed for application
in geophysics \citep{backusandgilbert1968,backusandgilbert1970}. \citet{dog1985} described how it
could be used for solar data. In the following discussion, we will assume that we want to
determine the sound-speed differences by inverting Eq.~(\ref{eq:inv}).

OLA inversions involve determining coefficients $c_i$ such that the sum
\be
\mathcal{K}(r_0,r)=\sum_i c_i(r_0)K^i_{c^2,\rho}(r)
\label{eq:avker}
\ee
is well localised around $r=r_0$. If $\int \mathcal{K}(r_0,r) \dd r=1$ then
\be
\left< {\delta c^2\over c^2}\right > (r_0) = \sum_i c_i(r_0){\delta\omega_i\over\omega_i}
\label{eq:cinv}
\ee
is the inversion result at radius $r_0$ as long as
\be
\mathcal{C}(r_0,r)=\sum_i c_i(r_0)K^i_{\rho, c^2}
\label{eq:cross}
\ee
is small, and the surface term contribution
\be
\mathcal{F}=\sum_i c_i(r_0){F_{\rm surf}(\omega_i)\over E_i}
\label{eq:olasurf}
\ee
is small as well. As is clear, the error in the solution will be
${e}^2(r_0)=\sum_i c_i^2(r_0)\sigma_i^2$, where $\sigma_i$ is the error
associated with the frequency of mode $i$.

 $\mathcal{K}(r_0,r)$ is the ``averaging kernel'' or ``resolution kernel'' at $r=r_0$.
The width of the resolution
kernel is a measure of the resolution of the inversions since the kernels represent the
region over which the underlying function is averaged. $\mathcal{C}(r_0, r)$ is the
``cross-term kernel'' and measures the contribution of the second
 variable (in this case density)
on the inversion results. The aim of OLA inversions is to obtain the narrowest possible averaging kernels that the 
data allow, keeping the uncertainty in the solution to acceptable levels, at the same time minimising the
effect of the cross-term kernel on the solution.

There are two widely used implementations of OLA. The first is the original one, which
helioseismology community often calls the \emph{Multiplicative} Optimally
Localised Averages, or MOLA. MOLA is computation-intensive and 
requires a large matrix to be set up and inverted at reach target radius $r_0$. A 
 less computationally intensive implementation, the \emph{Subtractive} Optimally
Localised Averages (SOLA) was introduced by \citet{pijpersmjt1992}. SOLA requires just one
matrix inversion.

\subsubsection*{Implementing OLA}
\label{para:ola_impl}

The inversion coefficients for MOLA are determined by minimising
\be
\int{\left(\sum_i c_i K^i_{c^2,\rho}\right)}^2 J(r_0, r)\dd r
+\beta\int\left(\sum_i K^i_{\rho,c^2}\right)^2 \dd r + \mu\sum_{i,j}c_ic_jE_{ij}
\label{eq:mola}
\ee
where, $J=(r-r_0)^2$ (often $J$ is defined as $12(r-r_0)^2$) and $E_{ij}$ are elements of the
error-covariance matrix. If the uncertainties in the mode frequencies are
uncorrelated, then $E_{ij}=\sigma_i\delta_{ij}$. This implementation is called the `Multiplicative' OLA because
the function $J$ multiplies the averaging kernel in the first term of Eq.~(\ref{eq:mola}).
The parameter $\beta$ ensures that the cross-term kernel is small, and $\mu$ ensures that the
error in the solution is small. The constraint of unimodularity of the averaging kernel
$\mathcal{K}$ is applied through a Lagrange multiplier.
The surface term is expressed as 
\be
F_{\rm surf}(\omega)=\sum_{j=1}^{\Lambda}a_j\Psi_j(\omega),
\label{eq:ola_surf}
\ee
$\Psi_j(\omega)$ being suitable basis functions and $\Lambda$ is the number of
basis functions. Eq.~(\ref{eq:mola}) is minimised subject to the
condition that the contribution of $F_{\rm surf}(\omega)$ to the solution is small. This is done
by adding constraints of the form
\be
\sum_i c_i {\Psi_j(\omega_i)\over E_i}=0,\; j=1,\ldots,\Lambda.
\label{eq:surfcon}
\ee

SOLA proceeds through minimising
\be
\int \left( \sum_i c_i K^i_{c^2,\rho}-\mathcal{T}\right )^2 \dd r +
\beta\int\left(\sum_i K^i_{\rho,c^2}\right)^2 \dd r + \mu\sum_{i,j}c_ic_jE_{ij},
\label{eq:sola}
\ee
where $\mathcal{T}$ is the target averaging kernel. The difference between the
averaging and target kernels in the first term of Eq.~(\ref{eq:sola}) is the reason why this
implementation is called `Subtractive' OLA.
The condition of unimodularity for the averaging kernels and the
surface constraints have to be applied as before. 

The choice of the target kernel is often dictated by what the purpose
of the inversion is. It is usual to use a Gaussian, as was done by \citet{pijpersmjt1992}.
They used
\be
\mathcal{T}(r_0,r)={A}\exp\left(-\left[{r-r_0\over \Delta(r_0)}\right]^2\right),
\label{eq:Tpijpers}
\ee
where $A$ is a normalisation factor that ensures that $\int \mathcal{T} \dd r =1$. However, this form has the
disadvantage that for small $r_0$,
$\mathcal{T}$ may not be equal to 0 at $r=0$ where the kernels $K_{c^2,\rho}$ and
$K_{\rho, c^2}$ are zero. This leads to a forced mismatch between the target and the resultant
averaging kernel. One way out is to force the target kernels to go to zero at $r=0$ using a modified
Gaussian such as
\be
\mathcal{T}(r_0,r)=Ar\exp\left(-\left[{r-r_0\over \Delta(r_0)}+{\Delta(r_0)\over 2r_0}\right]^2\right),
\label{eq:tar}
\ee
where again, $A$ is the normalisation factor that ensures that the target is unimodular.

The width of the
averaging kernel at a given target radius $r_0$ depends on the amount of information present for that
radius, the widths are usually smaller towards the surface than towards the core. This
is a reflection of the fact that (for p~modes at least) the amplitudes of the kernels are
larger towards the surface. Thus the target kernels need to be defined such that they have a
variable width. The usual practice is to define the width $\Delta_f$ at a fiducial
target radius $r_0=r_f$ and specify the widths at other locations as
$\Delta(r_0)=\Delta_fc(r_0)/c(r_f)$, where $c$ is the speed of sound. This inverse variation of the
width with sound speed reflects the ability of modes to resolve stellar structure \citep[see][]{mjt1993}.

The equations that result from the minimisation of Eqs.~(\ref{eq:mola}) and Eq.~(\ref{eq:sola})
and applying the constraints can be written as a set of linear equations of the form
\be
A\mathbf{c}=\mathbf{v}.
\label{eq:olamat}
\ee
For $M$ observed modes the matrix elements $A_{ij}$ for SOLA inversions are given by
\be
A_{ij}={\left\{ \begin{array}{ll}
\int K_{c^2}^iK_{c^2}^j \dd r + \beta\int K_\rho^i K_\rho^j \dd r + \mu E_{ij} & (i,j \le M)\\
\int K_{c^2}^i \dd r & (i\le M, j=M+1)\\
\int K_{c^2}^j \dd r & (j \le M, i=M+1)\\
0 & (i=j=M+1)\\
\Psi_j(\omega_i)/E_i & (i\le M, M+1 < j \le M+1+\Lambda)\\
\Psi_i(\omega_j)/E_j & (M+1 < i \le M+1+\Lambda, j \le M)\\
0 & (\rm otherwise)
  \end{array}
\right. }
\ee
The vectors $\mathbf{c}$ and $\mathbf{v}$ have the form
\be
\mathbf{c}={\left( \begin{array}{c}
c_1\\c_2\\ \cdot \\ \cdot \\ \cdot \\ c_M\\ \lambda \\ 0\\ \cdot \\ \cdot \\ 0
\end{array}
\right)}, \qquad\hbox{and}\quad
\mathbf{v}={\left( \begin{array}{c}
\int K_{c^2}^1\mathcal{T}\dd r \\ \int K_{c^2}^2\mathcal{T}\dd r \\ \cdot \\ \cdot \\ \cdot \\ 
\int K_{c^2}^M\mathcal{T}\dd r\\ 1 \\ 0\\ \cdot \\ \cdot \\ 0
\end{array}
\right)}
\ee

For MOLA, the elements of $A$ for $i,j \le M$ change, and are given by
\be
A_{ij}=(1-r_0)^2\int K_{c^2}^iK_{c^2}^j \dd r + \beta\int K_\rho^i K_\rho^j \dd r + \mu E_{ij}.
\ee
And as far as the vector $\mathbf{v}$ is concerned, the $(M+1)$th element for MOLA is the same as
that for SOLA, but all other elements are equal to 0. The column vector $\mathbf{c}$ is identical
in both cases. Note that in the case of MOLA, $A$ is a function of the target radius $r_0$. As a result,
the matrix has to be set up and inverted at every point where we need a result. This makes MOLA
very computationally expensive.

\epubtkImage{}{%
\begin{figure}[htb]
\centerline{\includegraphics[height=13pc]{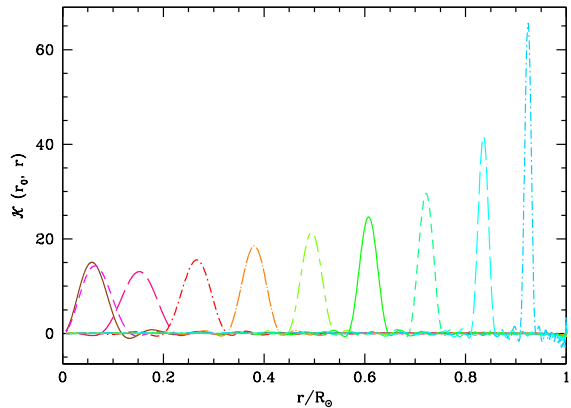}}
\caption{A sample of averaging kernels for sound-speed inversions obtained with 
the SOLA method. The averaging kernels were obtained for inversions of the frequency
differences between Model~S and the solar dataset BiSON-13 of \citet{basuetal2009}.
The results are for the inversions reported in that paper.
}
\label{fig:avker}
\end{figure}}

Choosing inversion parameters for OLA is a bit more difficult than that for RLS.
For MOLA inversions we need to determine at least three parameters -- the number of
surface terms $\Lambda$, the error suppression parameter $\mu$ and the
cross-term suppression parameter $\beta$. A fourth parameter, the width of the target averaging kernels $\Delta(r_f)$
at a fiducial radius of $r_f$, is needed for SOLA. 
\citet{cristina1999} and \citet{basu2014} give extensive discussions on how parameters can be
selected. In Figure~\ref{fig:avker} we show a few sample averaging kernels from the inversions
reported in \citet{basuetal2009}.

As in the case of RLS inversions, the surface term is the first parameter to be determined.
It is done by fitting different numbers of basis functions $\Psi$ to the
frequency differences to be inverted. The fit is done to the scaled differences. The number
of functions is increases till there is no large-scale structure discernible when the
residuals are plotted against frequency.

In the case of SOLA, the width of the target averaging kernel is such that
the mismatch between the obtained and target averaging kernel is minimised.
The mismatch can be defined as:
\be
\chi(r_0)=\int {\left[ \mathcal{K}(r_0,r)-\mathcal{T}(r_0,r)\right]}^2 \dd r.
\label{eq:avker_error}
\ee
The target kernels as defined by either Eq.~(\ref{eq:Tpijpers}) or Eq.~(\ref{eq:tar}) are
all positive. However, if the selected target width is too small, the resultant averaging kernels
have negative side-bands. Too narrow averaging kernels also result high uncertainties.

In the case of MOLA, the error-parameter $\mu$ and the cross-term parameter
$\beta$ determine the averaging kernel. While no mismatch can be defined there, one can
try to minimise the negative sidebands by ensuring that the quantity
\be
\chi'(r_0)=\int_0^{r_A} \mathcal{K}^2(r_0,r)\dd r + \int_{r_B}^1 \mathcal{K}^2(r_0,r)\dd r
\label{eq:mola_error}
\ee
is small. Here, $r_A$ and $r_B$ are defined in such a way that the averaging
kernel $\mathcal{K}$ has its maximum at
$(r_A+r_B)/2$ and its full width at half maximum is $(r_B-r_A)/2$.

In practice, small negative side lobes can be an advantage. Since the solutions at all radii are
obtained from the same set of data,
the error in the inversion result at one radius is correlated with that
at another. The error correlation between solutions at radii $r_1$ and $r_2$ are given by
\be
E(r_1,r_2)={\sum c_i(r_1)c_i(r_2)\sigma_i^2 \over
{\left[\sum c_i^2(r_1)\sigma_i^2\right]}^{1/2} {\left[\sum c_i^2(r_2)\sigma_i^2\right]}^{1/2}}.
\label{eq:errcor}
\ee
The error correlation has values between $\pm 1$. A value of $+1$ implies complete correlation
while value of $-1$ implies complete anti-correlation.
Correlated errors can introduce features into the solution on the scale of the order of the
correlation function width \citep{howemjt1996}.
While wide averaging kernels reduce $\chi(r_0)$ and $\chi'(r_0)$ and also reduce uncertainties
in the results, they can increase error correlations and give rise to spurious features
in the inversions.

It should be noted that it is almost impossible to reduce error correlations for density
inversions. The conservation of mass condition forces the solution in one part of the star
to be sensitive to the solution at other parts of the star. There is
an anti-correlation of errors between the core (where density is the highest) and the
outer layers (where density is the lowest), and the cross-over occurs around the radius
at which $r^3\rho$ has the largest value.

For fixed values of the cross-term suppression parameter $\beta$, and the error-suppression
parameter $\mu$, changing the width of the
averaging kernel changes the cross-term kernels, and 
and hence the contribution of the second function on the
inversion results of the first function.
The contribution of the cross-term kernels can be gauged using the quantity
\be
C(r_0)={\sqrt{\int \mathcal{C}^2(r_0,r)\dd r}}.
\label{eq:crossm}
\ee
We need to aim for small $C(r_0)$ in order to get a good inversion.
The requirement that the
error-correlation be small acts against the other requirements.

There are two other parameters that need to be determined, the error-suppression parameter $\mu$ and the
cross term suppression parameter $\beta$. Increasing $\mu$ decreases the error correlation
$E(r_1,r_0)$ and the uncertainty in the solution $e(r_0)$, but increases $\chi(r_0)$
 the mismatch between the
target kernels and the averaging kernel. The influence of the cross term as measured by
$C(r_0)$ also increases. Increasing
$\beta$ increases $E(r_1,r_0)$, $e(r_0)$, and $\chi(r_0)$, but decreases $C(r_0)$. 
Given that the cross term parameter reduces
systematic errors while the error-suppression parameter reduces random errors,
it is usual to try and reduce $C(r_0)$ even if that results in a somewhat
larger uncertainty in the solution. 

Given the complexity of the parameter finding process, it is advisable 
to start the inversion process by first
to examining the behaviour of the solutions obtained by inverting frequency differences
between two models using the observed modeset.

\subsubsection{Ensuring reliable inversions}
\label{subsubsec:reliable}

Since inversion results do depend on the parameters chosen,
the most reliable results are obtained when RLS and OLA results agree since
the two methods are complementary \citep{sekii1997}.

\epubtkImage{}{%
\begin{figure}[htb]
\centerline{\includegraphics[height=20pc]{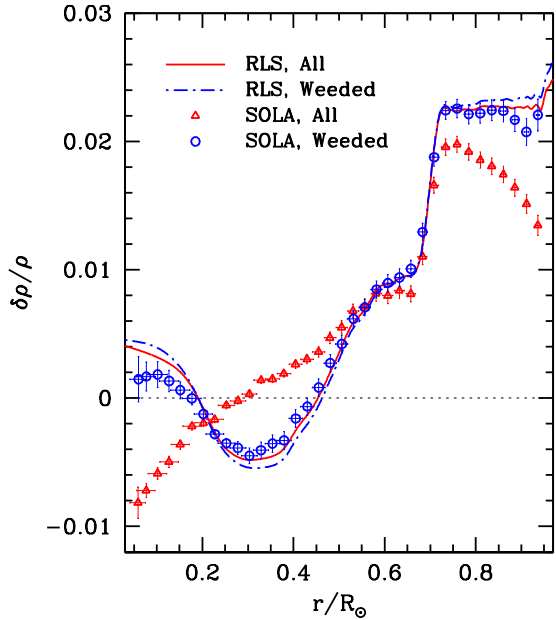}}
\caption{
Relative density differences between the Sun and model BP04 of \citet{bahcalletal2005} obtained
using both RLS and SOLA techniques. The results marked `All' were obtained
using the modeset GOLF1low (see text) while the results marked
`Weeded' where obtained when the $\ell=0,n=3$ and $\ell=3, n=5$ modes
were removed from the set. Note that RLS results remain the same, but SOLA
results change drastically. The inversion results are from \citet{basuetal2009}.
}
\label{fig:outlier}
\end{figure}}

OLA inversions are notoriously prone to systematic errors caused by outliers, while RLS results are
relatively insensitive. Unless both types of inversions are done, one often cannot detect the error due to
bad data points. An example of this is shown in Figure~\ref{fig:outlier}. The figure
shows the inverted density difference between the Sun and model BP04 of \citet{bahcalletal2005}.
The modeset was a combination of $\ell= 0, 1, 2$, and $3$ modes with frequencies less than $1800\ \mu\mathrm{Hz}$ 
obtained by GOLF as listed in \citet{bertelloetal2000} supplemented with modes from the
first year of MDI \citep{schouetal1998}. This was the so-called `GOLF1low' modeset
of \citet{basuetal2009}. 
RLS inversions show extremely large residuals for the $\ell=0,n=3$ and $\ell=0,n=5$ modes.
Almost identical results were obtained with RLS inversions whether or not these two modes are
included in the modeset that is inverted.
SOLA inversion results on the other hand changed
drastically, especially at very low radii. This discrepancy would not have been
noticed had both RLS and SOLA inversions had not been performed, and erroneous results
about the solar core would have been drawn.

The inversion results are also only as correct as the kernels used for the inversion.
It used to be customary to determine the difference between the density
profiles of the Sun and models using kernels of
$(\rho,Y)$ that were derived from kernels of $(\rho,\Gamma_1)$ using
the expansion given in Eq.~(\ref{eq:gamexp}). Such a transformation can be
done if one assumes that the equation of state of the model is the same as that for the
Sun. \citet{basujcd1997} realised that this could give rise to systematic errors
in the results. In the early days of helioseismology, the
systematic errors were smaller than the uncertainties caused by data errors,
but as uncertainties in the frequencies have become smaller, these systematic 
errors have become significant.
An example of this 
effect is shown in Figure~\ref{fig:rho_y} where we show the frequency difference
between the Sun and Model~S of \citet{jcdetal1996}. The inversions were performed using
both $(\rho,\Gamma_1)$ and $(\rho, Y)$ kernels. Note that the results differ
significantly in the core. One result (the one with $\rho,\Gamma_1$) implies
that the solar core is denser than that of Model~S, the other implies the 
opposite. Of these the results with $\rho,\Gamma_1$ kernels are more reliable
given that no assumption about any of the microphysics of the Sun went into its
derivation.

\epubtkImage{}{%
\begin{figure}[htb]
\centerline{\includegraphics[height=20pc]{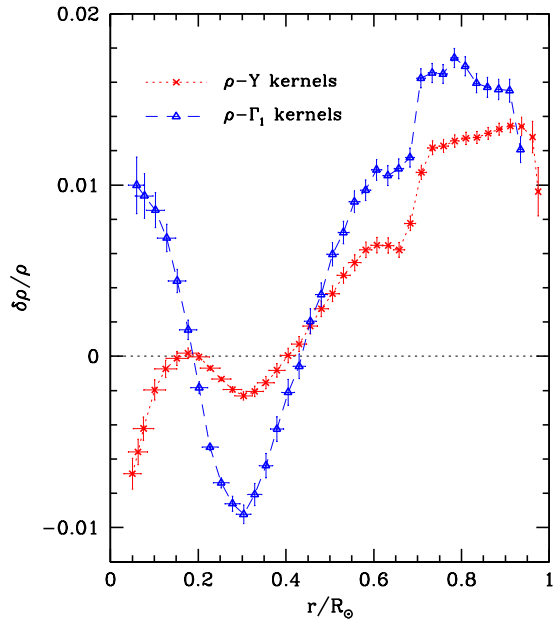}}
\caption{The relative difference in the density between the Sun and
reference model~S \citep{jcdetal1996} obtained by inverting the 
``Best set'' of \citet{basuetal1997}. 
Two results are shown, one obtained using the $(\rho,Y)$ kernel
combination, the other using the $(\rho,\Gamma_1)$ kernel
combination. Note the striking differences between the two results,
especially in the core.}
\label{fig:rho_y}
\end{figure}}

$(\rho,Y)$ and ($u,Y$) kernels had been used extensively in early inversion
because $Y$ kernels are non-zero only in the helium ionisation zone. This
 made it very easy to suppress the cross-term while keeping the error low
even with the early, limited data sets. The price paid was the systematic error
caused by differences between the equation of state of 
solar material and that used to construct solar models.
\citet{basujcd1997} showed that it is still possible to 
use $(\rho,Y)$ and $(u,Y)$ kernels without introducing errors by adding another term
in Eq.~(\ref{eq:gamexp}) that accounts for the intrinsic differences 
between the equations of state. Thus
\bea
{\delta\Gamma_1\over\Gamma_1}&=&\left({\delta\Gamma_1\over\Gamma_1}\right)_{\rm int}
+ \left({\partial\ln\Gamma_1\over\partial\ln P}\right)_{\rho,Y}{\delta P\over P}
+\left({\partial\ln\Gamma_1\over\partial\ln\rho}\right)_{P,Y}{\delta\rho\over\rho}
+\left({\partial\ln\Gamma_1\over\partial Y}\right)_{\rho,P}\delta Y\nonumber\\
&=& \left({\delta\Gamma_1\over\Gamma_1}\right)_{\rm int}+
\Gamma_{1,P}{\delta P\over P}+\Gamma_{1,\rho}{\delta\rho\over\rho}+\Gamma_{1,Y}\delta Y,
\label{eq:gamexp_int}\eea
where $(\delta\Gamma_1/\Gamma_1)_{\rm int}$ is the intrinsic difference between the 
equations of state. This term is a second cross term, and if its influence is minimised,
the systematic errors due to the uncertainties in the equation of state become 
negligibly small. There is a price to pay though, the uncertainty in the
solution becomes as large as those obtained when $(\rho,\Gamma_1)$ kernels are
used for density inversions, eliminating any advantage that using $Y$ kernels
may have given.

Although structure inversions rely on a linearisation of the oscillation
equations around a reference model, given the quality of current solar models,
this does not appear to add any errors. \citet{basuetal2000b} examined the influence
of the reference model on the estimated sound-speed, density and $\Gamma_1$ profile of the
Sun. They found that the effect is negligible, much smaller than the uncertainties caused
by data errors, when one uses modern solar models.

\newpage

\section{Results from Structure Inversions}
\label{sec:struc}

\subsection{Solar structure and the solar neutrino problem}
\label{subsec:neutrino}

Inversions of solar oscillation frequencies have allowed us to determine the
solar sound-speed, density and $\Gamma_1$ profiles 
\citep[see e.g.,][]{ jcdetal1989M, dziembowskietal1990, dappenetal1991, antiaandbasu1994,
dogetal1996, basuetal1997, turckchiezeetal1997, basuetal2000,basuetal2000b, basuetal2009}.
Inversions using solar p~mode data usually reveal the structure of the
Sun from $0.06\rsun$ to $0.96\rsun$ \citep[e.g.,][]{basuetal2009}. The lack of reliable
high-degree modes makes probing the near-surface regions more difficult.
The inner bound is set by the number of low-degree modes.

\epubtkImage{}{%
\begin{figure}[htb]
\centerline{\includegraphics[height=17pc]{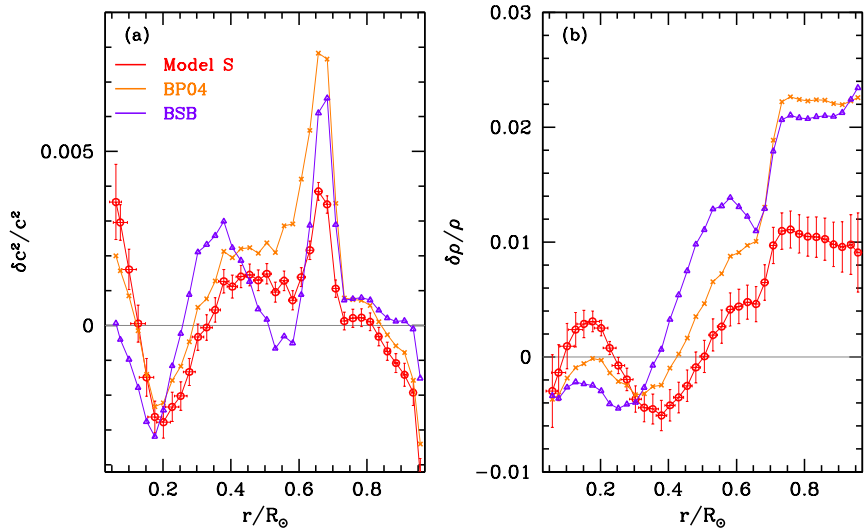}}
\caption{The relative difference in the squared sound speed and
density between the Sun and three solar models used earlier in Figure~\ref{fig:modeldif}.
The difference are in the sense (Sun\,--\,Model).
The results are from
\citet{basuetal2009} obtained using their BiSON-13 data set. In the figure
the vertical error-bars show $3\sigma$ uncertainties in the results caused
by errors in the input frequencies. The horizontal error-bars are a
measure of the resolution of the inversions. They show the distance between
the first and third quartile points of the resolution kernels. Error-bars are only shown 
for one model for the sake of clarity, they are the same for all models.
}
\label{fig:csq_rho}
\end{figure}}

The inversion results show that modern standard solar models agree quite well with the
Sun, certainly by usual astronomical standards. 
In Figure~\ref{fig:csq_rho} we show the relative sound-speed and density differences
between the Sun and three standard solar models. Note that the sound-speed difference
is fractions of a percent, while the density difference is about 2\% at the maximum.
There are however, statistically significant differences in some regions. In the case of sound speed,
we find a large difference just below the base of the convection zone. The
peak in the sound-speed difference has been attributed to the steep gradient in
the helium and heavy-element abundances just below the convection zone of the models. 
All models shown in the figure include diffusion. Diffusion increases the abundance
of helium and metals just below the base of the convection zone, which
in turn decreases the sound speed in the models (since $c^2\propto 1/\mu$, $\mu$
being the mean molecular weight). This localised difference can be reduced if some
mixing is included below the base of the convection zone \citep[e.g.,][]{richardetal1996,
basuetal2000b, brunetal2002}. There are indeed some helioseismic investigations
that suggest that the increase in $Y$ and $Z$ below the base of the 
convection zone is not as steep as that predicted by standard solar models
\citep{sasha1996, hmasmc1998}. Inversions for solar rotation 
\citep[see e.g.,][and references therein. More in Section~\ref{sec:rot}.]{mjtetal1996,schouetal1998b} 
show that there
is a strong shear layer at the base of the convection zone that is
usually referred to as the ``tachocline''. This shear layer could easily cause
mixing in the radiative-region just below the convection zone, thereby smoothing out
abundance gradients \citep[e.g.,][]{zahn1992}.
Models that include rotationally-induced mixing do indeed have a smoothed out profile
\citep[see e.g.,][]{richardetal1996}.
 The dip in the sound-speed difference around 0.2\rsun\ has not 
been completely explained yet, though it is often interpreted as being caused
by extra mixing in the early life of the Sun that is not present in the models
\citep{dogetal1996}. Density differences are more difficult to interpret since the
conservation of mass condition implies that a large difference in one part of the
Sun has to be compensated by an opposite difference elsewhere. 

The extremely small difference between the structure of standard solar models and the
Sun gave the first hints to the solution of the so-called solar neutrino
problem. Solar neutrino measurements started with a Chlorine based detector
\citep{davis1964,davis1994}. Other early experiments
included the water Cerenkov experiments Kamiokande \citep{totsuka1992,suzuki1995} and
Super Kamiokande \citep{super1991}
and gallium experiments GALLEX \citep{anselmann1995} and SAGE \citep{sage1994}.
The solar neutrino problem arose in the 1970s when
it was found that the observed flux of neutrinos from the Sun using the
Chlorine-based detector was only about a third of the neutrino flux predicted by models. 
The problem was confirmed by later observations using other detectors based on
water and gallium.
The small difference in structure between solar models and the Sun led
many helioseismologists to conclude that
the solution to the solar neutrino problem must lie with the standard model of
particle physics which postulates mass-less neutrinos. Early claims were based purely
on frequency comparisons \citep{elsworthetal1990} and comparisons of small separations
\citep{elsworthetal1995a}; later claims were based in inversion results
\citep[e.g.,][etc.]{bahcalletal1997, takatashibahashi1998, watanabe2001}. 
In fact if a non-standard solar model is constructed so that the neutrino constraints
from the Chlorine experiment are satisfied, the sound-speed difference between
the model and the Sun at the core would be about 10\% \citep{bahcalletal1998}, which
is much larger than what is seen for standard solar models. 
The solar neutrino problem had a second component, it was found that the number of neutrinos detected
by the Chlorine detector was not consistent with those detected by the water-based
detectors, which in turn were inconsistent with the gallium-based measurements. 
\citet{hmasmc1997} showed that it was not possible to construct a solar model that
satisfied all three solar neutrino constrains (i.e., those given by the Chlorine, water
and Gallium experiments) simultaneously even if some of the solar constraints were relaxed. 
\citet{hata1994,haxton1995,castellanietal1997, heeger1996} made similar inferences
using the observed data alone, without involving solar models.
Thus it was evident that the solution to the solar neutrino problem did not lie in 
deficiencies of modelling the Sun, but in the assumption regarding properties of neutrinos.
The particle-physics solution to the solar neutrino problem has since been
confirmed by results from the Sudbury Neutrino Observatory \citep{ahmadetal2002}, and the
solar neutrino problem is one of the few examples of fully solved problems in
astrophysics.

\subsection{Some properties of the solar interior}
\label{subsec:details}

While inversions have allowed us to determine solar structure in general, the finite
resolution of the inversions has often meant that specialised techniques have had to
be applied to determine some of the finer details about the solar interior, these
include determining the exact position of the base of the solar convection zone ($r_b$), the amount of
overshoot below the solar convection zone as well as the amount of helium in the
convection zone ($Y_s$).

\epubtkImage{}{%
\begin{figure}[htb]
\centerline{\includegraphics[height=17pc]{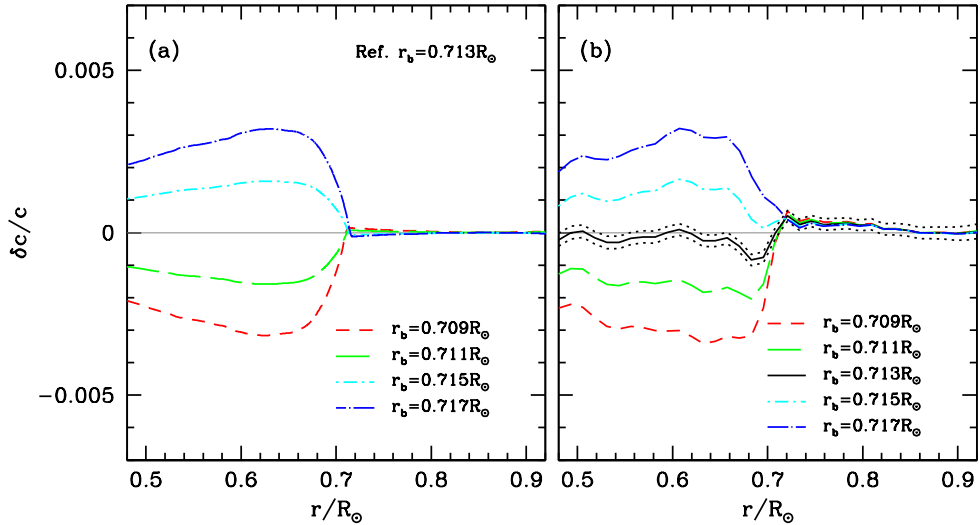}}
\caption{Panel (a): The relative sound-speed difference between a solar envelope
model with a convection zone base $r_b$ at 0.713\rsun\ and other
models with different values of $r_b$. All other physics inputs to the
models are identical. Panel (b): The sound-speed difference between the
Sun and the different models in Panel (a). The solar sound-speed results are 
from \citet{basuetal2009}. The dotted lines show the 2$\sigma$ uncertainties
on the solar results.
}
\label{fig:cz}
\end{figure}}

\subsubsection{The base of the convection zone}

The base of the convection zone is defined as the layer where the temperature gradient
changes from being adiabatic to radiative. This definition means that $r_b$ for standard
solar models is defined unambiguously. For solar models with overshoot below the base of the
convection zone, this definition results in some ambiguity. When overshoot is
modelled as a well-mixed region as well as a region which is adiabatically stratified,
$r_b$ defines the edge of the overshooting region. In models where overshooting is
merely a well mixed region, $r_b$ is the true base of the convection zone.
The abrupt change in the temperature gradient, from adiabatic to radiative, at 
$r_b$ results in a large change in the
sound-speed difference between two models that have different values of $r_b$
and this is illustrated in Figure~\ref{fig:cz}. The change in sound speed
can be used to determine the position of the convection-zone base of the Sun. 

\citet{ulrichandrhodes1977} and \citet{rhodesetal1977} compared early helioseismic data
with models to show that the Sun has a deeper convection zone than the models of that
era, and that the solar $r_b$ was between 0.62\rsun\ and 0.75\rsun. \citet{ber1980}
estimated the depth of the solar convection zone to be 200Mm. Precise estimates of
solar $r_b$ had to wait for better data sets.

\citet{jcdetal1991} used asymptotic sound-speed inversion results to determine the
dimensionless sound-speed gradient $W(r)$ given by
\be
W(r)\equiv \frac{r^2}{Gm} \frac{\dd c^2}{\dd r}.
\label{eq:wr}
\ee
$W(r)$ is nearly constant, and equal to $-0.67$, in the deeper part of the convection zone.
It increases abruptly below the convection-zone base. \citet{jcdetal1991} used the
position of this increase to find that the base of the solar convection zone
was located at $r_b=0.713\pm0.001\rsun$. \citet{fedorova1991} obtained similar results.
\citet{sbhma1997} did a detailed study of the systematic errors involved in the 
process. They used the function $H_1(w)$ (Eq.~\ref{eq:h1h2}) obtained between the
Sun and models with different values of $r_b$ to determine position of the
solar convection-zone base. Their
results, despite using much better data, was identical to that of \citet{jcdetal1991}.
\citet{basu1998} studied whether or not errors in measurement of the
solar radius would have any effect on $r_b$ determinations, she found none, and
using data from the GONG and MDI projects she obtained $r_b=0.7133\pm0.0005\rsun$.
\citet{sbhma2004} examined whether the adopted value of the solar heavy-element abundance
has any effect on the estimated value of solar $r_b$ and found none. Thus the position
of the solar convection zone base is known sufficiently precisely to provide a strong 
constraint that solar models need to satisfy.
The question of whether the base of the solar convection zone has any latitudinal
dependence has also been studied by different groups.
\citet{dogagk1995} inverted for $\delta u/u$ between the Sun and a
spherically symmetric model using BBSO data to find that the equator may be
somewhat deeper than that at the poles, but stated that
the difference did not exceed 0.02\rsun. However, neither
\citet{mario1998} nor \citet{sbhma2001} found any such latitudinal variation.

\epubtkImage{}{%
\begin{figure}[htb]
\centerline{\includegraphics[height=17pc]{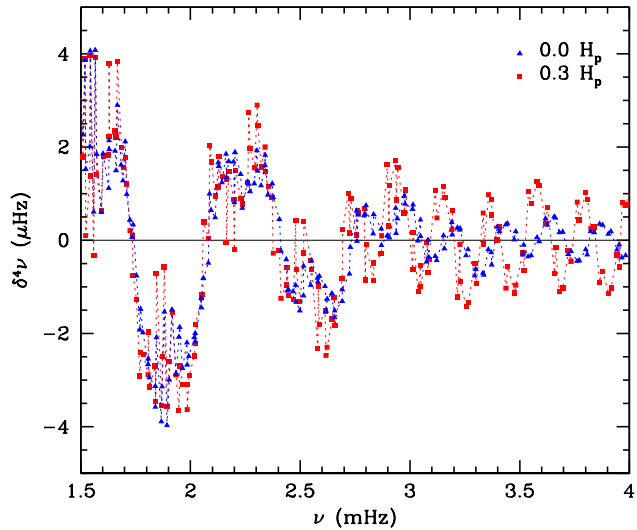}}
\caption{The signature of the acoustic glitches in two solar models, one without
overshoot (blue triangles) and one with an overshoot of $0.3H_p$ (red squares). The signature has been enhanced by taking
the fourth differences of the frequencies. Dotted lines joining the points are meant merely to
guide the eye, the jaggedness of the lines reflects the $\ell$-dependence of the signature.
Note that there are two distinct oscillatory patterns, the larger-wavelength one is the
signature of the He\,{\sc ii} ionisation zone, the smaller wavelength one is from the
base of the convection zone. Note that the amplitude of the smaller-wavelength pattern
is larger for the model with overshoot. 
}
\label{fig:ov}
\end{figure}}

The abrupt change of the temperature gradient at the convection-zone base is an
acoustic glitch that leaves its signature on the frequencies 
[see Section~\ref{sec:surf}, Eq.~(\ref{eq:glitch})], and this signature can be isolated to
study the region. The amplitude of the signal is a function of
the size of the glitch. Simple models of convective overshoot model the
phenomenon by extending the adiabatic temperature gradient 
artificially and then changing over to the radiative gradient below the overshoot
region. This increases the size of the acoustic glitch and that can be
used to determine the extent of overshoot below the solar convection zone, an
idea put forward by \citet{dogsekii1993}. In Figure~\ref{fig:ov} we show this
for two solar models.
Assuming that this model of overshoot is correct, then overshoot below
the solar convection-zone base is very small \citep[e.g.,][]{monteiroetal1994, basuetal1994, sbhma1994, 
iwrsvv1994, basu1997}, and \citet{basu1997} put an upper limit
of $0.05H_p$ on the amount of overshoot. Of course, such a simple model of overshoot is unlikely to
be correct and there have been attempts to modify the change of the temperature
gradient to make it more realistic \citep{jcdetal1995}, however, what is realistic is
still a matter of debate \citep[see, e.g.,][]{rempel2004}. 
Two-dimensional numerical simulations
suggest that overshoot could lead to a slightly extended mildly adiabatic temperature
gradient beneath the convection zone up to $0.05H_p$ followed by a 
rapid transition to a strongly sub-adiabatic region \citep{rogersetal2006}.
There are few 3D simulations that are focused on studied the overshooting region,
and there are issues with those \citep[see][for an account]{canuto2011}.

\subsubsection{The convection-zone helium abundance}

Another important property of the Sun is its convection-zone helium abundance $Y_s$. 
Since the Sun is a cool star, $Y_s$ cannot be measured spectroscopically. Yet,
this is an important parameter that controls solar evolution. The convection-zone
helium abundance is expected to be lower than the initial helium abundance $Y_0$ of the
Sun because of diffusion and gravitational settling, but nevertheless it can constrain models of the current Sun.
Helium leaves its signature on solar oscillation frequencies because it changes the
mean molecular weight (and hence sound speed), it also contributes an acoustic glitch
in the form of the He\,{\sc ii} ionisation zone (the He\,{\sc i} ionisation zone for the Sun overlaps
with the H\,{\sc i} ionisation zone and as a result causes large systemic error when determining
the helium abundance, and hence it is not used for such studies). It is the second 
effect that is usually exploited in different ways to determine
the solar helium abundance.

The first attempts to determine $Y_s$ \citep{dappendog1986,dappenetal1988b} were 
handicapped by the lack of high-precision data. More successful attempts were
made after the publication of the \citet{libbrechtetal1990} frequencies. \citet{dappenetal1991}
did a full inversion using kernels for $(u, Y)$ to obtain $Y_s=0.268\pm 0.01$;
\citet{dziembowskietal1991} on the other hand found $Y_s=0.234\pm 0.005$.
\citet{kosovichevetal1992} found $Y_s=0.232\pm 0.005$; they also investigated the
reason for the large variation in the estimated values of $Y_s$ and found that 
a major source of systematic error is the equation of state needed to
convert the $(c^2,\rho)$ kernels to the $(u, Y)$ kernels. 
\citet{agk1993M} used low degree data from the IPHIR instrument \citep{toutain1992},
and used the $A^*$--$Y$ kernel combination to examine whether low-degree
modes are enough to determine $Y$; $A^*$ is negligible in the CZ and hence its
contribution to the inversion should be small. He found that $Y_s=0.251$ and $0.256$
for the two data sets he used. 
The spread in the values of $Y_s$ remained even when better data from GONG and MDI
became available. Using reference models constructed with the MHD 
equation of state \citet{richardetal1998} estimated $Y_s$ to be $0.248\pm 0.002$
which using models constructed with the OPAL equation of state
\citet{dimauroetal2002} determined it to be $0.2539\pm 0.0005$.

\epubtkImage{}{%
\begin{figure}[htb]
\centerline{\includegraphics[height=17pc]{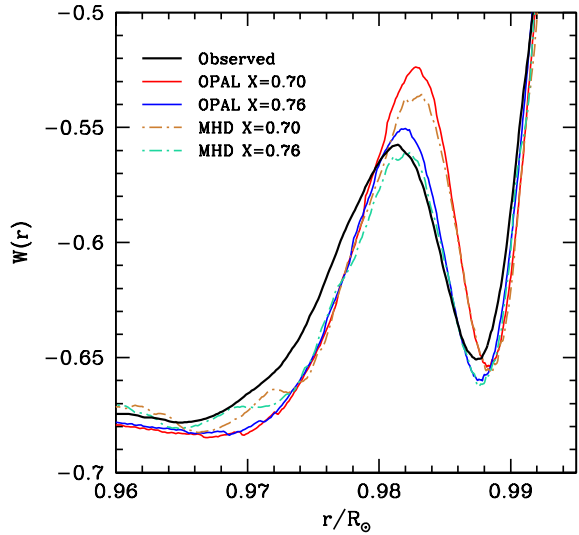}}
\caption{A comparison of the observed dimensionless sound speed gradient
of the Sun in the region of the He\,{\sc ii} ionisation zone
(black line) with that of different models constructed with different
equations of state and different convection-zone helium abundances.
The heavy-element abundance is the same for all models. Note
that the height of $W(r)$ depends on the helium abundance for a given
equation of state, but for a given helium abundance it depends on the
equation of state.
}
\label{fig:wrobs}
\end{figure}}

Performing direct inversions is only one way to determine the solar helium
abundance. \citet{svvetal1992} by using an asymptotic description of the
oscillation equations concluded
that $Y_s=0.25\pm 0.01$. \citet{guzikandcox1992} did a straightforward 
frequency comparison to find $Y_s=0.24\pm 0.005$. \citet{hmasb1994} used a
different method altogether -- following the idea of \citet{dog1984MmSAI}
they used the dimensionless sound speed gradient $W(r)$ 
(Eq~\ref{eq:wr}) to determine $Y_s$. 
The height of $W(r)$ in the helium ionisation
zone is a function of the helium abundance, unfortunately it also
depends on the equation of state (see Figure~\ref{fig:wrobs}). They found that $W(r)$ for models
constructed with the EFF equation of state did not match observations at all. Using models
constructed with MHD equations of state they estimated $Y_s$ to be $0.252\pm 0.003$.
Once the OPAL equation of state was made available, \citet{sbhma1995} showed that
$Y_s$ estimates using calibration method, when either $W(r)$ or $H_1(w)$ were
calibrated against models of know helium abundance, were less sensitive to
equation-of-state effects than full inversions. Using the \citet{libbrechtetal1990} data
they obtained $Y_s=0.246$ using MHD models, and $Y_s=0.249$ using
OPAL models. Taking into account systematic errors due to the techniques
and equation of state effects, they estimated $Y_s$ to be $0.249\pm 0.003$.
\citet{basu1998} used data from the GONG and MDI projects to revise the number
to $Y_s=0.248$. \citet{sbhma2004} revisited the issue. Instead of looking at $Y_s$, they
looked at the hydrogen abundance $X_s$ and found $X_s=0.7389\pm 0.0034$ using
models with $Z/X=0.0171$ in the convection zone.

Most of the effort into determining the solar helium abundance has been directed at
determining the present-day helium abundance of the Sun. In terms of stellar evolution though,
it is the initial helium abundance that is important. As a result of
gravitational settling, we cannot directly estimate what the solar initial helium abundance
was. However, using current constraints, it is possible to work backwards to
estimate $Y_0$ for the Sun. \citet{serenellibasu2010}
determined how the initial and present-day helium abundances depend on the 
 parameters used to construct solar models, and then they determined what the
limits on the initial helium abundance of the Sun should be given our knowledge of
the current structure of the Sun. They found 
that when only standard solar models are considered, the estimate of the initial
helium abundance for the Sun is $Y_0=0.278\pm 0.006$, independent of the
solar model or its metallicity. If non-standard models with turbulent mixing below the
convection-zone base are used, then $Y_0=0.273\pm 0.006$.

\subsection{Testing input physics}
\label{subsec:inp_phys}

Helioseismic inversion results can be used to test input physics. In Figure~\ref{fig:difnodif}
we showed the frequency-differences between the Sun and two solar models, one that included the
diffusion and settling of helium and heavy elements, and one that did not. We had not
been able to conclude which of the two models is the better one. The
situation changes immediately if we invert the frequency difference to determine the
relative sound-speed and density differences between the Sun and the two models. The
results of such an inversion are shown in Figure~\ref{fig:dif_nodif_inv}.

\epubtkImage{}{%
\begin{figure}[htb]
\centerline{\includegraphics[height=17pc]{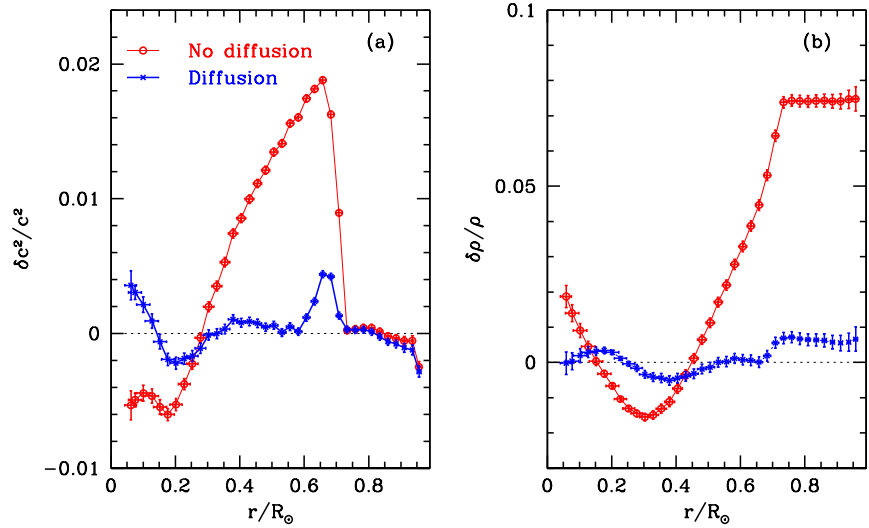}}
\caption{The relative differences in the squared sound speed [Panel (a)] and
density [Panel (b)] between the Sun and two solar models obtained by
inverting the frequency differences shown in Figure~\ref{fig:difnodif}. Note the
the model without diffusion has much larger differences with respect to the
Sun than the model with diffusion.
}
\label{fig:dif_nodif_inv}
\end{figure}}

As can be seen from Figure~\ref{fig:dif_nodif_inv}, the model without diffusion is much more discrepant
than the one with diffusion. The primary
reason for the large discrepancy of the non-diffusion model is that it has a much shallower 
convection zone than the Sun. Additionally, models without diffusion also have a much higher value of
$Y_s$ than the Sun. As a result, modern standard solar models include the diffusion and 
gravitation settling of helium and heavy elements.

\epubtkImage{}{%
\begin{figure}[htb]
\centerline{\includegraphics[height=17pc]{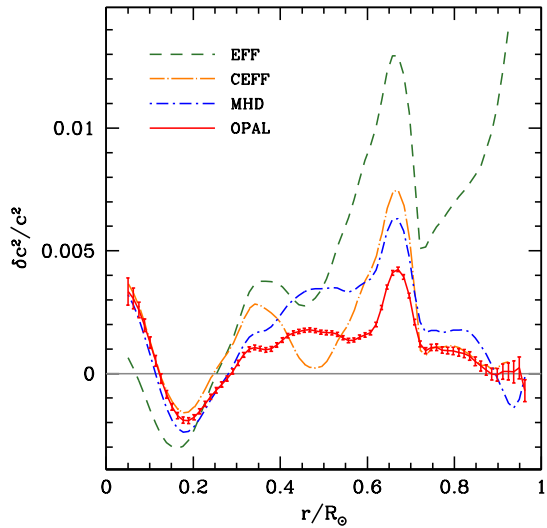}}
\caption{The relative squared sound-speed difference between the Sun and
standard solar models constructed with different equations of state. The results were obtained
by inverting the frequency-differences shown in Figure~\ref{fig:eos}.
}
\label{fig:csq_eos}
\end{figure}}

\epubtkImage{}{%
\begin{figure}[htb]
\centerline{\includegraphics[height=17pc]{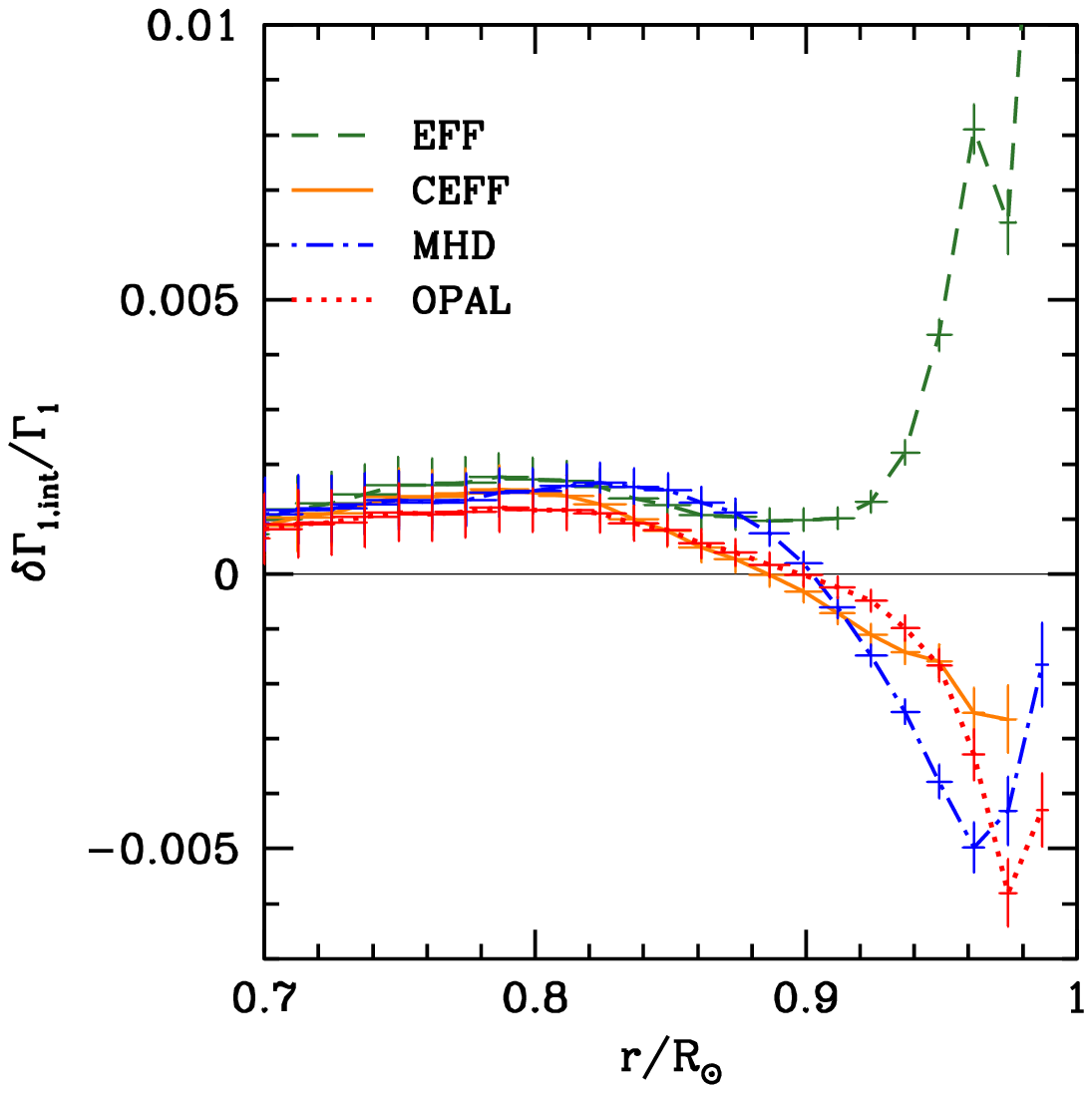}}
\caption{The intrinsic $\Gamma_1$ differences between four different equations of state and the
Sun. These were obtained by inverting Eq.~(\ref{eq:gamexp_int}).
}
\label{fig:eos_int_inv}
\end{figure}}

In Figure~\ref{fig:eos} we showed the frequency differences between models constructed
with different equations of state and the Sun and were overwhelmed by the
surface term. In Figure~\ref{fig:csq_eos} we show the sound-speed difference between the
Sun and the different solar models. It is pretty clear that EFF is discrepant, and
it appears that OPAL works the best. 

Solar models are constructed solving a set of coupled equations 
and the equation of state affects the structure in multiple ways. As a result, just
looking at the sound-speed difference between between models constructed
with a given equation of state and the Sun does not tell us the full story. A
better way is to look at the adiabatic index $\Gamma_1$ in the convection zone
where we expect adiabatic stratification that depends only on the equations of state.
\citet{basujcd1997} had shown that it is possible to isolate the effect of the
equation of state by inverting Eq.~(\ref{eq:gamexp_int}). The results of such an 
inversion are shown in Figure~\ref{fig:eos_int_inv}. Note that it is clear 
that EFF is discrepant. The OPAL equation of state seems to do better in
the deeper layers than the MHD equation of state, however, closer to the
surface MHD appears to be better. \citet{basuetal1999} examined why this is so
and concluded that it was a result of how the internal partition
function of hydrogen was treated in the two equations of state.

\citet{elliott1998} examined the equation of state for conditions present in
the solar core. They determined the difference in $\Gamma_1$ between models constructed
with the MHD and OPAL equations of state and the Sun and found a discrepancy in the core. 
The cause of the discrepancy was identified as the use of the non-relativistic approximation 
to describe partially degenerate electrons instead of the relativistic Fermi--Dirac integrals. 
The deficiency has since been corrected for both OPAL \citep{rogersandnayfonov2002} and 
MHD \citep{gongetal2001} equations of state.

Somewhat more indirect means have been used to test opacities and nuclear energy
generation rates. The stellar oscillation equations do not depend directly on these
inputs. What we look for are the effects of changes in these inputs on solar models
and how these models look in comparison with the Sun.
The earliest tests of opacities were simple. \citet{guzikandcox1991} compared
frequencies of models constructed with the then available opacity tables \citep[those of][]{coxandtabor1976} 
to suggest that the opacities were too low. \citet{saio1992} put opacity results on a more
quantitative basis by assuming that the sound-speed differences between the models and the
Sun were caused by opacity alone. He found that the discrepancy between
his models and the Sun can be reduced if the Los Alamos opacities \citep{weissetal1990}
were increased by 20\,--\,50\%. The publication of the OPAL opacity tables \citep{rogersandiglesias1992}
and the OP tables \citep{badnelletal2005} confirmed the results of \citet{saio1992}.
Similar methods have been used to look at differences between OPAL opacities and
the opacity of the solar material. \citet{tripathyandjcd1998} using the assumptions of \citet{saio1992}
calculated kernels linking sound-speed changes to opacity changes. \citet{jcdetal2009}
used this method to calculate the opacity changes needed to make low-$Z$ solar models,
more specifically models constructed with the \citet{ags05} metallicities,
agree with the Sun and found that changes of up to 30\% may be required. \citet{sbhma1997} looked at
density differences rather than sound-speed differences, and found that the density 
difference between solar envelope models that have the correct convection-zone depth and
helium abundance could be used to look at changes in opacity. They concluded that 
the OPAL opacities were consistent with seismic constraint. It should be noted though
that their results were for models with the then accepted solar metallicity of 
$Z/X=0.0245$. The authors used the same method to later quantity opacity changes
needed for models constructed with $Z/X=0.0171$ and found that an increase
of about 20\% would be needed to match the density profiles \citep{sbhma2004}. 

There have also been attempts to constrain the rate of the $p$-$p$ reaction. This is 
the reaction which controls energy generation inside the Sun. Given that the
rate of this reaction has not been measured, many helioseismologists
turned to seismic data to put constraints on this reaction
\citep[see e.g.,][]{hmasmc1998,hmasmc1999, hmasmc2002, degl1998, schlattl1999, tc2001}.
For instance, the seismic constraint on the cross-section of the
$p$-$p$ reaction obtained by \citet{hmasmc2002} \citep[see also][]{brunetal2002}
is $S_{11}=(4.07\pm0.07)\times10^{-25}\mathrm{\ MeV\ barns}$. This should be
compared to $(4.00\pm 0.03)\times10^{-25}\mathrm{\ MeV\ barns}$ \citep{adelbergeretal1998}
and $(4.01\pm 0.04)\times10^{-25}\mathrm{\ MeV\ barns}$ \citep{adelbergeretal2011} that has been
recommended by nuclear physicists.

Other ``secondary'' inversion results are those for the abundance profiles
inside the Sun. These inversions are done under the assumption that we know
the opacity of stellar material perfectly. \citet{kag1995} determined
$\delta X/X$ between the Sun and the model of \citet{jcdetal1993} and 
found that the results indicated that the Sun has a smoother gradient of $Z$ below the
convection zone than the model. This was confirmed by others. For example, \citet{shibahashiandtakata1996} assumed
$Z/X=0.0277$ and $Y=0.277$ in the solar envelope and varied the $X$ abundance
below to determine the $X$ abundance. \citet{hmasmc1998} did a similar analysis.
Both results showed that the hydrogen abundance below the convection-zone base
is smoother than that found in standard solar models. Models constructed with these
``seismic'' abundance profiles do not have the sharp difference in sound speed
at the base of the convection zone that we see in Figure~\ref{fig:csq_rho} for standard solar
models.

\subsection{Seismic models}
\label{subsec:seismic}

The fact that the structure of SSMs do not agree completely with that of the Sun have resulted
in the construction of so-called ``seismic models.'' These models are constructed so that 
their structure agrees with that of the Sun. The primary motivation behind early
seismic models was better predictions of the solar neutrino fluxes and estimating
other properties of the Sun.

There are two types of seismic models. The more common type are seismic models of the
present day Sun. These are obtained by solving the stellar structure equations using the
helioseismically determined solar sound-speed and density profiles as constraints.
The other type consist of models obtained by the usual evolution from the zero-age main sequence,
but in these models one or more physics inputs are tuned so that the structure
of the final model agrees with that of the Sun. The second type of seismic model is
not particularly common; the models of \citet{tcetal2001} and \citet{couvidat2003}
fall into this category.

One of the earliest examples of the first category of seismic models is that of \citet{fedorova1991}. 
\citet{dziembowskietal1994} and \citet{dziembowskietal1995} soon followed with their own models.
\citet{hma1996} constructed a seismic model using an elaborate iterative technique. Seismic models
were also constructed by \citet{takatashibahashi1998}, \citet{watanabe2001}, \citet{hiromoto1998}.
 Such models are still being constructed as helioseismic data
get better; these include the seismic model of \citet{choubey2001} which was used to estimate
properties related to mixing of different neutrino flavours.

\subsection{The solar abundance issue}
\label{subsec:abun}

In Figure~\ref{fig:csq_rho} we showed the sound-speed and density differences between some solar 
models and the Sun. All the differences were small and we could assume that we are quite good
at constructing solar models. 
All models shown is Figure~\ref{fig:csq_rho} however, were constructed assuming that the
Sun has a relatively high heavily element abundance. Model~S was constructed using
$Z/X=0.0245$ as per the estimates of \citet{gn93}. The other two models were constructed
with $Z/X=0.023$ as per \citet{gs98}. 

The solar heavy-element abundance is one of the
most important inputs to solar models. Since heavy elements increase opacity, the heavy-element 
abundance affects the structure of solar models. The solar heavy-element abundance is also
used as a standard for calibrating the abundances of other stars, as a result, there have been
many attempts to determine what the solar metallicity is, and some of the more recent 
estimates have generated a lot of debate and discussion in the community.
In a series of papers \citet{carlos2001, carlos2002}, and
\citet{asplund2004, asplund2005e, asplund2005} revised the spectroscopic
estimates of the solar photospheric composition downwards. The
main feature of their analysis was the use of three-dimensional model atmospheres
that included the dynamics of the gas and hence obviated the use of micro- and macro-turbulence
parameters; they also incorporated some non-LTE effects in their analysis.
That resulted in lowering the carbon, nitrogen and oxygen abundances
in the Sun by 35\% to 45\% of those listed in \citet{gs98}.
The revision of the oxygen
abundance leads to a comparable change in the abundances of neon and
argon since these abundances are generally measured through the abundances
ratio for Ne/O and Ar/O.
Additionally, \citet{asplund2000} also determined a somewhat
lower value (by about 10\%) for the photospheric abundance of silicon compared
with the value of \citet{gs98}.
As a result, all the elements for which abundances
are obtained from meteoritic measurements have seen their abundances
reduced by a similar amount. These measurements have been summarised
by \citet{ags05}. The net result of these changes
is that $Z/X$ for the Sun reduces to $0.0165$ (or $Z\sim 0.0122$), about 28\% lower than
the previous value of \citet{gs98} and almost 40\% lower than the old value of
\citet{ag89}. These changes in the solar abundance led to very large changes in 
the structure of solar models. Models constructed with these abundances do not agree
well with the Sun. The models have very shallow convection zones and low convection-zone
helium abundances. 

\epubtkImage{}{%
\begin{figure}[htb]
\centerline{\includegraphics[height=17pc]{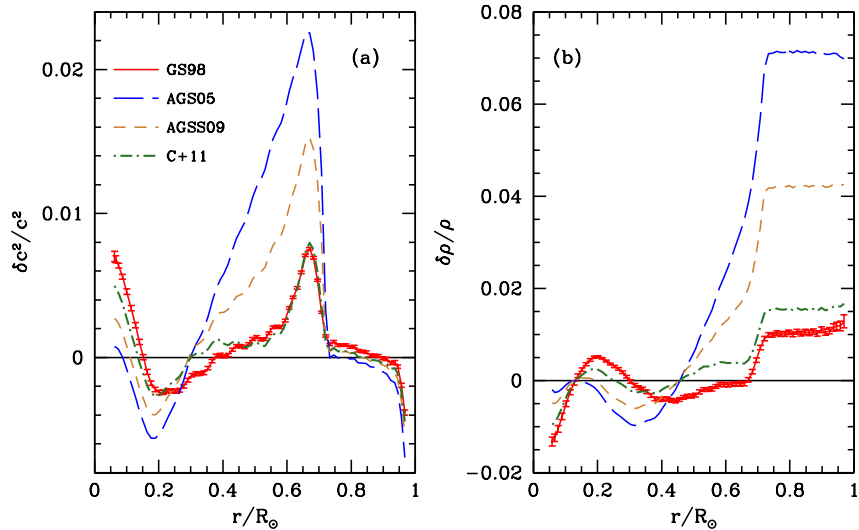}}
\caption{The relative sound-speed and density differences between the Sun and four standard
models constructed with different heavy-element abundances but otherwise identical physics
inputs. The model GS98 was constructed with \citet{gs98} abundances, AGS05 with
\citet{ags05} abundances, AGSS09 with \citet{agss09} abundances, and models
C+11 was constructed with \citet{caf2010,caf2011} abundances supplemented with
abundances from \citet{lod2010}.
}
\label{fig:abun_all}
\end{figure}}

The mismatch between the low-$Z$ solar models and the Sun led to numerous attempts
at modifying the models, and their inputs. There have also been attempts to 
determine solar abundances from helioseismology. These attempts and their results 
have been reviewed thoroughly by \citet{sbhma2008}. What we give below is an
update of the situation.
With further improvement in analysis, abundances were updated by \citet{agss09} to $Z/X = 0.018$.
While this improved the models somewhat, they were by no means as good as those
with higher values of $Z/X$ \citep[see e.g.,][]{serenelli2009}. There were other, independent, attempts
to determine solar heavy element abundances using three-dimensional model atmospheres,
and this led to $Z/X=0.0209$ \citep{caf2010,caf2011}. In Figure~\ref{fig:abun_all} we show the
relative sound-speed and density differences between the Sun and solar models 
constructed with different heavy-element abundances. It is clear that the low-$Z$ models
do not match. Some of the properties of these models are listed in Table~\ref{tab:props}.
\citet{sbhma2013} used solar envelope models and density inversion results to determine
the amount how much opacity increase would be needed to make the models
in Figure~\ref{fig:abun_all} agree with the Sun; they found changes of 6\% for the model
with \citet{caf2010,caf2011} abundances, 
16.5\% for the \citet{agss09} model and 26.5\% for the \citet{ags05} models. A recent 
review of the issue can be found in \citet{bergemann2014}.

\begin{table}[htb]
\caption{Properties of standard solar models shown in Figure~\ref{fig:abun_all}}
\label{tab:props}
\centering
\begin{tabular}{lcccc}
\toprule
Mixture & $Z/X$ & $R_{\rm CZ}$ & $Y_{\rm CZ}$ & $Y_0$ \\
\midrule
Helioseismic & -            & $0.713\pm0.001^a$ & $0.2485\pm 0.0034^b$ & $0.273\pm 0.006^c$ \\
GS98          &  0.023   &  0.7139     &  0.2456  &  0.2755 \\
AGS05        & 0.0165  &  0.7259     &  0.2286  & 0.2586 \\
AGSS09      & 0.018    &  0.7205     &  0.2352  & 0.2650 \\
C+11        & 0.0209  &  0.7150     &   0.2415  &  0.2711 \\
\bottomrule
\end{tabular}

{$^a$} {\small Basu \& Antia (1997)}; {$^b$} {\small Basu \& Antia (2004)};
{$^c$} {\small Serenelli \& Basu (2010)}\hfill\\
\end{table}

Since the \citet{sbhma2008} review was published there have been more investigations to
see whether low-$Z$ models can be brought in agreement with the Sun without too many
changes to the inputs. One such attempt was by \citet{zhang2014} who showed that
including the dynamical effects of convection, in particular the kinetic energy
flux, could make low-$Z$ models agree with the Sun. While the scope of that
investigation was very limited and not applied to SSMs but only
to envelope models, and it is not clear if the kinetic energy profile
used is realistic, it is certainly an interesting, physically motivated idea.
Another, somewhat unconventional, idea was proposed, and ultimately rejected,
by \citet{vincent2013}. They examined the impact of particles
with axion-like interactions with photons and whether they affect the spectroscopically
determined values of the solar heavy-element abundance. They found that these
interactions could in principle resolve the problem by inducing a slight increase in the
continuum opacity at line-forming heights in the solar
atmosphere. This in turn would reduce the computed equivalent
widths of solar absorption lines for any given elemental abundance, which would
result in an underestimation of the abundances. However, the authors
found that the coupling necessary to obtain the required changes are
ruled out by current experiments.

There have been more efforts to determine the solar abundance independently
of spectroscopy. \citet{svvetal2013}, using solar envelope models and the adiabatic
index $\Gamma_1$ claim that helioseismic data are consistent with low (and even very low)
metallicities in the range $Z=0.008$\,--\,$0.013$. This contradicts the results of 
\citet{hmasb2006}. These low-$Z$ results also contradict what was found
by \citet{villante2014} who did a joint statistical analysis of helioseismic
and solar neutrino data to find that the solar abundances of oxygen and iron are
consistent at the $\sim1\sigma$ level with (high) values of \citet{gs98}. Solar
$Z$ values as low as $0.008$ are also inconsistent with current models of Galactic
chemical evolution.

Thus the solar abundance issue is still alive and well! Reliable high-degree mode
frequencies will help in resolving the near-surface layers of the stars where the 
ionisation zones of different elements leave their imprint. Until then we have to
look at multiple diagnostics and determine whether we can reach a consensus on
this issue.

\subsection{How well do we really know the structure of the Sun?}
\label{subsec:gmodes}

Inversions of solar p~mode frequencies available today allow us to
determine the structure of the Sun between about $0.05$ to $0.96\rsun$ reliably.
There are a few \citep[e.g.,][]{dimauroetal2002} that go closer to the surface, 
but these results are geared towards studying near-surface features.
The interior limit is set by the available low degree modes, while the near-surface limit
is set by the lack of reliable high-degree modes. For the intervening layers,
different data sets give consistent sound-speed inversion results.
As a result,
we believe that we know the solar sound-speed profile quite well.
This is not the case for density -- 
improvements in the mode sets, particularly in the low-degree range, change the
density inversion results in a statistically significant manner.
This can be seen in Figure~\ref{fig:modesetchange}. The reason for this is the
conservation of mass condition. Since the models are constructed to have the same
mass as the Sun, the integrated density difference between the Sun and the
models should be zero. As a result when mode sets improve, the density results
change at all radii. Improvements of low-degree modes are particularly important
as they give better results for the core. Since density is highest in
the core, a small change in the inferred densities in that regions results in a
large change elsewhere.

\epubtkImage{}{%
\begin{figure}[htb]
\centerline{\includegraphics[height=15pc]{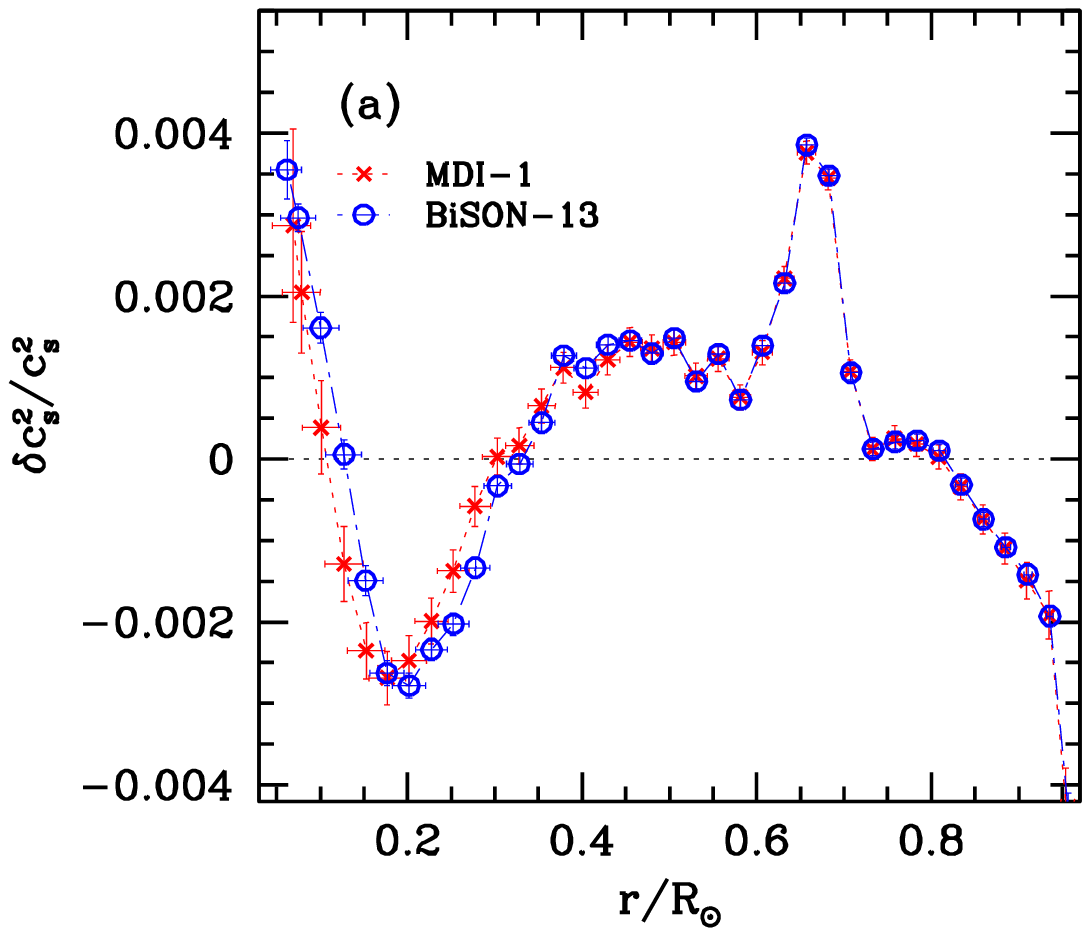}\includegraphics[height=15pc]{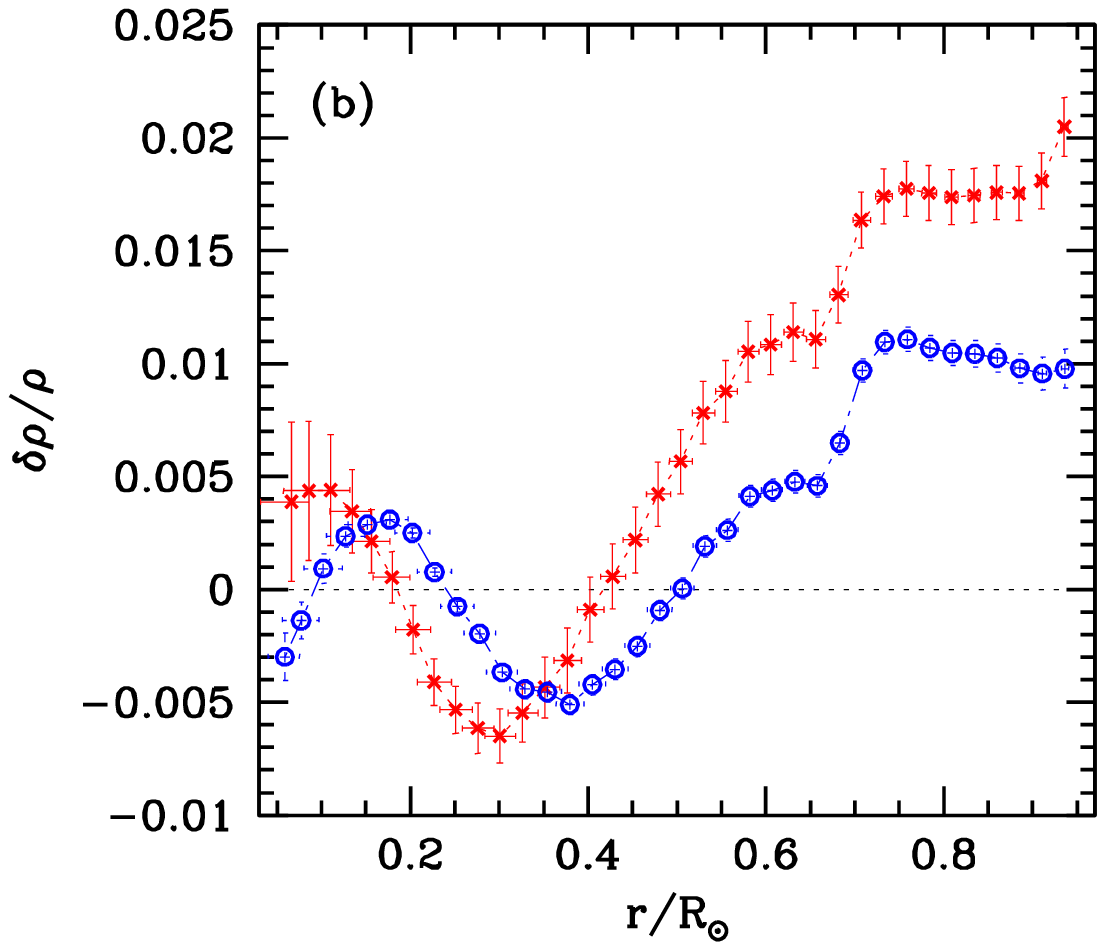}}
\caption{The relative sound-speed and density differences between the Sun and Model~S
obtained by inverting two different data sets that only differ in their $\ell=0$, 1, 2, and
3 modes. The results are from \citet{basuetal2009} and the two mode sets, labelled
MDI-1 and BiSON-13 are described in that
paper. Note that while there is very little change in the sound-speed differences, the
density differences change significantly.
}
\label{fig:modesetchange}
\end{figure}}

Better results for the solar core can be obtained with g~modes \citep[see e.g.,][]{agk1988}. These modes have
high amplitudes in the core and hence, sample the core well. However, as mentioned in
Section~\ref{sec:theory}, g-mode amplitudes decay in convection zones and the
Sun has a fairly deep one, making the task of detecting g~modes very difficult.
There is a long history of trying to detect solar g~modes and a short
report of the rather frustrating quest can be found in \citet{gmode2013}. For a
more technical review see \citet{gmode2010}.

\citet{rafa2007} claim that they detect periodic structures
that was consistent with the expected splitting for dipole
g~modes in periodograms constructed with GOLF data. That work
did not list individual frequencies or frequency-splittings. \citet{rafa2008} showed that
there was statistically significant power in the same frequency range in the VIRGO data.
There is no consensus yet as to whether or not individual g~mode peaks can be
detected with the data we have today, and for that matter, whether or not their
frequencies can be measured. \citet{rafa2010} claims tentatively that the individual modes
can be seen.
This is the only group that has claimed to have measured frequencies
of individual g~modes and not merely their signature in modern, high-precision data. If their results are
confirmed independently, and the frequencies are measured to a good precision,
we should be able to get a better handle on the structure of the solar core and
the density of all layers of the Sun. 

The other part of the Sun which we have not probed very well yet is close to the solar surface, i.e.,
the near-surface layers.
The reason is the dearth of reliable high-degree modes. Most helioseismic data sets
have mode frequencies of up to $\ell=200$ (GONG) or $\ell=250$ (MDI and HMI). These modes
do not probe the outer layers very well. Having high-degree modes assists
the inversions by better defining the surface term by resolving differences
close to the surface \citep{cristina2000}. The reason for the lack of modes is technical:
above about $\ell=200$, individual modes have large line-widths 
due to their short lifetimes, and they merge into ridges
when combined with the observational spatial leakage.
The spatial leakage is the spherical-harmonic transform equivalent of the side lobes seen in Fourier
transforms. The leakage occurs because a 
spherical harmonic decomposition is not orthogonal when using data over the visible
solar disc, which is less than half a hemisphere. This results in 
modes of similar degrees and azimuthal orders leaking into the estimate of a specific
degree and azimuthal order. Other reasons of leakage include line-of-sight projection effects, 
distortion of modes by differential rotation, etc.
For low-$\ell$ models the leakages are separated in frequency domain,
but the large widths of high-degree modes means that the leakages and the central peak
form a ridge. Asymmetries in the leakage could mean that the highest point of the
ridge may not be the eigenfrequency of the Sun. Thus determining frequencies
of high degree modes is challenging \citep[see, e.g.,][]{rhodes2003,reiter2003,korzennik2013}.
At lot of effort is being put into this, however, progress has been slow.

\newpage

\section{Departures from Spherical Symmetry}
\label{sec:rot}

The Sun is spherical in shape -- the measured difference between the polar and equatorial
radii is only about 1 part in $10^5$ as per the analysis by \citet{kuhn1998} of MDI data. Updated results
obtained with data from the PICARD satellite puts the number slightly lower,
a difference of 6.1~km, i.e., about 9 parts in $10^6$ \citep{picard2014} or perhaps even lower
\citep[$5.7\pm0.2\mathrm{\ km}$,][]{meftah2015}. However, although spherical, the Sun is not spherically symmetric.
The main deviation from spherical symmetry in the solar interior is caused by rotation;
on the solar surface, deviations from spherical symmetry are also seen in magnetic fields.

Rotation, magnetic fields and anything else that breaks spherical
symmetry causes oscillation frequencies to split and lifts the degeneracy between 
frequencies of modes that have the same degree $\ell$ and radial order $n$ but that have different
values of the azimuthal order $m$ \citep{ledoux1949, ledoux1958, cowling1949}. 
These splittings can be used to determine solar 
rotation. 
The frequency splittings $\nu_{n,\ell,m}-\nu_{n,\ell,0}$ can be separated into
two components, one that is odd in $m$ and one that is even in $m$. The
odd component arises from rotation, and in particular from advection and Coriolis
force. The component even in $m$ is caused by centrifugal forces, large-scale
magnetic fields or any non-spherically symmetric distortion of the star.

As mentioned in the introduction, it is customary to write the frequencies
and their splittings as a polynomial expansion in $m$, i.e.,
\begin{equation}
{\nu_{n\ell m}\over2\pi}=\nu_{n\ell m}=\nu_{n\ell}+\sum_{j=1}^{j_{\max}} a_j(n\ell )\mathcal{P}^{n\ell }_j(m),
\label{eq:nusplit}
\end{equation}
where $\mathcal{P}^{n\ell }_j(m)$ are related to Clebsch--Gordon coefficients.
In this representation, the odd-order $a$-coefficients are sensitive to rotation. This is
a result of the structure of the kernels that relate the frequency splittings to rotation
\citep[for a description of the kernels see][]{jcd2003}. The $a_1$ coefficient encodes the rotation
rate averaged over all latitudes; $a_3$ and higher order coefficients describe differential
rotation.
The even-order
coefficients are sensitive to the second-order effects of rotation (i.e., the centrifugal forces)
as well as magnetic fields and other deviations from spherical symmetry. 
The actual values of the $a$ coefficients at a given epoch depends on the polynomials $\mathcal{P}$ used
in the expression given in Eq.~(\ref{eq:nusplit}). The most common practice today is to
use either the Clebsch--Gordon coefficients or the \citet{ritzwoller1991} formulation, or 
polynomials proportional to those. Connections between some of the more
common definitions of the coefficients may be found in \citet{pijpers1997}.

In the case of the
Sun, the first odd-order coefficient $a_1$, is much larger than the first even-order
coefficient (see Figure~\ref{fig:acoef}) indicating that rotation is the primary
agent for breaking spherical symmetry in the Sun.

\epubtkImage{}{%
\begin{figure}[htb]
\centerline{\includegraphics[height=15pc]{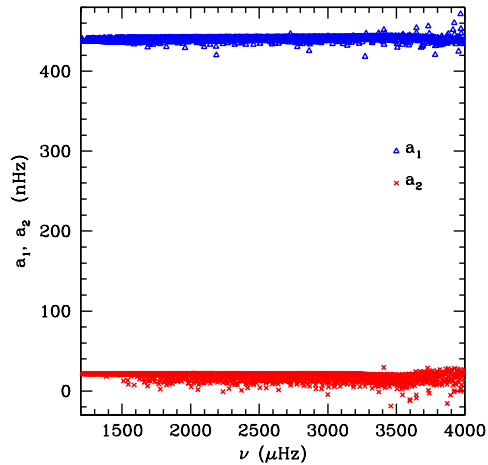}}
\caption{The dominant odd-order splitting component ($a_1$; blue triangles) and the dominant even-order
component ($a_2$; red crosses) for the Sun. Data are from a MDI frequency set constructed
with a one-year time series from observations that began on May~1, 1996.
Note that the $a_1$ component is much larger than the $a_2$ component, indicating that
rotation is the main agent for breaking spherical symmetry in the Sun.
}
\label{fig:acoef}
\end{figure}}

\subsection{Solar rotation}
\label{subsec:odda}

For a detailed review of solar rotation, readers are referred to the 
\textit{Living Reviews in Solar Physics} article by \citet{howe2009}; 
we describe the basic features here.

\epubtkImage{}{%
\begin{figure}[htb]
\centerline{\includegraphics[height=17pc]{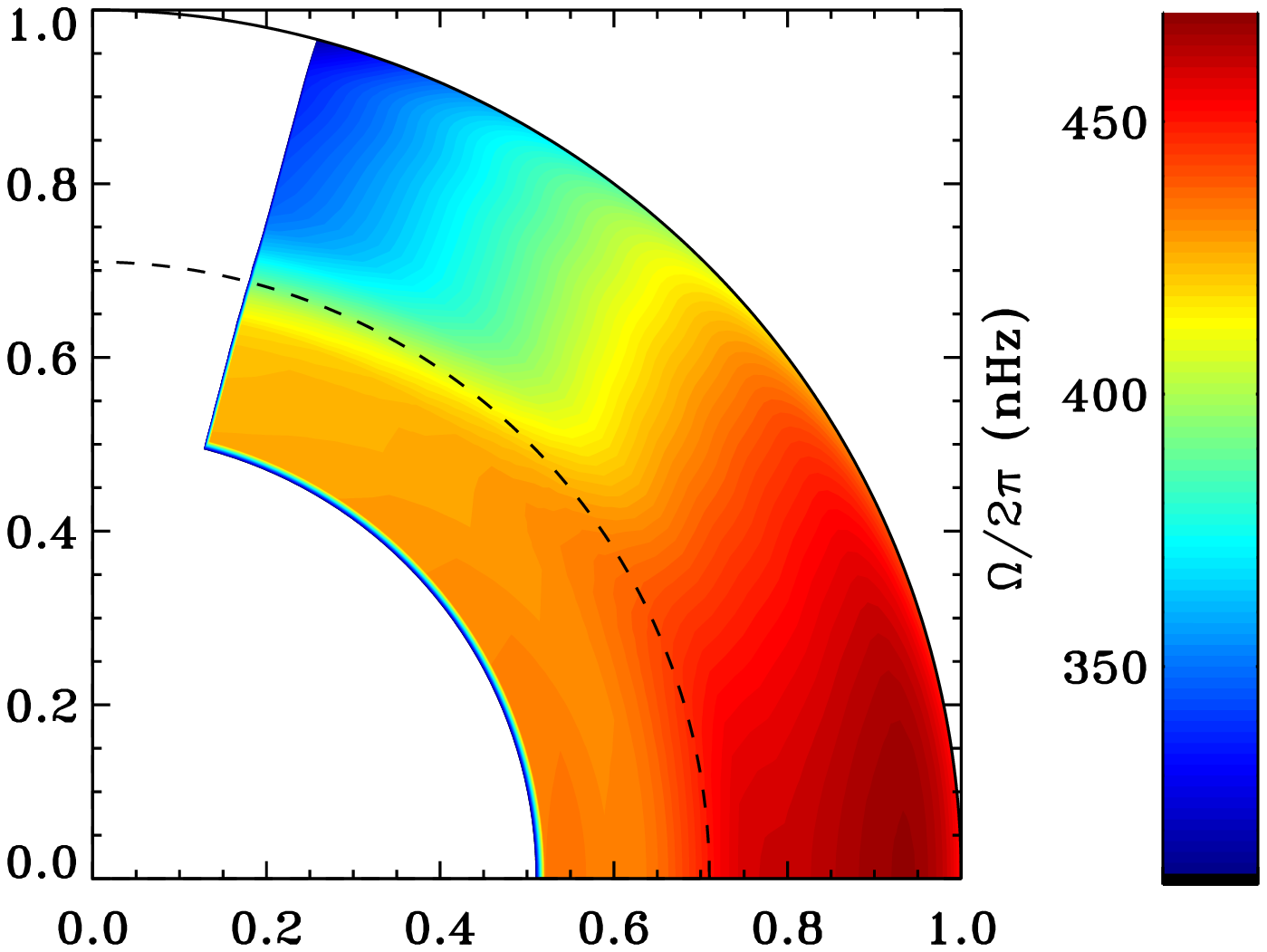}}
\caption{The mean solar rotation profile obtained with data collected by the GONG
project between mid-1995 to mid-2004. The
rotation rate in the deep interior has large uncertainties and hence, has not been
shown.
The inversion results are from \citet{howeetal_2005}. (Image courtesy of Rachel Howe.)
}
\label{fig:rot}
\end{figure}}

The motions of sunspots on the solar disk 
had revealed that the surface of the
Sun rotated differentially with the solar equator rotating once every 25 days and the
higher latitudes taking a longer time to rotate. This had been confirmed from
Doppler shifts of photospheric lines \citep[][etc.]{snodgrass1983, snodgrass1984,ulrichetal1988},
as well as measurements with other tracers \citep[e.g.,][]{duvall1980, snodgrass1990, kommetal1993}.
It had been predicted that the rotation would be
constant on cylindrical surfaces throughout the convection zone,
matching onto the observed surface rotation at the top of the convection zone, i.e., 
the Sun iso-rotation contours of the Sun will be cylinders. Helioseismology
revealed a very different, and more complicated, picture 
\citep[see e.g.,][and references therein]{mjtetal1996,schouetal1998b}.
The mean rotation profile of the solar interior is shown in Figure~\ref{fig:rot}.
A few features stand out immediately. The Sun rotates like a solid body below the
base of the convection zone, though the rotation in the innermost parts of the Sun is not
really known because it is so poorly constrained by the observations of
p~modes. Within the convection zone rotation changes with both
latitude and radius. For most of the convection zone, rotation is nearly 
constant on cones, i.e., at fixed latitude the rotation is almost independent of depth,
however, there is some hint of rotation
on cylinders in the near-equatorial regions of the convective envelope \citep{gilman2003}.
There are two distinct shear layers, one in the immediate sub-surface 
layers, and a more pronounced one at the base of the convection zone.
The shear layer at the base of the convection zone is now called the 
``tachocline.'' For a review of different methods of determining solar
rotation, see \citet{beck2000}.

The data that yield the solar rotation rate have also been used to determine
the total angular momentum, $H$, of the Sun, its total
rotational kinetic energy, $T$, as well as the gravitational
quadrupole moment, $J_2$. \citet{pijpers1998} showed that the error-weighted mean
of these quantities obtained with GONG and MDI data are:
\be
H=(190.0\pm1.5)\times 10^{39}\mathrm{\ kg\ m^{2}\ s^{-1}},
\ee
\be
T=(253.4\pm7.2)\times 10^{33}\mathrm{\ kg\ m^{2}\ s^{-2}},
\ee
\be
J_2=(2.18\pm0.06)\times 10^{-7}.
\ee
These were redetermined by \citet{antiaetal2000} and they obtained consistent, though slightly
different results.

\subsection{Other deviations from spherical symmetry}
\label{subsec:evena}

While solar rotation has been studied in great detail, other asphericities
in the Sun have not been the subject of many detailed studies. The reason for this is two-fold, the first 
being that the signature of such asphericities in the Sun is small, as is clearly seen 
in Figure~\ref{fig:acoef}. The second reason is perhaps the more fundamental one:
the even-$a$ coefficients have the disadvantage that the signal in the
coefficients  could be a result of sound-speed asphericities or magnetic fields or both.
\citet{zweibel1995} showed that even-order frequency splittings cannot be
unambiguously attributed to the effect of a magnetic field, and the effect
of a magnetic field on frequency splittings can be reproduced by a perturbation
of the sound speed. \citet{zweibel1995} argued that the ambiguity in the
signature of global models arises because the range of latitudes sampled by the
modes depends on their azimuthal order $m$. This allows both magnetic and
acoustic perturbations to produce the same frequency shifts. However,
the spatial configuration of a magnetic field that produces a given splitting
is not the same as the spatial configuration of an acoustic perturbation
that produces the same splitting. As a result, studies of asphericity
using frequency splitting proceed by assuming that the even-order splittings
or $a$ coefficients are caused  by  magnetic fields alone or  by aspherical
perturbations to structure alone. Most modern investigations
however, do correct for the second-order effects of rotation.
It should be noted that asphericities in the shape of the Sun can also 
produce even-order splittings \citep{dogtaylor1984}, however, the measured asphericity of the
Sun's limb is smaller than what is needed to explain the observed $a$ coefficients.
For a review on how magnetic fields can affect mode frequencies and
splittings see \citet{mjt2006}.

\subsubsection{Magnetic fields}
\label{subsubsec:magf}

One of the first estimates of solar interior magnetic fields  obtained with frequency splittings was
that of \citet{dziemgoode1989}. They developed a formalism relating the
frequencies to the magnetic field and used the data from \citet{libbrecht1989}
to find that there should be an axisymmetric quadrupole toroidal
magnetic field of $2\pm 1\mathrm{\ MG}$ centred near the base of the convection zone. They confirmed
the results using data from \citet{libbrechtandwoodard1990} \citep{dziemgoode1991}.
\citet{goodemjt1992} tested the hypothesis that the radiative interior of the
Sun could be hiding a large-scale magnetic field, which might not be
axisymmetric about the observed rotation axis and found that the strength
of an axisymmetric relic field must be less than 30~MG. Such
a field will cause an oblateness of $5$\,--\,$10\times 10^{-6}$.
\citet{goodedziem1993} revisited their older work and using updated data for the splittings found that 
the significance of the mega-Gauss field at the convection-zone base was lower than
what they had suggested in their earlier papers. Their result was revised further by
\citet{basu1997} who used different data (those obtained by the GONG project), and a somewhat different
method, to find an upper limit 0.3~MG on the toroidal magnetic field concentrated below 
the convection-zone base.

\citet{dogmjt1990} presented a perturbation method to calculate the effects of 
magnetic fields and rotation on the frequency splittings. Modified versions of that
formulation have been used extensively to estimate the magnetic fields in the
interior of the Sun. Using the precise $a$ coefficients obtained by the
GONG and MDI projects, \citet{hmaetal2000} used the \citet{dogmjt1990} formulation to do
 a forward analysis and showed that the even-order $a$ coefficients, when corrected
for the second-order effects of rotation, can be explained by a 20~KG field
located at a depth of 30\,000~km below the solar surface. They also found an upper
limit of 0.3~MG for a toroidal field at the base of the convection zone.
\citet{hma2002} used the data to put 
an upper limit of $10^{-5}$ on the ratio of magnetic pressure to gas pressure in
the solar core. \citet{baldneretal2009} did a more detailed analysis; they used
multiple data sets and found that the data are best fit 
by a combination of a poloidal field and a double-peaked near-surface toroidal field.
The toroidal fields are centred at $r_0=0.999\rsun$ and $r_0=0.996\rsun$ and have
peaks strengths of $380 \pm 30\mathrm{\ G}$ and $1.4 \pm 0.2\mathrm{\ KG}$ respectively. The poloidal field
is best described by a dipole field with a peak strength of $124 \pm 17\mathrm{\ G}$.
It should be noted that the idea of a shallow field goes back a long way -- in an IAU Symposium 
in 1986, Gough \& Thompson showed how the then-available data have hints of a shallow
perturbation \citep[see][]{dogmjt1988}.

\subsubsection{Acoustic asphericity}
\label{subsubsec:asph}

As mentioned earlier, even-order splitting coefficients can also be caused by aspherical acoustic 
perturbations, and hence, there have been a number of investigations aimed at
estimating the acoustic asphericity, if any, of the Sun. Early works
\citep{kuhn1988,kuhn1989} related the frequency splittings to temperature perturbations,
but the data were not good enough to make firm conclusions. Data from GONG and MDI changed
the situation somewhat, however, there are still very few studies of asphericity
of solar internal structure.

\citet{sbhma2001E} used a combination of the central frequencies and the even-order $a$ coefficients
to make frequency combinations pertaining to different latitudes and used those to
determine the position of the base of the convection zone as a function of
latitude. They did not find any statistically significant difference in the position
with latitude. \citet{antiaetal2001} did a detailed analysis of all the sets of splittings
available to determine the latitudinally dependent sound-speed profile
of the Sun by doing a two-dimensional inversion of the even-order $a$ coefficients. 
They found that it was possible to determine
structural asphericities within the convection zone, though the errors became progressively larger the
deeper they went. They found sound-speed asphericities of the order of $10^{-4}$. The 
most notable feature is a peak in the asphericity around $0.92\rsun$.  The other feature is that
sound-speed perturbation at the equator is negative, but  it is positive at mid-latitudes, a
results which can be interpreted as the equator being cooler than the mid-latitude region.
The amount of the temperature asphericity is similar to that found by \citet{dziemetal2000}
who inverted the different orders of the coefficients, i.e., $a_2$, $a_4$ etc., separately.

Asphericity in the near-surface layers is expected to be larger because the low
densities there make the layers susceptible to perturbations. However, the absence of reliable  high-degree
mode-frequencies makes it difficult to study these regions.

\newpage

\section{Solar-Cycle Related Effects}
\label{sec:cycle}

One of the more interesting features of solar oscillations is that the oscillation
frequencies change with solar activity. This was noted even in early data
\citep{woodardandnoyes1985, woodard1987,libbrecht1989,palle1989,elsworth1990, libbrechtandwoodard1990,
anguerra1992}.
Modern data have confirmed the early results.
The solar-cycle related changes affect the central frequencies and frequency
splittings of low degree modes \citep{salabertetal2004, toutain2005,chaplinetal2007} as well as intermediate degree 
modes \citep{bachmann1993}. Although not
too many high-degree data sets are available, it is clear that there are changes in the frequencies of
high-degree modes too \citep{cristina2006, cristina2008,rhodesetal2011}. 
The frequency changes  appear to be functions
of frequency alone when mode inertia is taken into account, and is reminiscent of
a ``surface term'' (see Figure~\ref{fig:solarcycle}).
Frequency splittings also show solar-cycle related shifts. 

\epubtkImage{}{%
\begin{figure}[htb]
\centerline{\includegraphics[height=15pc]{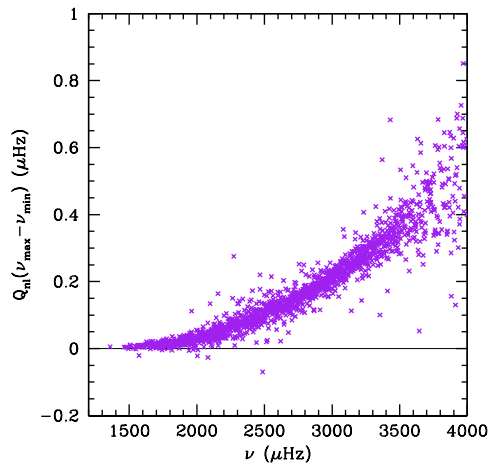}}
\caption{The scaled frequency differences between two sets of solar data, one
obtained at solar maximum and the other at solar minimum. The frequencies were
obtained by MDI, each from a 72-day long time-series.
The one corresponding to solar maximum was from observations that began in June 2001 (MDI set 3088), the one 
for the minimum was from observations that began in May 1996 (MDI set 1216). (Data courtesy of Tim Larson.)
}
\label{fig:solarcycle}
\end{figure}}

\citet{woodardetal1991} showed that 
changes in solar frequencies were strongly correlated with the changes in the  magnetic
field strength. This was subsequently confirmed by 
\citet{bachmann1993}, \citet{elsworth1994} and \citet{regulo1994}. The variation of the frequencies with the changing
magnetic activity is now well established and its nature being looked into in 
detail. Although solar oscillation frequencies increase with increase
in solar activity, the rise and fall does not always follow the same path.
\citet{chano1998} analysed low-degree cycle~22 data to find a ``hysteresis'' effect. This
was also seen by \citet{tripathyetal2001} in intermediate-degree data. 
Like \citet{chano1998}, they found that some of the solar activity indices
also show this effect. However, the interesting feature shown by \citet{tripathyetal2001} is
that while magnetic indices
(like KPMI, the Kitt Peak Magnetic Index and MPSI, the Magnetic Plage Strength Index)
show the same type of hysteresis as solar frequencies, some of the radiative indices of solar activity 
do not. For the 10.7-cm flux as well as the equivalent width of the He~10830~{\AA} line, the ascending and descending
branches cross each other.

The relationship between solar frequency shifts and solar activity indices is however,
not always simple. Most investigations assume a linear relationship between the
different indices and frequency shifts, there are however,  significant deviations and
anomalies. \citet{tripathyetal2007} looked at the correlation between frequency shifts and
magnetic indices over different time periods and showed that the correlation 
varies from their to year. They
also demonstrated that there were significant differences between long-term and short-term variations
and that the correlation complex. Deviations from a simple trend were also
found by \citet{rachel2006} who
studied at the frequency shifts for modes of $\ell=0$\,--\,$2$  observed by BiSON, GONG and MDI, and
found that even after a linear solar-activity dependence is removed, there are
significant variations can are seen in the different data sets. The 
authors conclude that the variations are likely to be related to the stochastic excitation of the modes.
\citet{chaplinetal2007} looked at three solar cycles worth of data from BiSON and
compared the behaviour of mean frequency shifts of low-$\ell$ modes with several proxies of global solar
activity to determine, among other things, which activity proxy follows the frequency shifts most closely.
They assumed a linear correlation and concluded that the He\,{\sc i} equivalent width and the Mg\,{\sc ii} core-to-wing data
had the best correlation, though the 10.7-cm flux also fared well. This led to an in-depth examination
of the shifts of intermediate degree modes by \citet{jain2009}. Using data from GONG and MDI
they examined the correlation between the frequency shifts and nine different 
activity indices for cycle~23. Their updated results are summarised in Figure~\ref{fig:jain}. They also 
divided their data into a rising phase (1996 September~22 to 1999 June~26), a high activity phase
(1999 June~27 to 2003 January~12) and a declining phase (2003 January~13 to 2007 July~26) and
found that the correlation between frequency shifts and solar
activity changes significantly from phase to phase except in the case of  the 10.7-cm flux. In most
cases, the rising and the declining phases were better correlated than the high-activity
phase.  The different degrees of correlation of the different activity indices with the frequency
shifts should, in principle, inform us about changes in the Sun, however, interpreting
the correlations has proved to be difficult. 

\epubtkImage{}{%
\begin{figure}[htb]
\centerline{\includegraphics[height=30pc,angle=90]{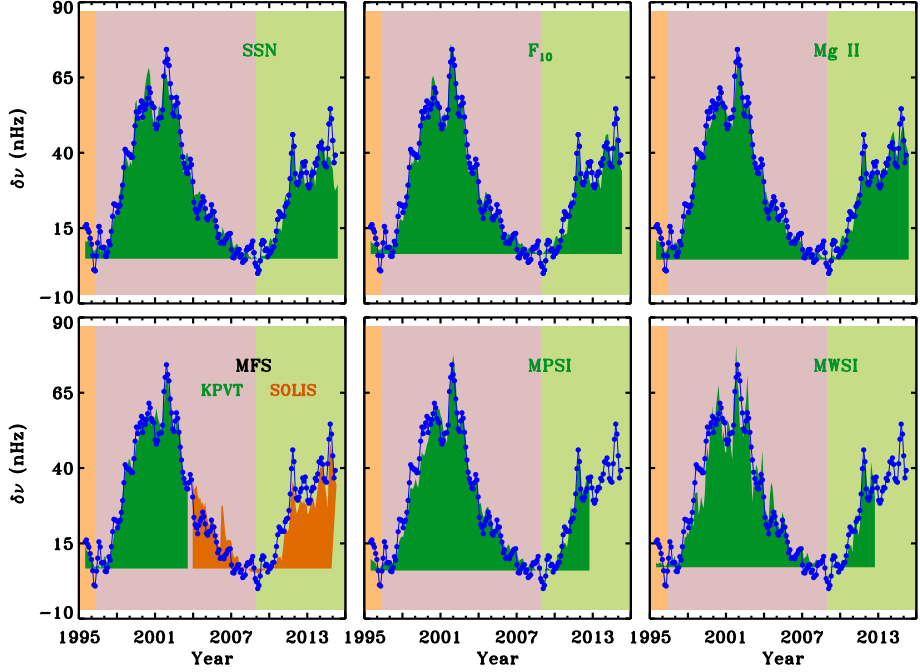}}
\caption{The time-evolution of the GONG frequency shifts (points) plotted with different
activity proxies (filled regions) scaled to fit on the same axis. The frequency shifts were calculated
with respect to the temporal mean of the frequencies. 
The orange, pink and light-green backgrounds denote solar cycles 22, 23 and 24 respectively.
Figure courtesy of Kiran Jain.
}
\label{fig:jain}
\end{figure}}

\epubtkImage{}{%
\begin{figure}[htb]
\centerline{\includegraphics[height=20pc]{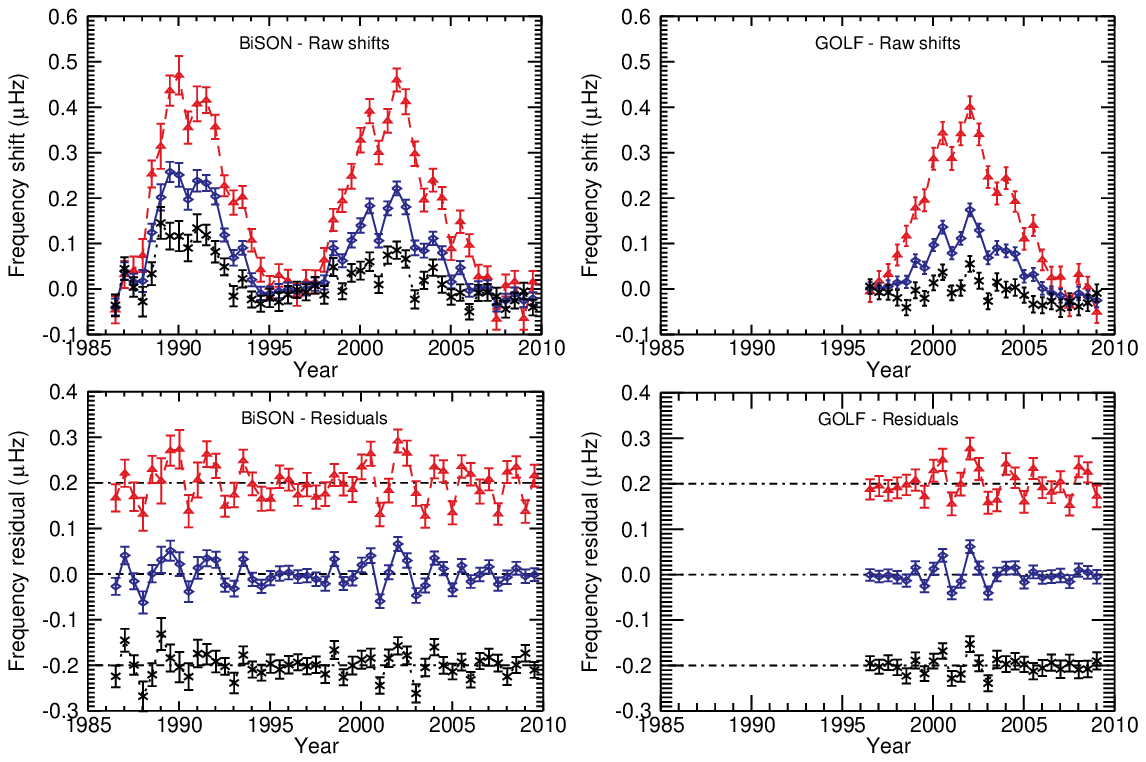}}
\caption{The upper panels show the shifts in frequencies of low-degree modes from BiSON (left) and
GOLF(right) as a function of time. The blue solid line is for
all modes with frequencies between 1.88 and 3.71~mHz. The black and red lines
are subsets: black for 1.88 to 2.77~mHz and red from 2.78 to 3.71~mHz.
The lower panels show the residuals when the dominant solar-cycle trend is removed. The
black and red curves have been displaced by $0.2$ and $-0.2\mathrm{\ mHz}$ respectively for the
sake of clarity.
(Image reproduced with permission from \citealp{fletcher2010}, copyright by AAS.)
}
\label{fig:twoyear}
\end{figure}}

The Sun also  exhibits variability  of time scales much shorter than
the usual 11-year solar cycle. Over the last few
decades, it has been suggested that the Sun shows significant variability
on a shorter time scale of about two years. \citet{benevolenskaya1995} and \citet{benevolenskaya1998}
showed this using low-resolution magnetic field observations taken at the Wilcox Solar Observatory.
\citet{mursula2003} reported periods between 1 and 2 years in various heliospheric parameters
such as solar wind speed, interplanetary magnetic fields, geomagnetic activity etc.
\citet{valdes2008} showed that such 1\,--\,2 year variabilities are also seen
in coronal holes and radio emissions. 
Readers are referred to the \textit{Living Reviews in Solar Physics}
article of \citet{hathaway2015} for details of the quasi-biennial period
in different observations. 
Such short-term variations are also seen
in solar frequencies. 
\citet{fletcher2010} showed that solar frequencies are modulated
with a two year period. This was seen in data from both BiSON and GOLF (see Figure~\ref{fig:twoyear}).
Note that the amplitude of the modulation is almost
independent of the frequency of the mode.  This is quite unlike the main solar-cycle related
modulation where high-frequency modes show a larger change of frequencies than low-frequency
modes. However, the amplitude of the modulation shows a time-variation: the amplitude is larger at
solar maximum than at solar minimum. \citet{fletcher2010} and \citet{broomhalletal2011} speculated
that the two-year signal is that from a  second dynamo whose amplitude is modulated by the 
stronger 11 year cycle. The almost frequency-independent nature of the two-year signal points to
a second dynamo whose seismic effects originate at layers deeper than those of the
primary dynamo. The presence of two different types of
dynamos operating at different depths has already been proposed
to explain the quasi-biennial behaviour observed in other proxies
of solar activity \citep{benevolenskaya1998, benevolenskaya1998A}. However, not everybody 
agrees at the second signal is that of a second dynamo; \citet{simoniello2013}
claim that the signal could result from the beating between a dipole and quadrupole magnetic configuration of 
the solar dynamo.

\subsection{Were there solar cycle-dependent  structural changes in the Sun?}
\label{subsec:struc_change}

The apparent lack
of a degree dependence in the frequency shifts have led to the conclusion that
solar cycle related changes in the Sun are confined to a thin layer close
to the surface or even above the surface \citep[][etc.]{libbrechtandwoodard1990,goldreich1991, evans1992,
nishizawa1995}. This picture was confirmed by subsequent studies \citep[e.g.,][]{rachel1999,sbhma2000,
dziemgoode2005}. Theoretical work too suggested shallow changes \citep[e.g.,][]{goldreich1991,balmforth1996,
lietal2003}. There have been attempts to determine the location of the changes, and with data
available over more than a solar cycle, and it has indeed been possible to detect some
of changes.

\epubtkImage{}{%
\begin{figure}[htb]
\centerline{\includegraphics[height=17pc]{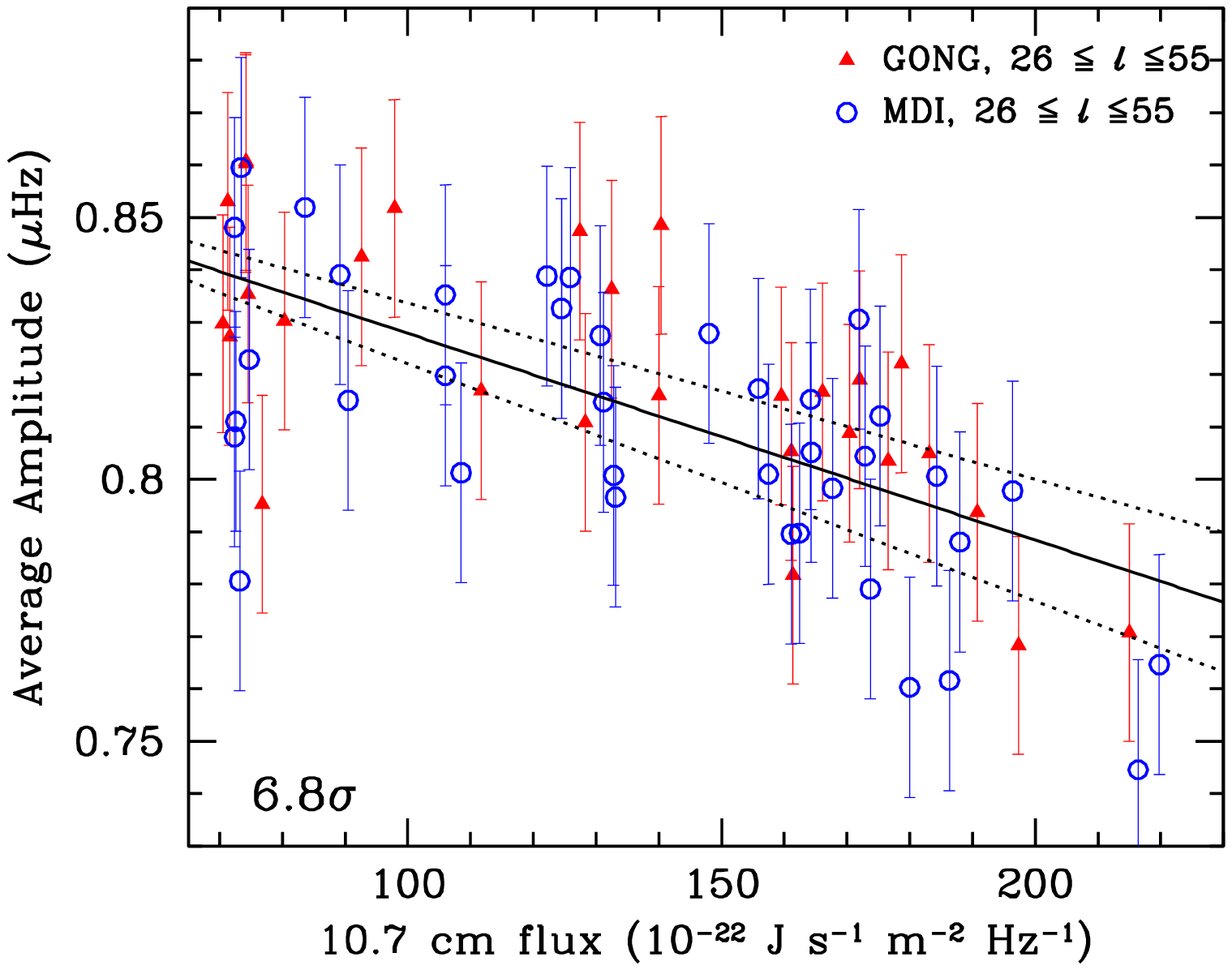}}
\caption{The average amplitude of the oscillatory signal in the 4th differences of
different MDI and GONG data sets plotted as a function of the 10.7-cm flux.
The average  was calculated in the frequency range 2\,--\,3.5~mHz. The results
were obtained for the degree range  $26\le\ell\le 55$.
The black continuous line is a least-squares
fit to all the data points. The dotted lines show the $1\sigma$ uncertainty because of
errors in the fitted parameters. The data used to plot the figure are from
\citet{mandel2004}.
}
\label{fig:4th}
\end{figure}}

\citet{goldreich1991} and \citet{gough2002} had suggested that there could be changes around the
second helium ionisation zone. This does seem to be the case. \citet{mandel2004} used 
the fourth-differences of the frequencies obtained by GONG and MDI,
i.e.,
\be
\delta^4\nu_{n,\ell}=\nu_{n+2,\ell}-4\nu_{n+1,\ell}+6\nu_{n,\ell}-4\nu_{n-1,\ell}+\nu_{n-2,\ell}.
\label{eq:fourthdif}
\ee
to enhance the signature of the acoustic glitch caused by the He\,{\sc ii} ionisation zone and showed that
the amplitude of the signal was a function of the 10.7-cm flux, and that the amplitude
decreased with increase in activity (see Figure~\ref{fig:4th}).
The changes in the signal from the He\,{\sc ii} ionisation zone were also
seen by \citet{gav2006} in BiSON data and by \citet{ballot2006} in GOLF data. Like \citet{mandel2004},
\citet{gav2006} also found that the amplitude of the signal decreased as activity increased.
\citet{mandel2004} interpreted  the change in
the ionisation zone as one caused by a change in temperature.  
\citet{dog2013} however, challenged these results. He analysed the potential significance of the 
results of \citet{mandel2004}, \cite{gav2006} and \cite{ballot2006} 
under the assumption that the temporal variation
of the amplitude of the He\,{\sc ii} signal is caused by a dilution by a broadly distributed magnetic
field of the ionisation-induced influence on the  wave propagation speed. He concluded that 
if the variation of the He\,{\sc ii} signal were a direct result of the presence of a 
temporally varying large-scale magnetic field, then the total solar cycle change
of the spatial average of the magnetic field $\langle B^2 \rangle$ in the vicinity of the
He\,{\sc ii} ionisation zone is about $(30 \pm 10)^2\mathrm{\ Tesla}^2$ which is greater than most estimates.
An examination of cycle~24 data should help in this regard.

Direct inversions of changes in the structure of the solar interior
probed by the spherically symmetric global modes have
yielded inconsistent results about changes inside the Sun. \citet{dziemetal2000},
looking at differences in mode frequencies between 1999.2 and 1996.4
claimed to detect a significant change at a depth of 25\,--\,100~Mm which they
interpreted as either being due to a magnetic perturbations of $(60\mathrm{\ kG})^2$ or
a relative temperature perturbation of about $1.2\times10^{-4}$. However, neither
\citet{basu2002} nor \citet{effdarwich2002} could  find
any measurable differences in the solar interior.
\citet{effdarwich2002} were able to put an upper limit of $3\times10^{-5}$ for
any change in the sound speed at the base of the convection zone.
\citet{chouser2005} and \citet{serchou2005} tried used a different technique -- they plotted the
smoothed scaled frequency differences as a function of the $w$ which is related
to the lower turning point (Eq.~\ref{eq:wrt}), and
showed that there was a discernible time-variation at $\log(w)\approx 2.7$. This
corresponds approximately to
the position of the convection-zone base. 
\citet{chouser2005} estimated the change to be
caused by a magnetic field variation of $(1.7\mbox{\,--\,}1.9)\times 10^5\mathrm{\ G}$ at the convection-zone base
assuming equipartition, or a temperature perturbation of about 44\,--\,132~K.

\citet{baldner2008} tried an altogether different approach. Instead of inverting data sets of different epochs
and then comparing the results, they took 54 MDI data sets covering the
period from 1996 May~1 to 2007 May~16, and 40 GONG data sets spanning 1995 May~7 to 2007 April~14
and did a principal component analysis (PCA) to separate the frequency
differences over solar cycle~23 into a linear combination of different time-dependent components.
They found that the dominant component was indeed predominantly a function of frequency, however, it also
showed a small dependence on the lower turning point of the modes, implying a time dependent change
in the interior. \citet{baldner2008} inverted the appropriately scaled first principal component
to determine the sound-speed difference between solar minimum and maximum. They found
a well-localised change of  $\delta c^2/c^2=(7.23\pm2.08)\times10^{-5}$
at $r=(0.712^{+0.0097}_{-0.0029})\rsun$.
The change near the convection-zone base can be interpreted as a change in the
magnetic field of about 290~kG. They also found
changes in sound speed correlated with surface activity for $r > 0.9\rsun$.
\citet{cristina2012} examined the surface variations further by determining 
the frequencies of modes with $\ell \le 900$ with MDI data centred around  2008 April (solar minimum)
and 2002 May (solar maximum) and using those to resolve the surface structure
better. She found a two-layer configuration -- the sound speed at solar maximum 
was smaller than at solar minimum at round 5.5~Mm, the minimum-to-maximum difference decreased until $\sim7\mathrm{\ Mm}$. 
At larger depths, the sound speed at solar maximum was larger than at solar minimum and the 
difference increases with depth until $\sim10\mathrm{\ Mm}$. She could not detect deeper changes.

One of the interesting features of solar-cycle related frequency changes is the nature of the
  change in the frequencies of f~modes. As mentioned in Section~\ref{subsec:props}, the frequencies of
f~modes are essentially independent of stratification and depend on the local value of
gravitational acceleration $g$. Given that we do not expect the mass of the
Sun to change as a function of the solar cycle, the f-mode changes
 led to investigations of changes in the solar ``seismic radius''.
\citet{dziemetal1998} looked into f-mode frequency changes in the period
between 1996 May~1 and 1997 April~25 and arrived at a value of 4~km.
However, \citet{goodeetal1998} looked at the time variation of the f-mode frequencies between 
1996 May~1 and 1997 April~25, to finding a relative radius difference
of $6\times 10^{-6}$ or 4~km. They did not find any monotonic trend, instead
they found a 1-year cycle in the frequencies, which of course is probably an artefact.
\citet{antiaetal2000S} examined the f-mode frequency changes from 1995 to 1998 to
conclude that if the changes were due to a change in radius, the change
correspond to a change of 4.7~km.

The f mode changes are $\ell$ dependent.  \citet{antiaetal2001E} looked into the $\ell$ dependence of the
f-mode frequency changes over the  period 1995 to 2000 and concluded that
the most likely reason for the change is not a change in radius, but effects
of magnetic fields. \citet{sofiaetal2005} however, argued
that the $\ell$-dependent perturbations to f-mode frequencies points to a
non-homologous change in the radius.
\citet{sandrine2007} inverted the f-mode differences to determine the change in solar radius
as a function of depth; their results showed a smaller change than was predicted
by the \citet{sofiaetal2005} model but nonetheless showed a clear variation. 
Surprisingly, the issue has not been investigated using HMI data and for that matter
reprocessed MDI data, and perhaps it is time that we do so!

\subsection{What do the even-order splittings tell us?}
\label{subsec:even_change}

The even-order frequency splittings and splitting coefficients change with time;
even early helioseismic data showed so. \citet{kuhn1988} had first suggested that
the even-order splittings showed a time variation; \citet{libbrecht1989},
\citet{libbrechtandwoodard1990} and \citet{woodardandlibbrecht1993} agreed. 
\citet{kuhn1988E}, \citet{libbrecht1989}, \citet{kuhn1989} studied the
temporal variation of the $a_2$ and $a_4$ coefficients and its relation
to changes in the latitudinal dependence of limb temperature
measurements, and \citet{kuhn1989} predicted that the same relationship should 
extend to higher order coefficients.
The inference from the inversions of the BBSO even-order $a$ coefficient data
carried out by \citet{libbrechtandwoodard1990} and \citet{woodardandlibbrecht1993}
was that most of the variation in the even
coefficients was localised close to the surface  at the active
latitudes, with a near-polar variation anti-correlated to the global
activity level. 
\citet{dogetal1993A} reported latitude-dependent variations in $u$
between 1986 and 1989. Although they did not give a quantitative estimate, their
figures suggest a change of a few parts in $10^{-4}$.

The temporal variations of the even-$a$ coefficients have been put 
on a much firmer basis with data collected by the
GONG, MDI and HMI projects.  In particular, one can show that the even $a$-coefficients
are extremely well correlated with the corresponding Legendre component of the
average magnetic field on the Sun \cite[see e.g.,][]{antiaetal2001}.
\citet{howeetal2002}
used the frequency changes of individual rotationally split
components within an ($\ell,n$) multiplet to carry out one-dimensional
inversions for frequency shifts as a function of latitude; they showed that the 
frequency shifts were closely associated with regions of higher magnetic
flux both as a function of time and latitude. There have been other
attempts to determine whether the
North-South symmetric latitudinal distribution of
sound speed changes as a function of time. \citet{antiaetal2001} did not find any time-variation
at least  below a radius
of $0.98\rsun$ which is where the most reliable results are obtained.
However, \citet{baldner2008} using a principal component analysis found significant
differences in the speed between solar maximum and minimum at latitudes below
$30^\circ$, their inversions were unstable at higher latitudes. They found an
increase in the sound speed at the lower latitudes during solar maximum compared
with solar minimum. 

\epubtkImage{}{%
\begin{figure}[htb]
\centerline{\includegraphics[height=20pc]{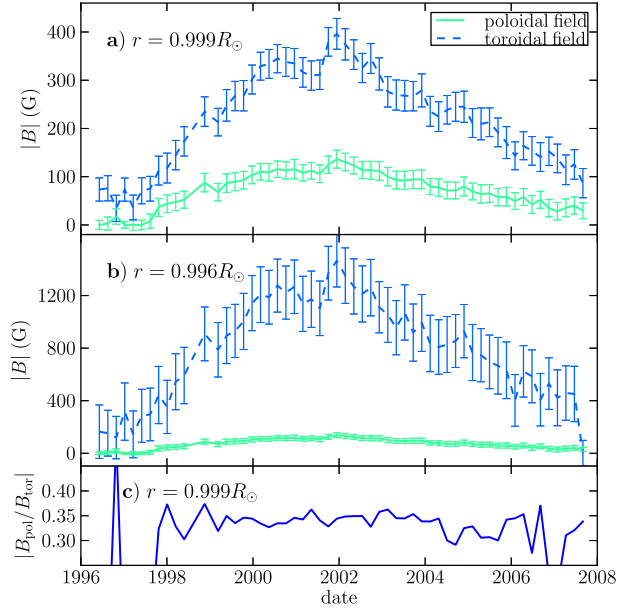}}
\caption{Strength of the inferred magnetic fields plotted as a function of time for
solar cycle~23. 
Panels (a) and (b) show the magnetic field
strengths at $0.999\rsun$ and $0.996\rsun$ respectively. Panel (c) 
shows the ratio of the poloidal to toroidal field at $0.999\rsun$, the
ratio looks similar at $0.996\rsun$. Thus while the strength of both poloidal and
toroidal fields change with time, their ratio does not evolve much. The
results in this figure are from \citet{baldneretal2009}.
(Image courtesy of Charles Baldner.)
}
\label{fig:magfield}
\end{figure}}

\epubtkImage{}{%
\begin{figure}[htb]
\centerline{\includegraphics[height=17pc]{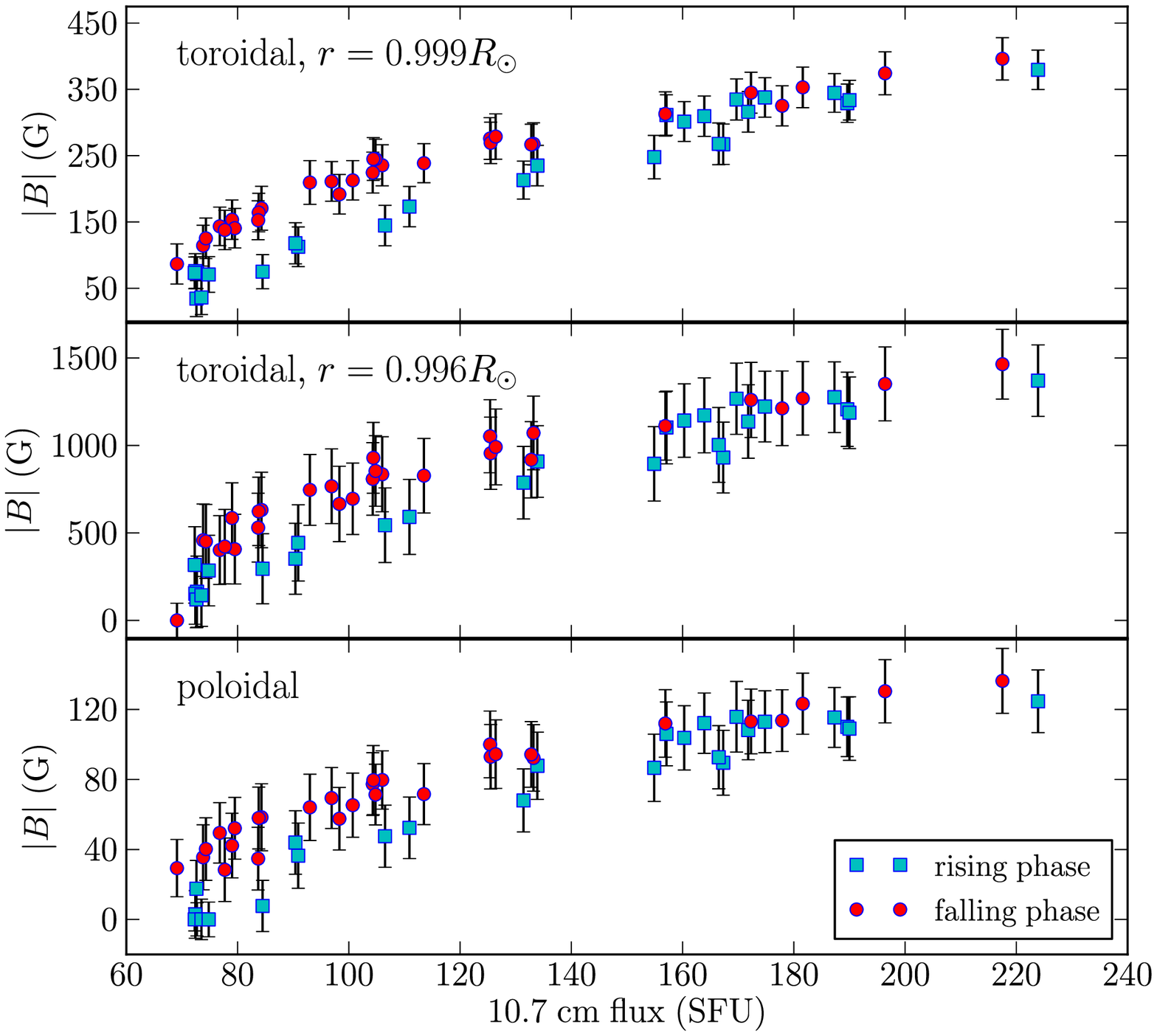}}
\caption{The strength of the solar interior magnetic field during cycle~23 plotted as a
function of the 10.7-cm flux. The top two panels show the toroidal field
at 0.999\rsun and 0.996\rsun. The lowest panel is the poloidal field. The cyan squares
are for the rising phase of the cycle while the magenta circles are for
the falling phase. Results are from \citet{baldneretal2009}.
(Image courtesy of Charles Baldner.)
}
\label{fig:activity}
\end{figure}}

As mentioned earlier in Section~\ref{subsec:evena}, the even-order splitting coefficients can be interpreted
as being either due to structural asphericities of a result of magnetic fields.
\citet{baldneretal2009} analysed the changes in these coefficients in terms of
magnetic fields. They found at that each epoch, they could explain the data
as being due to a combination of a poloidal and a near-surface toroidal field,
with a changing strength (see Figure~\ref{fig:magfield}). They also found that
the strengths of both components tracked the 10.7-cm radio flux, however, there
were hints of saturation at high activity (see Figure~\ref{fig:activity}). They
also found that the relation between the magnetic field strengths during the 
rising phase was different from that of the falling phase.

\subsection{Changes in solar dynamics}
\label{subsec:dyn_change}

The most visible solar-cycle changes revealed by global
helioseismic data are changes in solar dynamics. Readers are again
referred to the
\textit{Living Reviews in Solar Physics} article by \citet{howe2009}
for details, we describe a few highlights.

The most visible change in solar dynamics are changes in the ``zonal flows'' which
are alternating bands of prograde and retrograde motion that can be observed when the
time-average rotation rate is subtracted from the rotation rate at any one
epoch 
\citep[e.g.,][]{agkjs1997, js1999, antiabasu2000, antiabasu2001, 
antiabasu2010, antiabasu2013, howeetal2000, howeetal2009,
howeetal2011, howeetal2013L, howeetal2013b}, i.e.,
\be
\delta\Omega(r,\theta,t)=\Omega(r,\theta,t) - \langle\Omega(r, \theta, t)\rangle,
\label{eq:zon}
\ee
where $\Omega$ is the rotation rate, $r$, the radius, 
$\theta$ the latitude and $t$ time. $\langle\Omega(r, \theta, t)\rangle$ is the
time average of the rotation rate. From the zonal flow rate one can define the zonal flow velocity as
\be
\delta v_\phi=\delta\Omega\; r \cos\theta.
\label{eq:zonv}
\ee

The zonal flows are thought to be the same as the ``torsional oscillations'' 
seen at the solar surface \citep{howard1980, ulrichetal1988, snodgrass1992}.
At low latitudes, just like the torsional oscillations, the zonal-flow bands appear to migrate towards
the equator with time. At high latitudes these move polewards with time
\citep{antiabasu2001, ulrich2001}. As more and more data have been
collected, it has become clear that the zonal flows penetrate through
most of the solar convection zone \citep{vorontsovetal2002, basuantia2003,howeetal2006,
antiaetal2008}. At low latitudes, zonal flows seem to rise from deeper in the
convection zone towards the surface as a function of time. We show the changing nature of
solar zonal flows in Figures~\ref{fig:zon} and \ref{fig:zonlat}.

\epubtkImage{}{%
\begin{figure}[htb]
\centerline{\includegraphics[height=19pc]{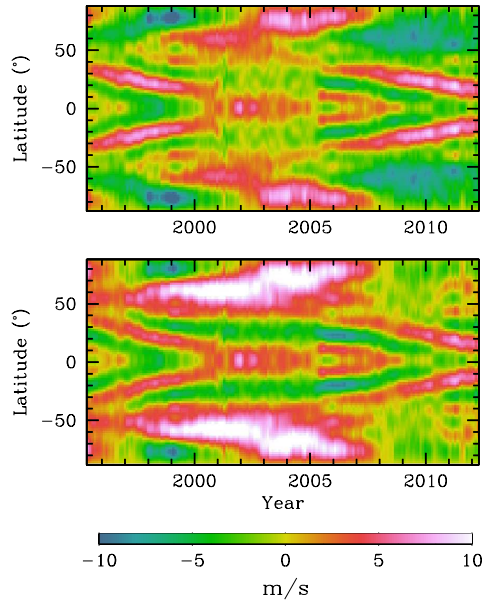}}
\caption{
The zonal flow velocity at $0.98\rsun$ as determined using GONG data plotted
as a function of latitude and time. The top panel was obtained by subtracting the average of the 
cycle~23 rotation velocities,
while the lower panel is that obtained by subtracting the cycle~24 average rotation between
2009 and 2012.
(Figure from \citealp{antiabasu2013}.)
}
\label{fig:zon}
\end{figure}}

\epubtkImage{}{%
\begin{figure}[htb]
\centerline{\includegraphics[height=17pc]{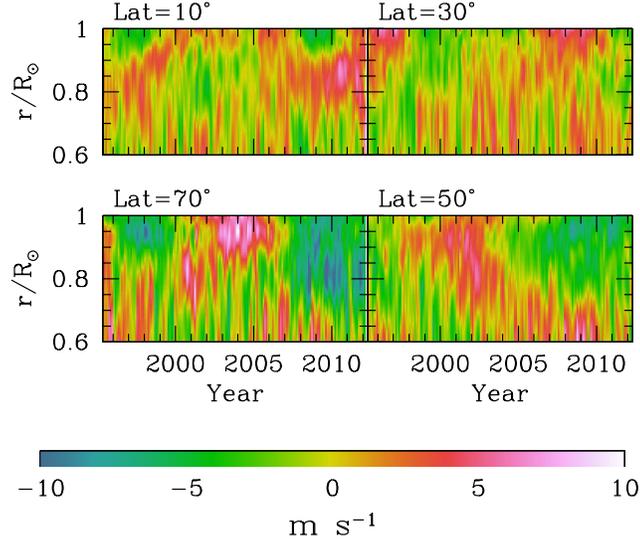}}
\caption{The zonal flow velocity plotted as a function of radius and time for a few latitudes. Results were
obtained using data obtained by the GONG project
(Figure from \citealp{antiabasu2013}.)
}
\label{fig:zonlat}
\end{figure}}

\epubtkImage{}{%
\begin{figure}[htb]
\centerline{\includegraphics[height=17pc]{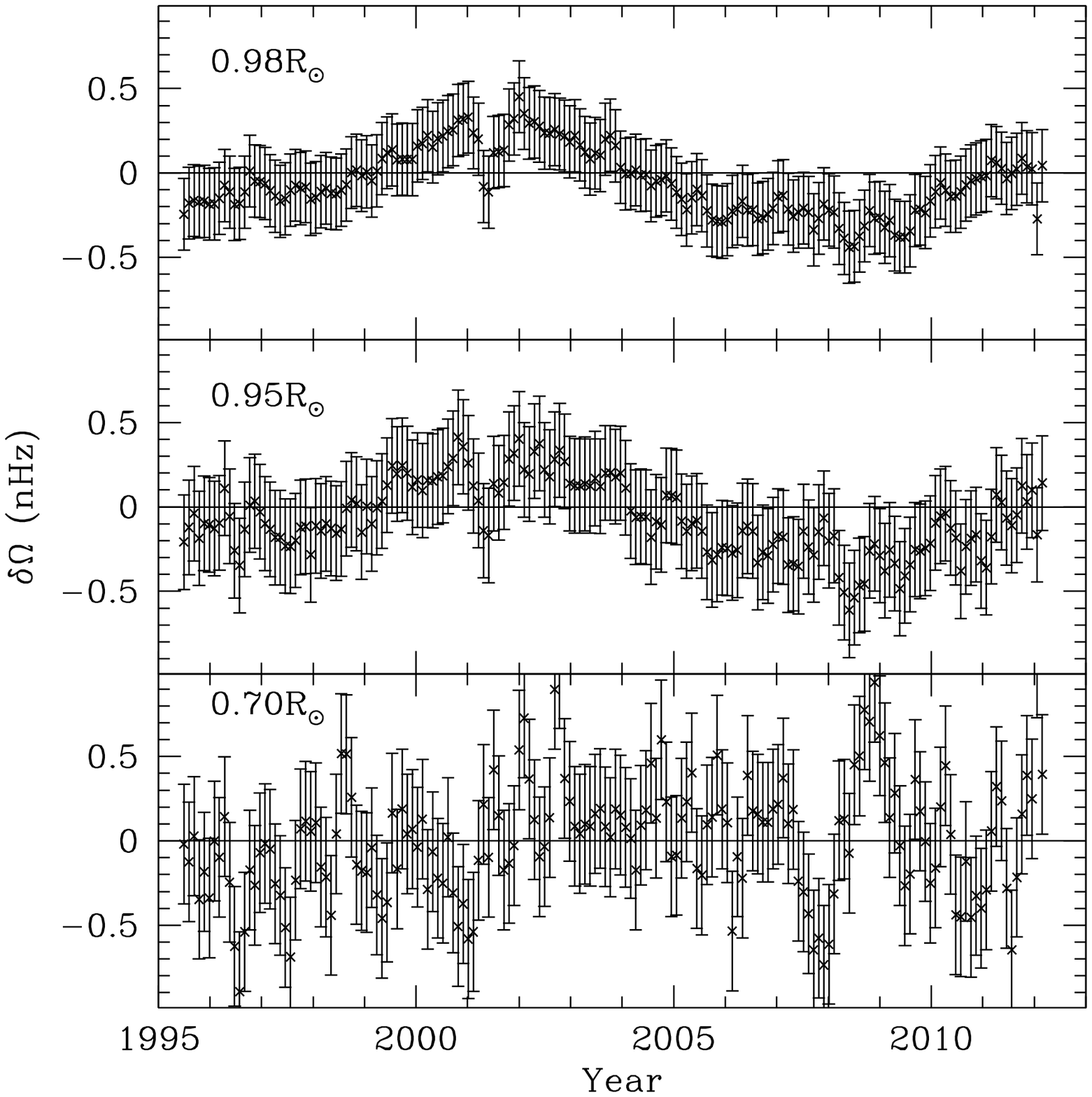}}
\caption{
The change in the latitudinally independent component of the solar rotation rate plotted
as a function of time
(Figure from \citealp{antiabasu2013}.)
}
\label{fig:om1}
\end{figure}}

While the most visible changes in solar dynamics are in the zonal flows, and which are highly latitudinally dependent,
there are also changes in the component of the rotation that is independent of latitude. We can see this
by inverting just the $a_1$ coefficient. In Figure~\ref{fig:om1} we show the changes in the
latitude-independent part of the solar rotation in the outer layers of the Sun as determined using
MDI data. The results were obtained by subtracting the time-averaged $a_1$ inversion
result from the $a_1$ result at any given epoch.
 Note that there
appears to be a solar-cycle related change, with the fastest rotation at  the cycle~23 solar maximum and
the slowest at the two minima. Rotation during cycle~24 also appears to be somewhat slower than 
that during cycle~23, at least in the outer layers. The results for deeper layers are too noisy
to draw any conclusion. There is significant uncertainty as to whether the dynamics of the solar
radiative zone shows any time dependence \citep{2013eff}.

While the dynamics within the convection zone clearly changes with time, the tachocline
does not show much change. \citet{hmasbtach2011} did a detailed
analysis of the then available GONG and MDI data and found that of the various
properties of the tachocline, the jump in the rotation rate across the tachocline
is the only parameter that shows a significant change with solar activity, but
even that change is small. 

\subsection{The ``peculiar'' solar cycle~24}
\label{subsec:peculiar}

Solar cycle~24 is showing all signs of being a very weak cycle. The minimum that preceded it was extremely long and
very quiet and unusual in its depth -- it had more sunspot free days and any recorded
in the space age, and the 10.7-cm flux was the lowest ever recorded. The polar fields
observed by the Wilcox solar Observatory  and
other instruments were also very weak \citep[e.g.,][]{lo2010,nat2012}. The structure of the solar
corona was  very different during the minimum -- instead of having a simple, equatorial
configuration as is normally seen during solar minima, the corona was bright at higher
latitudes too. Another
noteworthy feature is that the fast solar winds were confined to
higher latitudes \citep{manoharan2012}.
In fact the minimum was unusual enough to have multiple workshops dedicated to it and
many of the measured peculiarities are discussed in the workshop proceedings edited by \citet{aspcs2010}.

Of course one could ask if the minimum was really peculiar, and it is possible that
the lack of extensive data for earlier solar cycles
is fooling us. After all, the minimum between cycles 23 and 24 was the not the only
one with many sunspot free days, there were were sunspot free minima in 
 the early part of the 20th century (the minima of 1900 and 1910). Going even further
back in time, we can find a lot of other minima that were as deep at the cycle~24 minimum, at least as far as
sunspot numbers are concerned \citep[see, e.g.,][]{sheeley2010,basu2013J}. As mentioned
earlier, a peculiarity
of the minimum was the configuration of the solar corona. However, as \citet{judge2010} have
pointed out, there are eclipse records that show that the solar corona during the solar minimum
circa 1901 was very similar to that of 2009 as seen in eclipse photographs. Thus the seeming
peculiarity of the solar minimum is simply a result of not having enough data; after all
the Sun is 4.57~Gyr old, the sunspot data go back to only about 400~years and solar magnetic
field data to date back to only 40~years.

While whether or not cycle~24 itself is ``peculiar'' is a matter of definition, it is
certainly different from the cycles seen over the last 60 or so years for which we have
had more data than just the sunspot number.  Unlike earlier candidates for possibly
peculiar cycles,  
we have detailed helioseismic data for cycle~24 which allow
us to compare the internal structure and dynamics of cycle~24 with that of cycle~23,
and the indications are that the Sun, particularly solar dynamics, has a different
behaviour in cycle~24 compared to cycle~23.
 As  mentioned earlier, the latitudinally
independent part of solar rotation appears to be slower in cycle~24 compared to the
cycle~23. Details of the solar zonal flows are different too, as can be seen from
Figure~\ref{fig:zon} and Figure~\ref{fig:zonlat} and \citet{antiabasu2013} had found
that while
the zonal flows at all latitudes in cycle~23 can be fitted with three harmonics
of an 11.7 year period, one cannot simultaneously fit the mid-latitudes of cycle~24.

\epubtkImage{}{%
\begin{figure}[htb]
\centerline{\includegraphics[height=17pc]{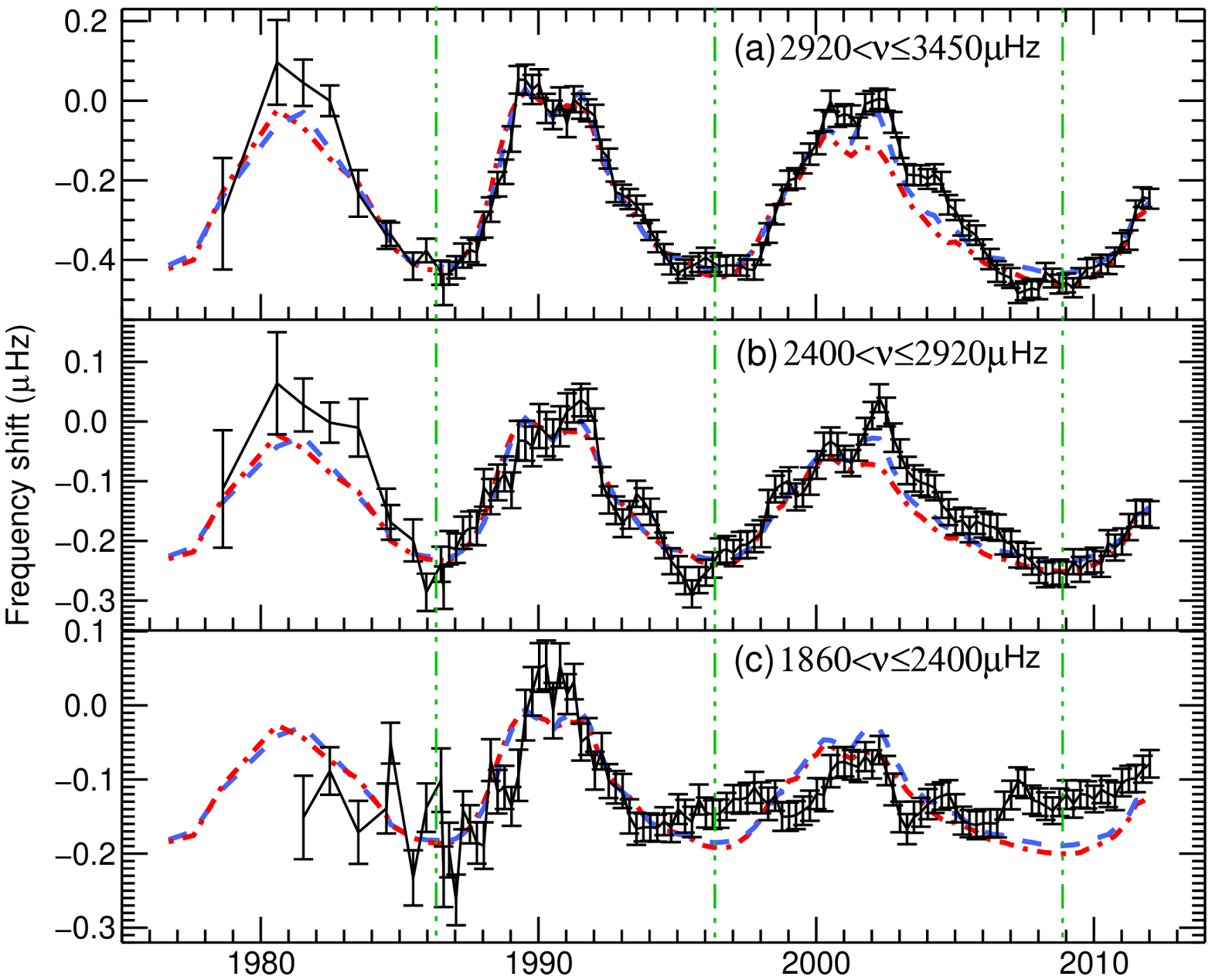}}
\caption{Average frequency shifts of solar oscillations observed by BiSON as a function 
of time for three different frequency ranges. The plotted shifts were calculated 
with respect to the averaged frequencies during the cycle~22 maximum, specifically 
over the period 1988 October to 1992 April. The black points in each panel show the average shift in frequency. 
The blue dashes show the 10.7-cm flux and the red dot-dashed line is the International Sunspot Number.
The green vertical lines mark the approximate positions of solar minima.
(Image courtesy of Anne-Marie Broomhall.)
}
\label{fig:bison}
\end{figure}}
\epubtkImage{}{%
\begin{figure}[htb]
\centerline{\includegraphics[height=17pc]{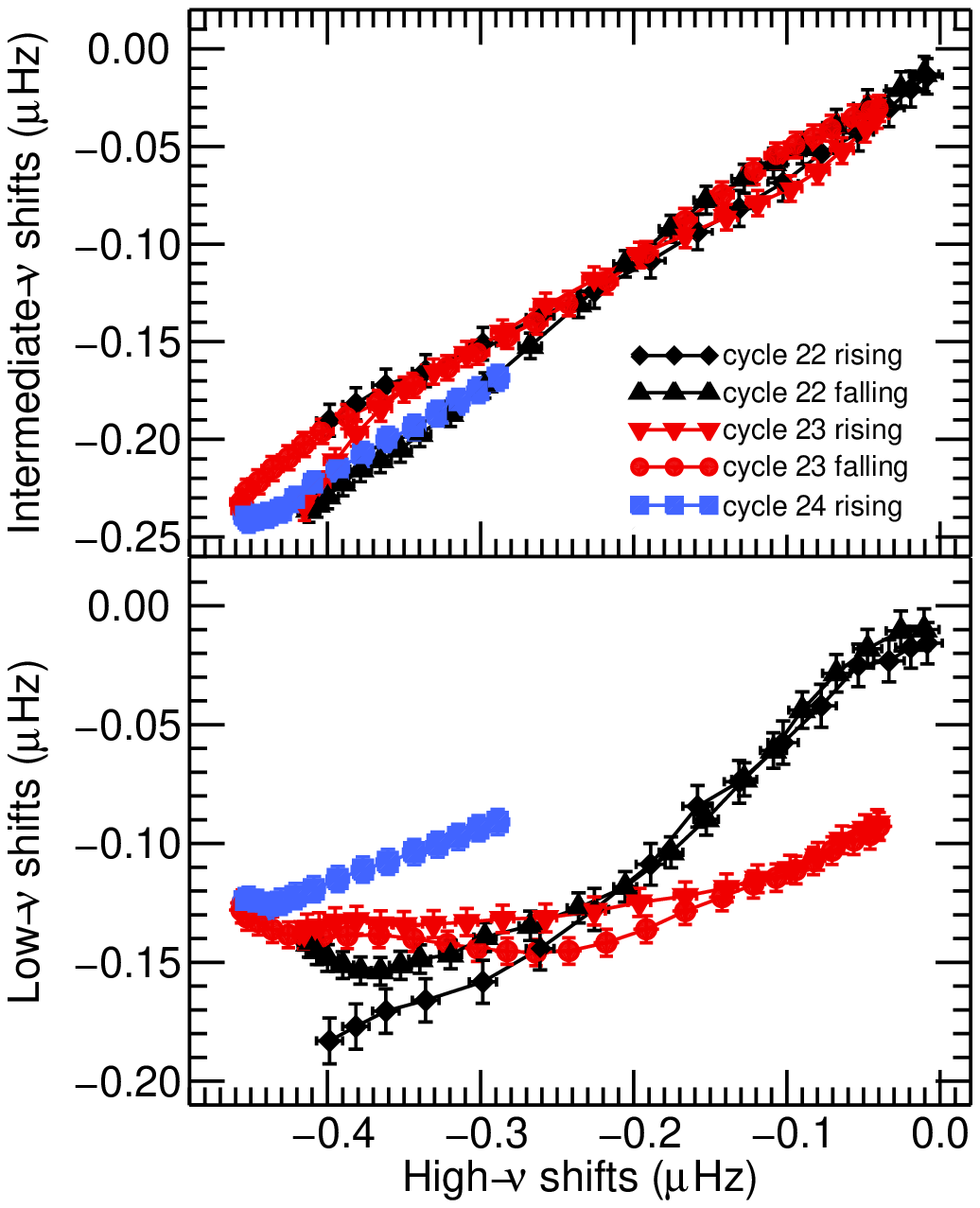}}
\caption{
Frequency shifts in the mid-frequency range (upper panel) and low-frequency range (lower panel) 
plotted as a function of the frequency shifts in the high-frequency range for cycles 
22, 23, and 24. The frequency shifts have been smoothed to remove the signature of the two-year cycle.
The frequency ranges are identical to those in Figure~\ref{fig:bison}.
(Image courtesy of Anne-Marie Broomhall.)
}
\label{fig:bison_shift}
\end{figure}}

There were early signs that the Sun was heading towards an unusual phase, unfortunately,
we missed a chance to predict the peculiarity and only saw the signs after the
onset of cycle~24 \citep{basuetal2012}. 
 BiSON has been collecting data for over thirty years
and its observations cover cycles 22 and 23 in their entirely and all of cycle~24 to date. They
also cover a few epochs of cycle~21. \citet{basuetal2012} found that the frequency-dependence
of the solar-cycle related shifts were different in cycle~23 compared with cycle~21. This can
be seen in Figure~\ref{fig:bison} where we plot the frequency shifts of low-degree
solar modes for three different frequency bands. The shifts are calculated with respect to the 
frequencies at the maximum of cycle~22. Note that the high- and intermediate-frequency bands
show a different pattern of the shifts compared with that of the low-frequency band. The difference
is particularly notable for cycle~23. Also note that the frequency shifts in the high- and intermediate-frequency
bands stop following the sunspot number (in red) and the 10.7-cm flux (in blue dashes) sometime just before
the maximum of cycle~23. This figure thus shows that cycle~23 itself was quite different from cycle~22,
and perhaps the ``peculiarity'' of cycle~24 had its seeds in the descending phase of the previous cycle.

Another way to show the differences between the cycles is to plot the shifts in the different frequency
ranges against one another, as has been done in Figure~\ref{fig:bison_shift}. Note that the shifts of the
intermediate frequency range appears to follow that of the high frequency range for all cycles shown.
The situation is different for the low-frequency modes. Cycle~23 is very different, and although cycle~24
the low-frequency shifts in cycle~24 have the same slope with respect to the cycle~22 shifts, they are
following a different path altogether. It is therefore quite possible that as far as the signatures
from the solar interior are concerned, each cycle is distinct in it own way and that
the assumption that we can have a ``typical'' solar cycle is simply a reflection of the
lack of data.

\newpage

\section{The Question of Mode Excitation}
\label{sec:excite}

All the results discussed in the earlier sections were obtained by
analysing and interpreting data on the frequencies of solar oscillation.
However, the frequencies are not the only observable -- the height of the
individual modes as well as the width of the peaks in the power spectra
can be measured as well. Using the information contained
within these observations requires us to understand how modes are excited
and damped.
The two possibilities for exciting the modes are (1) self excitation and 
(2) stochastic excitation
by turbulent convection of intrinsically damped oscillations.

In the solar case when  self-excitation due to over-stability%
\footnote{`Stable' modes are not
self-excited. When it comes to excitation,  terms 'unstable' and `overstable'
are often used interchangeably, but they have a subtle difference. Unstable
modes are  those whose amplitudes grow exponentially with time. Overstable modes
are a special class of unstable modes and correspond to harmonic oscillations, i.e.,
they are periodic, and the amplitudes of these oscillations increase with time.} 
was considered,  it  was argued that the
limiting amplitude could be determined by non-linear processes such as
mode coupling \citep[e.g.,][]{vandakurov1979, dziembowski1982}. 
\citet{andoandosaki1975} performed one of the first calculations to
determine whether or not solar modes are unstable, and found them to
be so. They found that the driving of oscillations is mainly due to the
$\kappa$ mechanism \citep{cox1967I} of the hydrogen ionisation zone with damping being
provided by radiative loses in the optically thin layers. However, their
results ignored the coupling of convection to pulsations. Similar
results were obtained by 
\citet{goldreichandkeeley1977a} who found that if damping by
turbulent viscosity was neglected, all modes with periods greater than about
6 minutes were unstable, and that the $\kappa$ mechanism  was
responsible for driving the modes. Modes with shorter
periods were stabilised by radiative damping. However, when they
included a model (albeit a crude one) of the interaction between
convection and oscillations, they found that the turbulent dissipation
of pulsational energy made all the modes stable.
There have been many investigations to determine whether or not
solar modes are stable \citep[e.g.,][]{gough1980, jcdfrandsen1983,
hmasmcdna1982, hmasmcdog1988}, the results have been contradictory
and have depended on details of how energy dissipation was treated.
\citet{goldreichandkeeley1977a} had concluded that the lack of a
reliable time-dependent theory of convection the process of
determining the stability of solar modes an uncertain one. 

The consensus today is that solar oscillation modes are stochastically
excited by turbulent convection. \citet{goldreichandkeeley1977b} argued that
the small line-widths and low amplitudes of low-order p modes makes it
hard to imagine non-linear processes that lead to the saturation of unstable
modes at the observed low amplitudes. If the modes are accepted as being
stable, then the source of excitation becomes the question. Guided by general
principles of thermodynamics that indicate that sources of 
excitation are intimately related to the sources of damping, and since
turbulent viscosity appears to be one of the main damping sources of
p modes, the question is how they might act as sources of excitation.
 \citet{goldreichandkeeley1977c} presented one of the first formulations of 
for the stochastic excitation of solar p modes. Although they were unable to
reproduce the observed p mode velocity amplitudes at that time, the idea
behind the work has remained the basis for other calculations.
Under this assumption, the oscillations can be described as a damped oscillator
whose amplitude $A$ satisfies the equation
\be
\frac{\dd^2 A}{\dd t^2}+2\eta \frac{\dd A}{\dd t}+\omega_0^2A=f(t),
\label{eq:excite}
\ee
where $\eta$ is the damping constant, and $f(t)$ is a random forcing function that is 
independent of $A$. The complexity of the model lies in determining the
damping term $\eta$ and the forcing function $f$. These depend on entropy fluctuations and 
fluctuations of the Reynolds stress produced by convection in stellar envelopes.

A tremendous amount of work has been done since the Goldreich--Keeley papers with
regards to mode excitation in the Sun. 
\citet{gough1980} showed that turbulent-excitation models predict the
right order of magnitude for the p-mode amplitudes and also explains
the excitation of many modes simultaneously. 
\citet{kumaretal1988} calculated how the mode energy should be distributed,
\citet{goldreichkumar1988} derived expressions for the emissivity and absorptivity of acoustic 
radiation by low Mach number turbulent fluids, and \citet{goldreichetal1994} investigated
the rates at which energy is supplied to modes as a function of their frequencies and 
degree.  
The stochastic excitation model has also been applied to solar-like
oscillations in other stars
Sun-like stars \citep[][etc.]{jcdfrandsen1983, balmforth1992, houdeketal1999, dupretetal2009,
samadiandgoupil2001, belkacemandsamadi2013}

Observations do guide us somewhat. For instance, even within the
rubric of stochastic mode excitation there was a question as to
whether the convective flux perturbations that are induced by
oscillations play a role.
\citet{jcddoglibb1989} used the then available
data on line-widths and showed that calculations that
took some account of the perturbation of convective
fluxes fared much better when compared with observations than calculations that
did not. The data \citet{jcddoglibb1989} used for their tests were 
crude by today's standards, but they were able to demonstrate the
diagnostic potential of p-mode line widths. \citet{elsworthetal1995} and 
\citet{chaplinetal1995} looked into the distribution energy in the modes,
and found the energy distribution followed a negative exponential form. These
results agree with the prediction of \citet{kumaretal1988}. However, both
\citet{elsworthetal1995} and \citet{chaplinetal1995} found some high-energy 
events that did not follow the trend.
\citet{chaplinetal1997} did an analysis of mode linewidths using BiSON data
and found that the linewidths predicted by theory \citep[in particular][]{goldreichandmurray1994,
balmforth1992} were in line with the BiSON results.

Another question that could  be asked about the excitation mechanism is 
exactly where the excitation sources lie. \citet{goodedogagk1992} analysed
high-$\ell$ data to show that the likely source of excitation are `exploding granules' about
200m below the base of the photosphere. \citet{kumarandlu1991} looked above the
acoustic cut-off frequency and place the sources somewhat lower, at $\log(P)=5.7$, deeper than
what would be suggested by the mixing length approximation. The Goode et al. results
were based on the mixing length approximation that postulates that upflows
and downflows are symmetric, simulations of convection have shown that this is not the case at all, and in
fact that the downflows are fast and cold and are what are observed as intergranular lanes, and that
upflows are slow and hot. Observations now suggest that the sources of excitation are not
exploding granules, instead they are located in the dark intergranular lanes
\citep{rimmeleetal1995} and are related to the collapse of these lanes \citep{goodeetal1999}.

One of the biggest obstacles in calculated mode-excitation related properties
of the Sun and Sun-like stars is the lack of a good description of convection.
This makes using the information in mode amplitudes and mode widths difficult to use.
The conventional mixing length approximation is clearly wrong when compared with simulations,
even variations such as that of \citet{cm1991} do not not describe the dynamics of 
convection. Non-local formalisms, even those that include the anisotropy of flows,
have been proposed \citep[e.g.,][]{dog1977a,dog1977}, but all such
formalisms have at least one, and sometimes more, free parameters and thus lack
true predictive power. As a result, it is becoming increasingly common to use
simulations of stellar convection to perform such calculations. \citet{steinandnordlund2001}
first derived expressions for the excitation of radial modes. Other works,
such as those of \citet{samadietal2003a,samadietal2003b, steinetal2004,samadietal2007}, etc.,  have
followed. \citet{steinetal2004} for instance have shown that they can reproduce the
excitation rates of modes of $\ell=0$\,--\,$3$. They have also applied their method
to a few other stars. Given the difficulty in modelling convection analytically,
we expect that the study of  mode excitation in simulations of convection to become the usual
way will study mode-excitation in solar-like oscillators.

\newpage

\section{Seismology of Other Stars}
\label{sec:stars}

Pulsating stars, such as cepheids and RR Lyr\ae\ stars have been known for a long time. These
stars lie on the instability strip and their pulsations are excited by a ``heat engine''
such as the $\kappa$-mechanism \citep[see e.g.,][]{cox1967I}. We shall not discuss these
stars here. This section is devoted to solar-like oscillations in stars, i.e., oscillations
in stars that, like in the case of the Sun, are excited 
stochastically by a star's convective envelope. For recent reviews of this
subject, readers are referred to \citet{2013ARA&A}, \citet{2016jcd}, etc.

There are three major differences between the study of the Sun and other stars using solar-like
oscillations:
The first difference is related to how stars are observed. Resolved-disc observations of stars other than the 
Sun are not possible. This  means the data sets for other stars are restricted to modes of low degree only. 
We had seen earlier the change in any scalar quantity can be written in terms of spherical harmonics
and a oscillatory function, as in Eqs.~(\ref{eq:xir}) and Eq.~(\ref{eq:pdash}). We can express the intensity 
of a mode at the observing surface in the same manner. Since we are looking at a physical quantity,
we use only the real part, thus,
\be
I(\theta,\phi,t)\propto\Re\left\{Y^m_\ell(\theta,\phi)\exp[-i(\omega t+\delta_0)]\right\},
\label{eq:int}
\ee
where, $I$ is the intensity and $\delta_0$ is a phase. In other words
\be
I(\theta,\phi,t)\propto P^m_\ell(\cos\theta)\cos(m\phi-\omega t + \delta_0).
\label{eq:reali}
\ee
We need to integrate $I(\theta, \phi, t)$ over the visible disc in order
to obtain the result of whole-disc observations. Neglecting the effects of limb darkening,
the integrated intensity is
\be
I(t)=\frac{1}{A}\int I(\theta,\phi,t)\dd A,
\label{eq:disc}
\ee
where $A$ denotes the area of the visible surface. If we have a spherically
symmetric system, we are free to choose our coordinate system. For ease of calculation
we assume that the polar axis is pointed towards the observer which makes
the integral 0 for all $m$ except $m=0$, and for $m=0$
\be
I(t)=\mathcal{S}_\ell I_0 \cos(\omega t-\delta_0),
\label{eq:it}
\ee
where, $S_\ell$ is the spatial response function given by 
\be
\mathcal{S}_\ell=\frac{1}{\pi}\int_0^{2\pi}\\d \phi
\int_0^{\pi/2}\sqrt{2\ell+1}P_\ell(\cos\theta)\cos\theta\sin\theta\dd\theta
=2\sqrt{2\ell+1}\int_0^{\pi/2}P_\ell(\cos\theta)\cos\theta\sin\theta\dd\theta.
\label{eq:spat}
\ee
On evaluating the integral for different values of $\ell$ one finds that $\mathcal{S}_\ell$ which  controls the
 amplitude of the signal, becomes small quite rapidly as $\ell$ is increased, and thus only very
low $\ell$ modes can be observed \citep[see][]{jcddog1982}.
The second difference between helioseismic and asteroseismic analyses is that unlike the Sun, we generally
do not have independent estimates of the mass, radius, luminosity and age of the the stars,
and all these quantities have to be determined from the seismic data.
The third difference is that
stars come in
many different masses and radii and are at different states of evolution, and as  a result, the
frequency spectrum of the stars can often be much more complicated than that of the Sun.

As had been mentioned earlier in Section~\ref{subsec:props}, the propagation of modes
of different frequencies in  star is controlled
by the Lamb and Brunt--V{\"a}is{\"a}l{\"a} frequencies. For a given mode, the profiles of these two frequencies
determine what type of a mode it is and where it can propagate. The propagation diagram for stars in different
stages of evolution can be very different from each other because of the how the Brunt--V{\"a}is{\"a}l{\"a}
frequency evolves with time. In Figure~\ref{fig:bv_star} we show the propagation diagram for
a $1\msun$ model at three stages of evolution -- the main sequence, the subgiant, and the red giant.
Note how the
buoyancy frequency gets larger and larger towards the core as the star evolves.

In un-evolved stars like the Sun, p- and g-modes occupy distinct oscillation cavities and 
occur in distinct frequency regimes.
The frequency regimes are not so distinct in evolved stars where the two cavities are in close proximity
(see Figure~\ref{fig:bv_star}) and hence, the cavities can couple.
 This results in modes that have p-mode like character in the outer layers and g-mode like
character in the interior with a small evanescent region in between.  These are the so-called `mixed modes'.
The widely observed pure p modes have their highest amplitudes close to the surface,
and hence carry less information about the core. Mixed modes, on the other hand, have high amplitudes in the
core (just like g modes), but also have high amplitudes at the surface (making them observable).
This dual nature of mixed modes make them ideal for studying stellar cores. Note that
since there are cannot be $\ell=0$ g modes, there can be no $\ell=0$ mixed modes.
Mixed modes are seen in subgiants. Most subgiant mixed modes are $\ell=1$ modes, though some $\ell=2$
mixed modes are also seen
The pulsation spectrum of red-giants on the other hand are extremely complicated.
Almost all non-radial modes have g-mode like character in the core
and p-mode like character in the envelope, however, only a few of all the possible modes are excited. The
excited modes are those with the lowest mode inertia \citep{dupretetal2009}.

The frequency of an observed mode is not sufficient to tell us whether
or not it is a mixed mode. Its frequency needs to be examined in the
context of the frequencies of its neighbouring modes, and in particular, the
frequency separation between neighbouring modes. For this
we make use of the fact that p modes are almost equally spaced in frequency, and that
 $\ell=0$ and $2$ modes have similar frequencies [Eq.~(\ref{eq:tassoul})],
as do $\ell=1$ and $3$ modes.  The usual practice is to divide the observed power spectrum
(or frequencies) of a star into frequency slices of width \dnu\ and stack the slices on top of each other.
The result is usually called an `\'echelle' diagram \citep{grec1983}.
If the \dnu\ used is correct, the $\ell=0$
modes line up, with the $\ell=2$ modes close by, the same happens to
$\ell=1$ and $3$ modes.  If
an avoided crossing is present, it shows up as a deviation in the vertical
structure of the modes of that degree in the \'echelle diagram. We show the \'echelle
diagram of three stars in Figure~\ref{fig:echelle}, note that for the Sun (Panel a)
all modes line up and for the subgiant (panel b) the $\ell=1$ modes show a break,
and the giant (panel c) shows many $\ell=1$ mixed modes.

\subsection{Asteroseismic analyses}

How asteroseismic data are analysed depends on the type of data available.
For  stars with good signal-to-noise characteristics
and long time series (just how long depends on the star in question),  individual
peaks in the oscillation spectra can be resolved. We can determine a set of
frequencies from each star. 
For such  stars, the
usual way of analysis to construct stellar models, calculate their frequencies
and see which model fits the data. This is described further in
Section~\ref{subsubsec:detailed}.

\epubtkImage{}{%
\begin{figure}[htbp]
\centerline{\includegraphics[width=20pc]{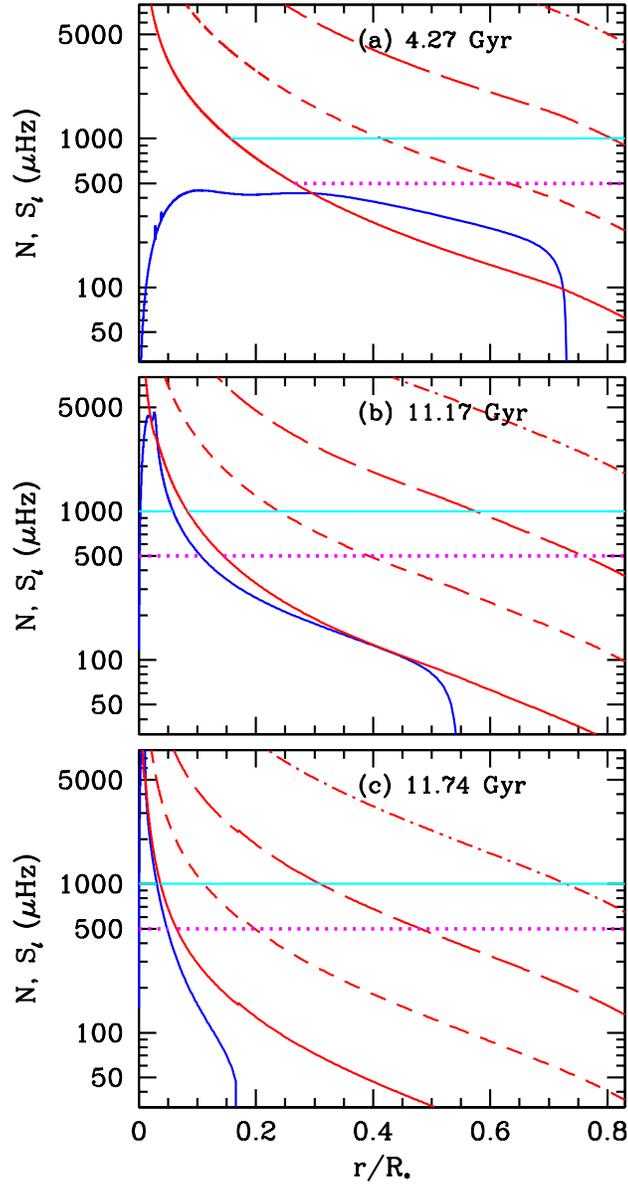}}
\caption{Propagation diagram for a $1\msun$ stellar model in the main-sequence (panel a),
subgiant (panel b) and redgiant (panel c) stages. The blue line is the
Brunt--V{\"a}is{\"a}l{\"a} frequency, while the red lines show the
Lamb frequency for $\ell=1$ (solid line), $\ell=5$ (small dash),
$\ell=20$ (long dash) and $\ell=100$ (dot-dashed line).
The horizontal cyan line shows the propagation cavity for an $\ell=1$, 1~mHz mode and
the horizontal dotted magenta line shows the cavity for an $\ell=1$, 0.5~mHz mode. 
Note that for the star in panel (a), the two modes can only exists as p~modes with a cavity in the
outer part of the star. In panel (b) and (c) each mode has two cavities where they can propagate.
The region in between is where the modes are evanescent. 
Note that
in Panels (b) and (c) the evanescent zones for the modes are small and hence the interior
and exterior cavities can couple giving rise to a mixed modes.
}
\label{fig:bv_star}
\end{figure}}

\epubtkImage{}{%
\begin{figure}[htb]
\centerline{\includegraphics[height=12pc]{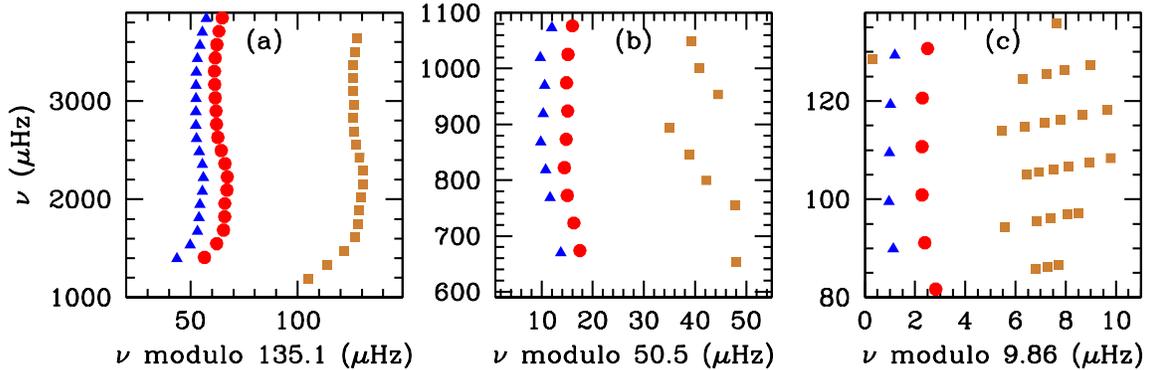}}
\caption{The frequencies of three stars shown in the form of an `\'Echelle' diagram.
\'Echelle diagrams are plotted by dividing the observed power spectrum
(or frequencies) of a star into frequency slices of width \dnu\
and stacking the slices on top of each other. Red circles are $\ell=0$ modes, brown squares are
$\ell=1$ and blue triangles are $\ell=2$ modes. Panel (a) is the Sun, Panel (b) is a subgiant and
Panel (c) is a red giant. Note that for the Sun, a main sequence star, all the modes
of a given degree line up. In the case of the subgiant, the $\ell=1$ modes show a break. This
break is an ``avoided crossing'' and
is a signature of the presence of mixed modes. Note the large number of $\ell=1$ modes in the
red giant. These are all mixed modes and the frequencies of these modes depend on details of
the buoyancy frequency profile. 
}
\label{fig:echelle}
\end{figure}}

\epubtkImage{}{%
\begin{figure}[htb]
\centerline{\includegraphics[width=20pc]{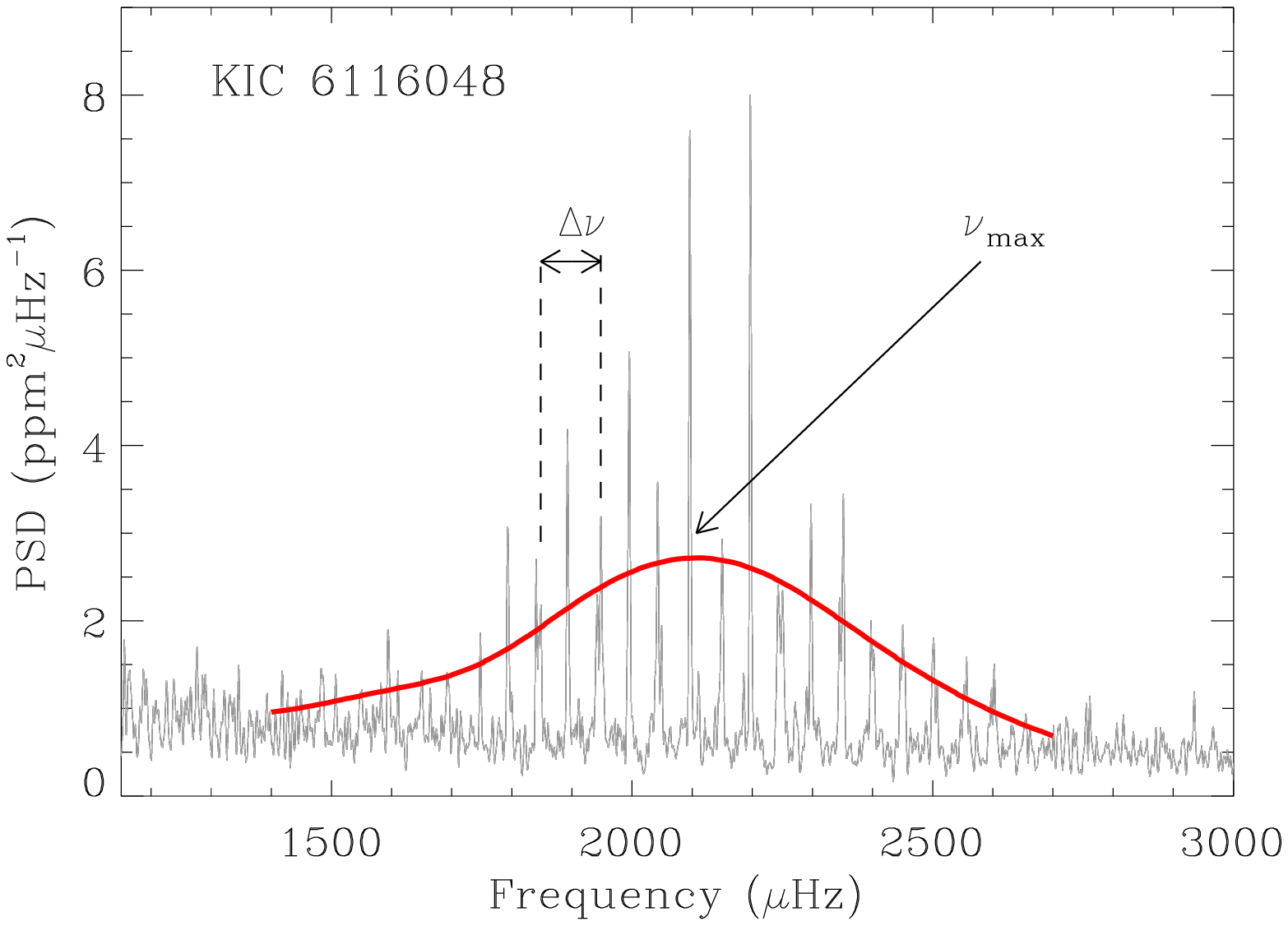}}
\caption{
The oscillation power spectrum of the star KIC~6116048 computed with \textit{Kepler} data obtained
over a period of one month. The red line is the envelope of power obtained by heavily smoothing
the power spectrum. The envelope has been multiplied by a factor of 5 to make it more visible.
(Image courtesy of William J.\ Chaplin.)
}
\label{fig:numax}
\end{figure}}

Estimates of frequencies of different modes are, however, available only
for a handful of stars. For most stars it is easier to determine two
average asteroseismic properties, the large frequency separation \dnu\ (Eq.~\ref{eq:tassoul})
and the frequency of maximum power \numax. As mentioned earlier, \dnu\ scales as
the square-root of density  \citep{ulrich1986,jcd1988}, thus
\be
 \frac{\Delta\nu}{\Delta\nu_{\odot}}\simeq\sqrt{\frac{M/M_{\odot}}{(R/R_{\odot})^{3}}}.
\label{eq:dnudnu}
\ee
Many methods have been developed to determine \dnu\ even when the modes are not resolved
\citep[see e.g.,][]{chaplin2008AN, stello2009, hekker2011}.

The mode amplitude (and power) are modulated by an envelope that is
roughly Gaussian in shape (red curve in Figure~\ref{fig:numax})
in stars (including the Sun) for which the  solar-like
oscillations have been observed. The maximum of this envelope is \numax.
The frequency at which power is the
maximum is \numax. \numax\ is related to the acoustic cut-off frequency
of a star \citep[e.g.,][]{hans1995,bedding2003} and scales
as
\begin{equation}
 {{\nu_{\max}}\over{\nu_{\max,\odot}}}\simeq{{M/M_{\odot}}\over
 {(R/R_{\odot})^2\sqrt{(T_{\rm eff}/T_{{\rm eff},\odot})}}}.
 \label{eq:numax}
\end{equation}
Again, \numax\ can be measure in low S/N spectra.
 \numax\ is an intriguing observable, and unlike the case
of \dnu, we do not as yet have a theory explaining the quantity. There are
some studies in this regard \citep[e.g., see][]{belkacemetal2011}, but the issue
has not been resolved. \numax\ carries diagnostic information on the excitation and damping of stellar modes, and
hence must depend on the physical conditions in the near-surface layers where the modes are
excited. Close to the
surface, the behaviour of the waves is influenced by the acoustic cut-off frequency
(Eq.~\ref{eq:cutoff}). \citet{brownetal1991} argued that $\nu_{\max}\propto\nu_{\rm ac}$ 
($\nu_{\rm ac}\equiv \omega_{\rm ac}/2\pi$)
because both
frequencies are determined by conditions in the near surface layers. \citet{hans1995} turned
this into a relation linking \numax\ to near-surface properties by noting that
Eq.~(\ref{eq:cutoff}) can be approximated as $\nu_{\rm ac}\propto gT^{1/2}_{\rm eff}$ and the
relation in Eq.~(\ref{eq:numax}) follows from this.

There have been numerous tests of the \numax\ scaling relation.
\citet{brunttetal2010} and \citet{bedding2014} compared asteroseismic and independently
determined properties (e.g., from binaries) of a selection of bright solar-type stars and red
giants. All the stars showed solar-like oscillations from ground-based or
CoRoT observations.  The estimated properties were found to
agree at the level of precision of the uncertainties, i.e., to 10\% or better.
The comparisons were extended to stars observed by \textit{Kepler}.
\citet{silvaaguirreetal2012} used asteroseismic
data on 22 of the brightest \textit{Kepler} targets with detected solar-like oscillations and found
agreement between stellar radii inferred from the scaling relations and those inferred from using
Hipparcos parallaxes at the level of a few percent. 
\citet{huberetal2012} combined interferometric observations of some of the brightest \textit{Kepler} and CoRoT
targets with Hipparcos parallaxes and
also found excellent agreement, at the 5\% level, with the  stellar radii inferred using the scaling
relation.
There have been other, more indirect, tests of the relation. For instance, \citet{silvaaguirreetal2015}
determined the properties of 33 \textit{Kepler} objects of interest (i.e., \textit{Kepler} targets suspected to have
an exoplanetary system) obtained from using
using combinations of individual frequencies
of those stars, as well as from using the \dnu\ and \numax\ only (but through grid-based
analyses with \dnu\ calculated from frequencies) and the agreement is very good, except in the case of
three outliers. For the rest, the median fractional difference and standard deviation
between the two estimates are $0.5\pm 1.2\%$ for density, $0.7\pm 0.8\%$ for radius and
$1.8\pm 2.1\%$ for mass. More recently \citet{coelhoetal2015} tested the \numax\ scaling relation
for dwarfs and subgiants and ruled out departures from the
$gT_{\rm eff}^{-1/2}$ scaling at the level of $\simeq 1.5\%$ over the full 1560~K range
in \teff\ that was tested. There is some uncertainty over the absolute calibration of the scaling, but again
of the same order.

\subsubsection{Grid based modelling}
\label{subsubsec:grid}

Grid based modelling is more correctly a grid based search to determine stellar
parameters. When \dnu, \numax\ and \teff\ are known, Eqs.~(\ref{eq:dnudnu}) and (\ref{eq:numax})
represents two equations in two unknowns ($M$ and $R$),
and  can be solved to obtain  $M$, $R$ (and hence surface gravity $g$ and mean density $\rho$)
in a completely model-independent manner.
However, the uncertainties in the results are often large. This is because,
although the theory of stellar structure and evolution tell us otherwise, the equations assume
that all values of $T_{\rm eff}$ are possible for a star of a given $M$ and $R$.
This freedom can often result in unphysical results.  The way out is to use an approach
that uses the  relation between $M$, $R$ and \teff\ implicitly using grids of stellar models.
For given values of \dnu, \numax\ and \teff, $M$ and $R$ (as well as any parameter, such as
age) is obtained by searching within a grid of models. Knowing a star's metallicity
helps in this.

Many grid based pipelines have been developed and used e.g., the Yale--Birmingham (YB)
pipeline \citep{basuetal2010A,gai2011,basuetal2012A}, RadEx10 of
\citet{creevey2013}, SEEK of \citet{quirion2010}, RADIUS of \citet{stello2009}, etc.
Most of these pipelines work by finding the maximum likelihood of the set
of input parameter data calculated with respect to the grid of
models. 
For example, the YB pipeline works in the following manner:
 For a given observational
(central) input parameter set, the first key step in the method is to
generate 10\,000 input parameter sets by adding different random
realisations of Gaussian noise to the actual (central) observational
input parameter set. For each realisation, we find all models in a
grid within $3\sigma$ of a minimum uncertainty  and use them to define the likelihood 
function as
\be
\mathcal{L}=\prod^{n}_{i=1}\left[\frac{1}{\sqrt{2\pi}\sigma_{i}}\times
 \exp(-\chi^{2}/2)\right], 
\label{eq:likelihood}
\ee
where
\be
 \chi^{2}=\sum^{n}_{i=1}\left(\frac{q^{\mathrm{obs}}_{i}-q^{\mathrm{model}}_{i}}{\sigma^{i}}\right)^{2},
 \label{eq:chi2}
\ee
where $q$ are the observations \dnu, \numax, \teff\ and [Fe/H].
The estimated property, such as radius or mass, is then the likelihood-weighted average
of the properties. The 10\,001 values of any given parameter, such as radius, 
estimated from the central value and the 10\,000 realisations
forms the probability distribution function that parameter. 
In the YB pipeline we adopt the median of the distribution as the estimated value
of the parameter, and we  use
$1\sigma$ limits from the median as a measure of the uncertainties.
At the price of 
making the results slightly model dependent, the method gives us results that
are physical.
Characteristics and systematic biases involved with grid-based analyses have been
investigated in detail by \citet{gai2011}, \citet{basuetal2010A}, \citet{bazot2012},
and \citet{gruberbauer2012}. The systematics related to determining \logg\ have
been investigated by \citet{creevey2013M}.

Grid-based modelling has been used to determine the global properties (mass, radius, density)
of all the short-cadence targets observed by \textit{Kepler} \citep{chaplinetal2014}.
This method has also being used to determine the global properties of many of the
red giants observed by \textit{Kepler} \citep{pinsonneault2014}.

\subsubsection{Detailed  modelling}
\label{subsubsec:detailed}

There are some stars for which frequencies of individual modes are available. In such 
cases one resorts to modelling the star, calculating the frequencies of models
and comparing the frequencies of the models with that of star. The star is
modelled to have the observed \teff\ and metallicity. This is, in a sense,
reminiscent of what used to be done in the early days of helioseismology. The biggest
difference is that unlike the Sun, we do not know the mass, radius, age and luminosity of
these stars independently. 

\epubtkImage{}{%
\begin{figure}[htb]
\centerline{
  \includegraphics[height=12pc]{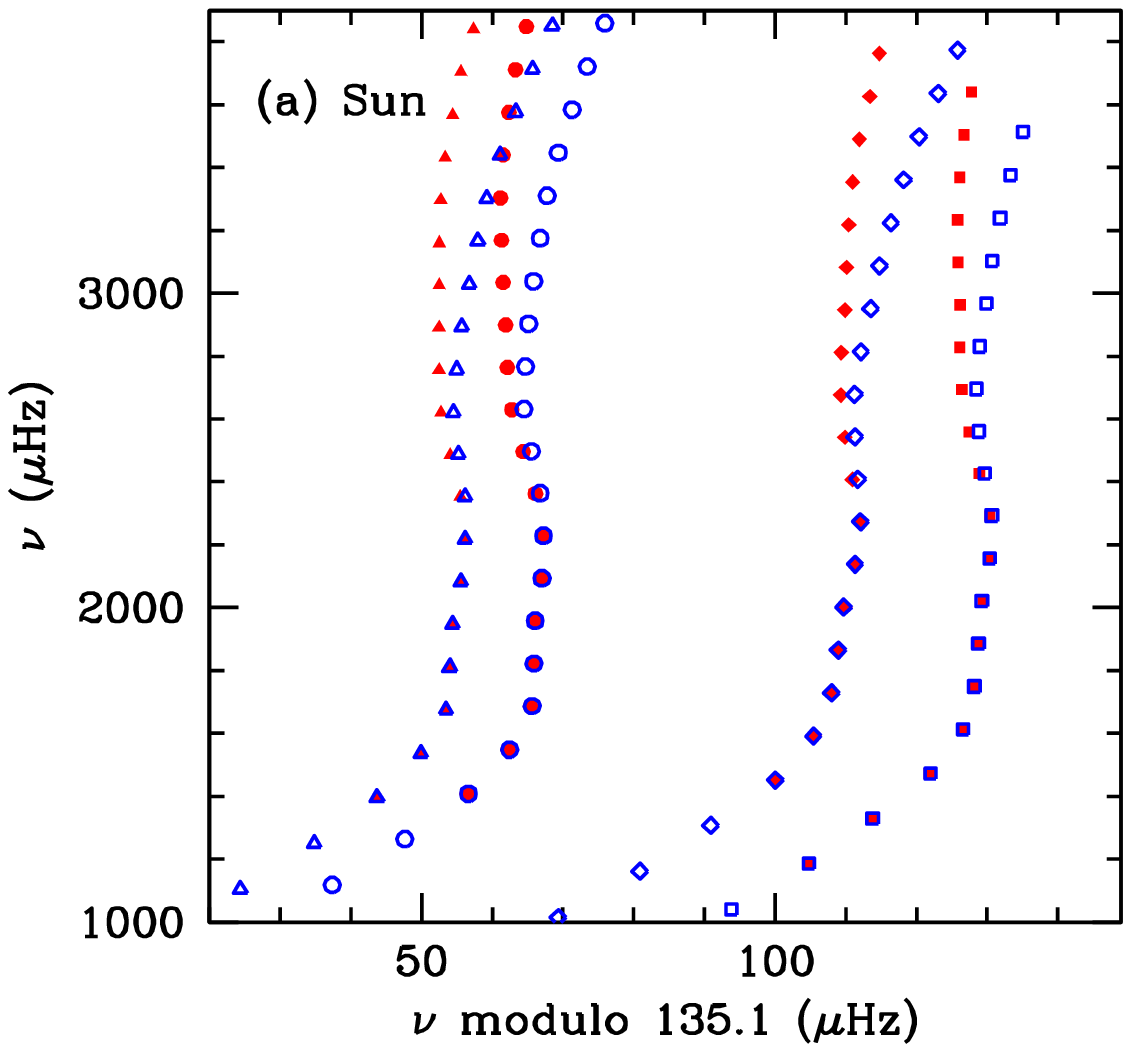}
  \includegraphics[height=12pc]{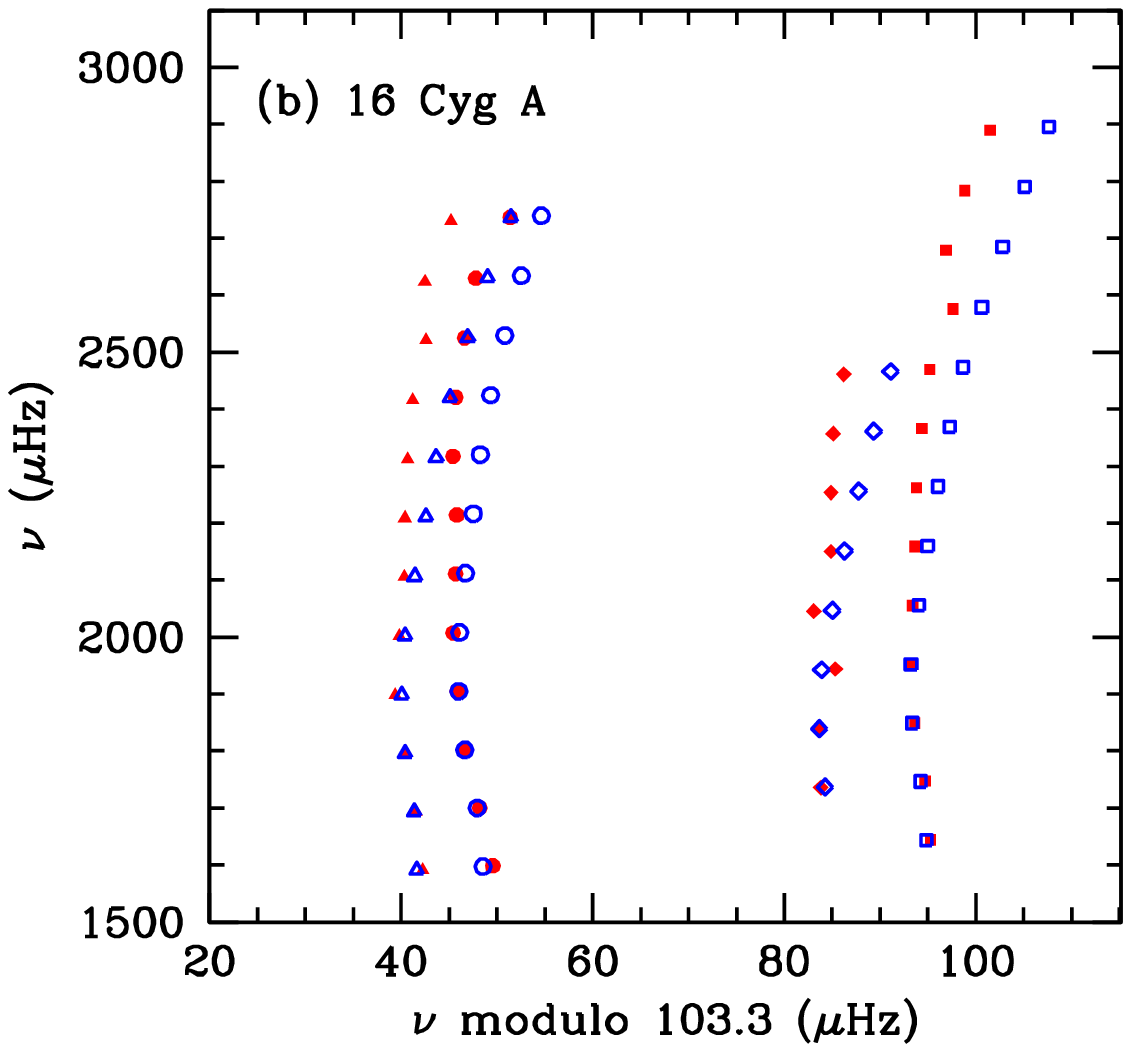}
}
\caption{\emph{Panel (a):}
The \'echelle diagram for low-degree solar modes (red) and Model~S (blue). \emph{Panel (b):} The \'echelle
diagram for 16~Cyg~A (red) and a model of that star (blue). Note that in both panels, the \'echelle
diagram of the model deviates from that of the star at high frequencies. This is nothing but the
surface term.}
\label{fig:16cyg}
\end{figure}}

One of the biggest issues in such an analysis is the surface term. In Figure~\ref{fig:16cyg}
we show the \'echelle diagrams of the Sun and 16~Cyg~A and their models. The familiar surface
term is seen as the frequency offset between the two stars and their models at
higher frequencies. As had been described earlier, the vexing issue of the surface term
led to the development of inversion techniques. In the case of the Sun, if we wished we could 
easily use a subset of the modes to define the surface term and use the rest to do the analysis,
though of course it is more usual to fit the surface term simultaneously. 
We do not have this luxury in the case of asteroseismic analysis. Thus ad hoc means are
usually used to correct for the surface term, usually by scaling the solar surface term.

The most commonly use correction for the surface term is that of
\citet{kjedlsen2008}. This correction is in the form of
a power law derived using solar models and their frequency
differences with respect to the solar frequencies. The scaling for the correction
is written in the form
\be
\nu_{\rm obs}(n)-\nu_{\rm ref}(n)=a\left(\frac{\nu_{\rm obs}}{\nu_0}\right)^b,
\label{eq:hans_sur}
\ee
where, $\nu_{\rm obs}$ are the observed frequencies, $\nu_{\rm ref}$ are the
frequencies of the model. To apply the correction one needs to
specify the exponent $b$ and the reference frequency $\nu_0$. The exponent is determined
using the solar surface term. The quantity $a$, though formally the frequency difference
at $\nu_{\rm obs}=\nu_0$ is often determined in an average manner. 
Variants of Eq.~(\ref{eq:hans_sur}) are also used, e.g., \citet{metcalfe2012} fix
$\nu_0$ to be \numax, reducing one degree of freedom. 
Other  ways that have been used widely depend on explicitly
scaling the  solar surface term; \citet{silvaaguirreetal2015} describe some  of these
methods.  
While the corrections are
widely used, it is unlikely that
the solar scaling can be used for all stars that show solar-like oscillations.

More recently, \citet{ballandgizon2014}, based on the work of \citet{dog1990},
suggested two surface term corrections: 
\be
\frac{c}{E}\left(\frac{\nu}{\nu_{\rm ac}}\right)^3,
\label{eq:ball1}
\ee
and
\be
\frac{1}{E}\left[c_3\left(\frac{\nu}{\nu_{\rm ac}}\right)+
c_{-1}\left(\frac{\nu}{\nu_{\rm ac}}\right)^{-1}\right],
\label{eq:ball2}
\ee
where, $E$ is the more inertia, and $c$, $c_3$ and $c_{-1}$ are constants
obtained by fitting the frequency differences. These forms are promising
and tests show that Eq.~(\ref{eq:ball2}) works very well over a large portion of the HR diagram
\citep{joey} and modellers are beginning to adopt this form.

Despite the ambiguity caused by the surface term, detailed modelling gives more precise
estimates of the properties of a star. Of course grid modelling is also affected by the
surface term, but to a lesser degree since the term gets reduced in the process of
calculating the average value of \dnu. Such detailed modelling has been reported
by \citet{metcalfe2010,metcalfe2012,metcalfe2014}, \citet{deheuvels2011}, \citet{savita2012}, 
\citet{victor2013}, \citet{silvaaguirreetal2015}, etc.
Because of the uncertainty caused by the surface term, investigators are looking into
using combinations of frequencies of different modes that should be relatively
insensitive to the surface term \citep[e.g.,][]{iwrsvv2003,iwrsvv2013,oti2005,iwr2014}.

\subsubsection{Exploiting acoustic glitches}
\label{subsubsec:other}

Asteroseismic analyses are not restricted to trying to fit the oscillation frequencies.
As in the case of the Sun, investigators are using signatures of acoustic glitches to
determine the positions of the helium ionisation zone and the convection-zone base.

\citet{mazumdar2012,mazumdar2014} determined the acoustic depths of the He\,{\sc ii} ionisation
zone of several stars. They were also successful in measuring the acoustic depths of the
convection-zone bases of some of the stars. More work needs to be done to translate the acoustic
depth to a physical radius that could help modellers.

\citet{basu2004M} had proposed that the amplitude of the He\,{\sc ii} acoustic glitch could be
used to determine the helium abundance of stars. \citet{verma2014} applied this idea to 16~Cyg~A and
16~Cyg~B using frequencies obtained from \textit{Kepler} observations. These two stars are solar
analogues -- only slightly higher masses than the Sun and they are slightly older.
The estimate of their current helium abundance is entirely consistent with the solar
helium abundance with the helium abundance of 16~Cyg~A in the range 0.231 to 0.251 and that
of 16~Cyg~B in the range 0.218 and 0.266.

\subsubsection{Analyses of internal rotation}
\label{subsubsec:stel_rot}

It is difficult to determine the rotation profile of a star using asteroseismic data. The
problems lie with the data themselves -- $\ell=0$ modes are not affected by rotation at all,
while $\ell=1$ modes split into at most three components and $\ell=2$ modes into at most five. Whether all
the splittings are visible also depend on the angle of inclination between the star and us
\citep[see e.g.,][]{gizonsolanki2003}. 
Thus data from which rotation can be determined are very limited.
The visibility of the components as a function of inclination has been used
successfully to study the spin-orbit misalignment in exoplanetary systems
\citep{huberetal2013,chaplinetal2013}. 

One of the first detailed investigations into the internal rotation profile of
a star other than the Sun was that by \citet{deheuvelsetal2012}. They compared the 
splittings of the mixed modes with those of the p modes of the star KIC~7341231 and
found clear evidence of a radius-dependent rotation rate. They showed that the
core of this star, which is just off the subgiant into the low end of the
red-giant branch, rotates at least five times faster than the surface. The
ratio between the core and surface rotation rates places constraints on models
of angular momentum transfer. \citet{deheuvelsetal2014} looked at a sample
of subgiants and found that the contrast between core and surface rotation
increases as the star evolves; their results suggest that while the cores
spin up, the envelope spins down. Data for at least two stars suggest a sharp
discontinuity in rotation between the core and the envelope located at the
hydrogen burning shell. The implications of these results are still being studied.

\citet{becketal2012} and \citet{mosseretal2012} have concentrated on determining
and interpreting the rotational splittings of red giants. Mode splittings in 
red giants are dominated by the rotation rate of the core. The data suggest that the
core rotates faster as it ascends the giant branch and it appears that there is
a slow-down in red-clump stars compared with stars on the ascending part of the
giant branch.

The most Sun-like stars that have been studied are the two components of the wide
binary system 16~Cyg. \citet{daviesetal2015} found that the data only allow
the estimation of an average rotation rate. 16~Cyg~A was found to have a rotation 
period of $23.8^{+1.5}_{-1.8}$~days, while the rotation rate of 16~Cyg~B is
$23.2^{+11.5}_{-3.2}$~days. 

The study of other Sun-like stars have not yet
yielded any notable results. There have however, been several investigations
into how one could study the internal rotation of these stars. Studies include
that of \citet{2014lund} who studied the the impact of different differential 
rotation profiles on the splittings of p-mode oscillation frequencies, and
\citet{2016hannah} who examined the sensitivity of asteroseismic
rotation inversions to uncertainties in the data. \citet{2016hannah2} also
examined whether one could find evidence of radial differential rotation in
an average-sense by inverting data for an ensemble of stars. This has not
yet been applied to real data yet.

\citet{gizonsolanki2004} had shown how one might be able to determine
differential rotation of a star, i.e., variation of rotation with latitude,
using seismic data, however results have been elusive. The data are 
not sensitive enough. This is not completely surprising
according to \citet{ballotetal2006}, but disappointing nevertheless.

A summary of recent developments in this field can be found in
\citet{aerts2015}.

\newpage

\section{Concluding Thoughts}
\label{sec:concl}

Helioseismology has revolutionised our knowledge of the Sun. It has allowed us
to  know the structure and dynamics of the solar interior very well. However,
what is clearly missing is better knowledge of the near-surface layers of the
Sun. Accurate and precise frequencies of  high-degree will resolve this issue.
Additionally, a concerted push to observe low-frequency low-degree p~modes and
g~modes will help clarify some of the issues that remain about the structure
and dynamics of the solar core.

Our database of helioseismic observations does not go very far back in time.
There have been continuous observations of low-degree modes for the last 30
years, and of intermediate degree modes for only 20 years. These observations
have given us tantalising glimpses of how the Sun changes over a solar cycle,
and on how one cycle can be very different from another. A 20-year database is
definitely not long enough to learn what the interior of the Sun behaves like
in a ``typical'' cycle, assuming of course, that a typical cycle can be
defined. We thus need to continue these observations as long as we can.

Asteroseismology is in an ascending phase. These are exciting times with newer
and better data being available each day. However, in the rush to exploit these
data, we seem to have forgotten that we need really good analysis techniques to
make optimum use of the data -- a case in point is the treatment of the surface
term. Now that the data stream from \textit{Kepler} has slowed down, and TESS
and PLATO are still a few years off, this is perhaps the time to figure out
newer and better techniques of analysing these data.

\section*{Acknowledgements}
I would like to thank the two anonymous referees who have helped improve this
review.  I would also like to thank Charles Baldner, Anne-Marie Broomhall,
William  Chaplin, Rachel Howe and Kiran Jain for preparing some of the figures
that appear in this review, and thanks are due to Tim Larson for early access
to reprocessed MDI data.

\newpage

\bibliography{ms}

\end{document}